\documentclass[trackchanges, twocolumn, tighten]{aastex701}

\AtBeginDocument{%
    \setlength{\parindent}{0pt}%
    \setlength{\parskip}{0.5\baselineskip}%
}

\usepackage{amsmath}

\usepackage{tikz}
\usetikzlibrary{arrows.meta,positioning,shapes.misc,shapes.symbols,shapes.geometric}

\tikzset{
  box/.style={draw, rounded corners, align=center, minimum width=6.8cm, minimum height=0.9cm, thick},
  arrow/.style={-{Latex[length=3mm]}, thick}
  }

\usepackage{xspace}
\usepackage{graphicx}
\usepackage{float}

\newcommand{\MS}{M\textsubscript{$\odot$}}    % Solar Mass
\newcommand{\RS}{R\textsubscript{$\odot$}}    % Solar Radius
\newcommand{\RE}{R\textsubscript{$\oplus$}}   % Earth Radius
\newcommand{\Mstar}{M\textsubscript{$\star$}} % Star Mass
\newcommand{\Rstar}{R\textsubscript{$\star$}} % Star Radius

\newcommand{\degrees}{\ensuremath{^{\circ}}}

\colorlet{MYCOLOR}{green!25!red!75!}
\colorlet{mycolor}{green!25!red!75!}
\newcommand{\NEMESIS}{\textcolor{mycolor}{\textsf{NEMESIS}}\xspace} % was a trailing space inside the braces

\AtBeginDocument{%
}

\newcommand{\NEMESISGitHub}{\url{https://github.com/daxfeliz/NEMESIS}}
\newcommand{\NEMESISDocs}{\url{https://nemesis-tess.readthedocs.io/}} % CHECK the link resolves; Read the Docs hostnames use hyphens
\newcommand{\leovetter}{\texttt{LEO-Vetter}}
\newcommand{\runin}[1]{\par\smallskip\noindent\textit{#1.}\enspace}
\newcommand{\DistMax}{100}

\newcommand{\NTargetList}{156{,}144}        % Table 1, Cut 5 = Table 7 first row
\newcommand{\NObsTargets}{115{,}384}        % Fig2_provenance.json; Figure 3 bars sum to it
\newcommand{\NObsStarSectors}{219{,}023}    % 213,658 + 4,482 + 883
\newcommand{\CoverageFrac}{97.6\%}          % 213,658 / 219,023
\newcommand{\NFailCutout}{4{,}482}
\newcommand{\NLCss}{214{,}541}
\newcommand{\NFailSearch}{883}              % = \NnoBLS + \NTLSskip
\newcommand{\NnoBLS}{872}
\newcommand{\NTLSskip}{11}
\newcommand{\NStarSectors}{213{,}658}       % searched star-sectors
\newcommand{\NSearchedTICs}{113{,}047}
\newcommand{\NMetricCompleteSS}{213{,}653}  % complete metric set; rate denominator
\newcommand{\NIndeterminate}{5}
\newcommand{\SearchPeriodRes}{\ensuremath{2\times10^{-3}}}
\newcommand{\FracSurveyTCE}{11\%}           % 23,339 / 213,658
\newcommand{\FracSurveySine}{0.4\%}

\newcommand{\MannMKsMin}{4.5}
\newcommand{\MannMKsMax}{10.0}
\newcommand{\MannMatchTol}{1}
\newcommand{\NStarSecMannConsistent}{211{,}346}
\newcommand{\FracMannConsistent}{98.9}

\newcommand{\MESthresh}{6.2}
\newcommand{\NTransitMin}{3}
\newcommand{\TCEs}{23{,}339}
\newcommand{\FAs}{22{,}599}
\newcommand{\FPs}{182}
\newcommand{\PCs}{558}                      % PC detections
\newcommand{\PCfrac}{2.4\%}                 % 558 / 23,339
\newcommand{\TCEsTIC}{18{,}933}
\newcommand{\FAsTIC}{18{,}451}
\newcommand{\FPsTIC}{145}
\newcommand{\PCsTIC}{527}
\newcommand{\NFAonly}{14{,}192}
\newcommand{\NFAandFP}{8{,}407}

\newcommand{\NCentroidWin}{3}
\newcommand{\CentroidHalfWindow}{1.5}
\newcommand{\NCentroidFlag}{2{,}835}
\newcommand{\NCentroidIndet}{2}
\newcommand{\NFluxPC}{610}
\newcommand{\NCentroidOnlyPC}{52}

\newcommand{\NSymFlagged}{6{,}722}
\newcommand{\NSymRedundant}{6{,}589}
\newcommand{\NSymUnique}{133}
\newcommand{\NSymPClost}{85}
\newcommand{\CDPPratioMin}{0.01}

\newcommand{\NfWeak}{0}
\newcommand{\NfNonUnique}{6{,}436}
\newcommand{\NfChases}{15{,}537}
\newcommand{\NfBadShape}{54}
\newcommand{\NfSingleEvent}{15{,}671}
\newcommand{\NfDMM}{5{,}839}
\newcommand{\NfSinusoidal}{557}
\newcommand{\NfUnphysDur}{3{,}344}
\newcommand{\NfAsym}{12}
\newcommand{\NfChi}{18{,}354}
\newcommand{\NfDataGap}{214}
\newcommand{\NfOddEven}{3{,}929}
\newcommand{\NfVshaped}{696}
\newcommand{\NfLarge}{1{,}164}
\newcommand{\NfSecondary}{1{,}966}
\newcommand{\NsecFired}{2{,}960}

\newcommand{\EMAX}{0.495}
\newcommand{\EMAXpct}{99}
\newcommand{\EMAXleakPct}{1}
\newcommand{\EMAXpf}{0.349}
\newcommand{\EMAXall}{0.415}
\newcommand{\EMAXphaseHalf}{0.315}          % 2 x 0.495 / pi
\newcommand{\NEBcal}{936}
\newcommand{\EBmorphMax}{0.5}
\newcommand{\NEBadmitted}{0.4}
\newcommand{\NsecExempt}{13}
\newcommand{\NsecOnly}{60}
\newcommand{\NsecOnlyHalf}{37}
\newcommand{\SecHalfWidth}{0.064}
\newcommand{\SecUniformPct}{13}             % 2 x 0.064

\newcommand{\BfAICmarginLow}{60}
\newcommand{\BfAICmarginHigh}{30}
\newcommand{\BfNtrSplit}{10}
\newcommand{\BfLowMESedge}{10}
\newcommand{\BfInjLowRej}{221}
\newcommand{\BfInjLowN}{234}
\newcommand{\BfInjLowPct}{94\%}             % 221 / 234
\newcommand{\BfInjAllPct}{14.7\%}
\newcommand{\KunBfInjPct}{1.6\%}
\newcommand{\BfBlindPC}{483}                % blind PC detections bad_fit would remove, of 558
\newcommand{\BfKnownCP}{12}                 % confirmed or known planets it would reject, of \KPnCPPC
\newcommand{\BfLeakCaught}{33}

\newcommand{\BfLeakLowMES}{28}
\newcommand{\BfLowMESmax}{9}
\newcommand{\BfInjLowRateMin}{93\%}
\newcommand{\BfHighMES}{12}
\newcommand{\BfHighLeakCaught}{5}
\newcommand{\BfHighLeakN}{9}
\newcommand{\BfHighInjPct}{3.6\%}

\newcommand{\PCsUnique}{525}                % signals; 530 reconciled minus 5 unconverged
\newcommand{\PCsMultiSector}{17}            % includes TIC 454291823 (S11, S12)
\newcommand{\PCsMultiCluster}{3}            % TIC 454291823 is one signal
\newcommand{\ReconNMax}{5}
\newcommand{\ReconTauP}{10^{-2}}            % math mode only
\newcommand{\ReconFT}{1}
\newcommand{\ReconSigPFloor}{10^{-3}}       % math mode only
\newcommand{\NMinEpochs}{2}
\newcommand{\ReconMaxGap}{700}
\newcommand{\ReconNLinks}{48}
\newcommand{\ReconPeriodOnlyLinks}{1}
\newcommand{\ReconNullTrials}{200}
\newcommand{\ReconNullMean}{3.65}
\newcommand{\NClosePairs}{2}

\newcommand{\NUnconverged}{5}
\newcommand{\NFitConverged}{525}            % candidates (the catalog)
\newcommand{\newPCs}{482}                   % candidates on stars with no TOI, CTOI, or Archive host (TIC match)
\newcommand{\NnewHosts}{479}                % stars hosting them
\newcommand{\NPCknownHost}{43}              % candidates on a TOI, CTOI, or Archive host
\newcommand{\NCatalogHosts}{522}            % 479 + 43
\newcommand{\NLarge}{125}                   % candidates with refit R_P > 8
\newcommand{\NmatchTOI}{34}                 % known-host candidates on TOI hosts; 2026-10-12 lists
\newcommand{\NmatchCTOI}{13}                % on CTOI hosts
\newcommand{\NmatchBoth}{4}                 % on stars hosting both; 34 + 13 - 4 = 43
\newcommand{\NmatchArchive}{26}             % on stars with an Archive transiting planet; none hosts one alone
\newcommand{\NCatalog}{\NFitConverged}

\newcommand{\NTriEvaluated}{\NFitConverged}

\newcommand{\NCatalogNew}{\newPCs}
\newcommand{\NCatalogNewHosts}{\NnewHosts}

\newcommand{\NCatalogKnown}{\NPCknownHost}
\newcommand{\NCatalogLarge}{\NLarge}

\newcommand{\minP}{0.51}                    % days; 525 candidates, jaxoplanet medians
\newcommand{\maxP}{9.10}
\newcommand{\minRP}{0.74}                   % R_earth
\newcommand{\maxRP}{28.71}                  % TIC 95112238 S6, a known host; not drawn in the period-radius diagram
\newcommand{\NSmall}{259}                   % refit R_P < 4 R_earth
\newcommand{\fracSmall}{49\%}               % 259 / 525 = 49.3%
\newcommand{\NSmallTLS}{318}                % TLS-depth R_P < 4 R_earth, adopted sector
\newcommand{\fracSmallTLS}{61\%}            % 318 / 525 = 60.6%
\newcommand{\NUSP}{57}                      % P < 1 d; refit and TLS agree
\newcommand{\MedRpMESlow}{3.3}              % median TLS radius, the 408 detections with MES 6.2 to 8
\newcommand{\MedRpMEShigh}{3.7}             % median TLS radius, the 150 detections with MES >= 8
\newcommand{\ExpFAfracRpClass}{63\%}        % share of expected false alarms with TLS R_P < 4 R_earth, 558 detections

\newcommand{\NOversample}{7}
\newcommand{\NChains}{4}
\newcommand{\NTune}{2{,}000}
\newcommand{\NDraws}{3{,}000}
\newcommand{\TargetAccept}{0.95}
\newcommand{\RhatMax}{1.01}
\newcommand{\ESSmin}{400}
\newcommand{\RhoFracErr}{15}
\newcommand{\LDsigma}{0.10}
\newcommand{\DurLogSigma}{0.20}
\newcommand{\SigLnPmin}{\ensuremath{10^{-4}}}
\newcommand{\SigLnPmax}{0.05}
\newcommand{\SigLnPdefault}{\ensuremath{10^{-3}}}
\newcommand{\StudentNu}{4}
\newcommand{\FitWindowDur}{6}
\newcommand{\FitWindowMinCad}{50}
\newcommand{\NCeilFinal}{104}               % R_P/R_star limited by its prior (Fit_flag 1)
\newcommand{\GrazExRpTLS}{1.92}             % TIC 266307162 S10, Rp_TLS 1.921 (table3_mrt, 11-14)
\newcommand{\GrazExRpFit}{9.67}             % same row, refit Rp 9.674

\newcommand{\NGalleryPanels}{21}            % 7 x 3 grid, 11-12 render
\newcommand{\NGalleryPerBin}{3}             % quota per MES bin before reallocation

\newcommand{\TriVersion}{1.0.20}            % installed triceratops package; page 7 subtitle shows wrapper 1.3.0
\newcommand{\NTriTrials}{20}
\newcommand{\TriMCsamples}{\ensuremath{5\times10^{6}}}
\newcommand{\TriExptime}{30}
\newcommand{\TriWindowDur}{3}
\newcommand{\TriWindowWide}{5}
\newcommand{\TriWindowPhase}{0.45}
\newcommand{\TriOOTdur}{0.75}
\newcommand{\TriRadiusCeiling}{8}
\newcommand{\FPPmax}{0.5}
\newcommand{\NFPPmax}{10^{-3}}              % math mode only
\newcommand{\NFPPnfp}{0.1}
\newcommand{\TriScatter}{0.002}
\newcommand{\NLikelyPlanet}{105}
\newcommand{\NLikelyNFP}{177}
\newcommand{\NTriNoDip}{1}
\newcommand{\NLikelyNFPall}{178}            % 177 + 1 whose target cannot produce the depth
\newcommand{\NFPfavored}{81}
\newcommand{\NTriUncertain}{161}            % 157 + 3 + 1
\newcommand{\NTriNearbyOpen}{157}
\newcommand{\NTriSplit}{3}
\newcommand{\NTriNegFlux}{1}                % TIC 401320705 S25
\newcommand{\NTriNearbyFrac}{64\%}          % (178 + 157) / 525
\newcommand{\NTriTOI}{33}                   % Section 5.3 test on the 525 (31 direct, 2 via CTOIs promoted to TOIs); TOI_catalog_2026-10-12-exofop.csv
\newcommand{\NTriTOIknown}{24}              % CP 22 + KP 2 among the \NTriTOI
\newcommand{\NTriTOIknownLP}{13}            % of those, likely planet
\newcommand{\NTriTOIknownNbr}{6}            % of those, inconclusive (nearby star not ruled out)
\newcommand{\NTriTOIknownNFP}{2}            % of those, likely nearby false positive
\newcommand{\NTriTOIknownFPF}{2}            % of those, false positive favored; 1 more has split trials
\newcommand{\NLargeLP}{10}                  % UNCONFIRMED: 10-20 tally, before the 10-21 rebuild and 11-01 merge; Batch 11 item 1
\newcommand{\NValidated}{5}                 % FPP < 0.015 and NFPP < 1e-3 at the median and 84th percentile
\newcommand{\NTriPlotted}{524}              % 525 minus the candidate with no finite trial
\newcommand{\TriPlotFloor}{10^{-5}}         % math mode only
\newcommand{\NNFPPzero}{62}
\newcommand{\TriMedFPPPlanet}{0.073}
\newcommand{\TriMedFPPNFP}{0.88}
\newcommand{\TriMedNFPPNFP}{0.40}
\newcommand{\TriDepthShallow}{13\%}
\newcommand{\TriDepthShallowOK}{10\%}
\newcommand{\NTriDepthMoreHosts}{108}
\newcommand{\NTriDepthFewerHosts}{19}
\newcommand{\TOIpubRpNEM}{5.19}
\newcommand{\TOIpubRp}{5.14}
\newcommand{\TOIpubRpErr}{0.16}
\newcommand{\NFavored}{262}                 % 105 + 157
\newcommand{\NNewFavored}{232}
\newcommand{\NNewFavoredClear}{89}
\newcommand{\NNewFavoredNbr}{143}           % 89 + 143 = 232
\newcommand{\NNewFavoredLowMES}{207}
\newcommand{\NNewFavoredHighMES}{25}        % 207 + 25 = 232
\newcommand{\NNewFavoredLarge}{23}

\newcommand{\NRepStars}{137}
\newcommand{\NRepTmagMin}{6.74}
\newcommand{\NRepTmagMax}{17.63}
\newcommand{\NRepSectors}{26}
\newcommand{\repFCRmax}{0.10}
\newcommand{\repSineMax}{15}
\newcommand{\repNbins}{30}
\newcommand{\repNquantiles}{5}
\newcommand{\FCRcutFrac}{65\%}

\newcommand{\minInjP}{1.0}
\newcommand{\maxInjP}{10.0}
\newcommand{\minInjRP}{0.5}
\newcommand{\maxInjRP}{11.0}
\newcommand{\NInjPeriodSteps}{12}
\newcommand{\NInjRadiusSteps}{12}
\newcommand{\NInjCells}{144}
\newcommand{\NInjTotal}{19{,}728}           % 12 x 12 x 137
\newcommand{\NInjPerCell}{137}
\newcommand{\NInjPerFigCell}{548}           % per displayed 2 x 2 cell of the coarse-grid figure
\newcommand{\WilsonHalf}{4.3 percentage points}
\newcommand{\fdrift}{1.0}
\newcommand{\BLSPeriodErr}{\ensuremath{10^{-4}}}
\newcommand{\SigmaPRatio}{twenty times}
\newcommand{\NztestPCnotRec}{1{,}914}
\newcommand{\NztestPeriodOK}{1{,}855}
\newcommand{\ZtestEpochMatch}{98.4\%}
\newcommand{\RecZtestBLS}{27.3\%}
\newcommand{\RecZtestTLS}{43.0\%}
\newcommand{\RecZtestGap}{15.7 percentage points}
\newcommand{\RecCohBLS}{41.4\%}
\newcommand{\RecCohTLS}{41.6\%}
\newcommand{\RecCohGap}{0.2 percentage points}
\newcommand{\NfdriftFlipTLS}{43}
\newcommand{\NfdriftFlipBLS}{28}
\newcommand{\NInjRecRad}{8{,}209}
\newcommand{\RadRecMed}{+1.4}
\newcommand{\RadRecLo}{-7.0}
\newcommand{\RadRecHi}{+5.7}
\newcommand{\RadRecBinMin}{-3.1}
\newcommand{\RadRecBinMax}{+2.2}
\newcommand{\RadRecBinRlo}{1.1}
\newcommand{\RadRecBinRhi}{11.0}
\newcommand{\RadRecSmallMin}{1.3}
\newcommand{\RadRecSmallMax}{3.1}
\newcommand{\RadRecSmallRmax}{2.4}

\newcommand{\ScrFrac}{10\%}                 % printed as "about \ScrFrac{}"
\newcommand{\NScrDone}{20{,}345}            % the scrambled sample
\newcommand{\NScrTCEsRaw}{6{,}310}
\newcommand{\NScrCleaned}{62}
\newcommand{\NScrTCEs}{6{,}248}             % 6,310 - 62
\newcommand{\NScrLeak}{37}
\newcommand{\MESbinWeakHi}{8}
\newcommand{\Effval}{0.9941}                % 6,211 / 6,248
\newcommand{\Efflo}{0.9930}
\newcommand{\Effhi}{0.9950}
\newcommand{\RFAval}{0.818}
\newcommand{\RFAlo}{0.785}
\newcommand{\RFAhi}{0.846}
\newcommand{\NPCweakbin}{408}
\newcommand{\RFAweakbin}{0.23}
\newcommand{\NPCstrongbin}{150}             % 408 + 150 = 558
\newcommand{\NexpFakePC}{329}
\newcommand{\NexpFakeStrong}{14}

\newcommand{\NInjTCErec}{7{,}696}
\newcommand{\NInjPCrec}{6{,}429}
\newcommand{\Cvetval}{0.835}                % 6,429 / 7,696

\newcommand{\CvetQLPpm}{69.9\%}
\newcommand{\CvetQLPemOne}{88.3\%}
\newcommand{\NDEparams}{12}
\newcommand{\NSplitRepeats}{5}
\newcommand{\SplitShrink}{85}

\newcommand{\NOutGrid}{76}                  % detections of the 525 signals outside the injection grid
\newcommand{\OccMinP}{\minInjP}             % whole injection grid [Dax decided 11-11]
\newcommand{\OccMaxP}{\maxInjP}
\newcommand{\OccMinRP}{\minInjRP}
\newcommand{\OccMaxRP}{\maxInjRP}
\newcommand{\IntOcc}{0.81}                  % weight R_FA (1 - FPP): 0.8096 +0.4223 / -0.4368; lower limit
\newcommand{\IntOccErr}{0.44}               % larger side; the +0.42 side is a literal in Section 5.2
\newcommand{\OccRate}{0.94}                 % weight R_FA alone: 0.9410 +0.4806 / -0.5155; lower limit

\newcommand{\KPnHostTCE}{341}
\newcommand{\KPnMatch}{127}                 % UNCONFIRMED: 1.5 h floor, test run kpa_floor_test_20261110_160231; Batch 11 item 3
\newcommand{\KPnCmp}{62}
\newcommand{\KPnCmpPC}{40}
\newcommand{\KPnCmpExp}{39.7}               % 39.7377
\newcommand{\KPnCmpSD}{3.0}                 % 2.9743
\newcommand{\KPcmpZ}{0.09}
\newcommand{\KPnCP}{50}
\newcommand{\KPnCPPC}{34}
\newcommand{\KPnCPExp}{33.2}                % 33.1818
\newcommand{\KPnCPSD}{2.6}                  % 2.6351
\newcommand{\KPcpZ}{0.31}
\newcommand{\KPratio}{1.01}                 % 40 / 39.7377
\newcommand{\KPratioErr}{0.07}              % 2.9743 / 39.7377 = 0.0748 (Batch 8, 11-07)
\newcommand{\KPhighMES}{15}
\newcommand{\KPinjHighFrac}{78}
\newcommand{\KPinjHighC}{0.90}
\newcommand{\KPinjLowC}{0.16}
\newcommand{\NknownFootprint}{190}          % 0.5 <= P <= 10 d; TFOPWG PC/APC/CP/KP TOIs + Archive planets on stars with no TOI or CTOI; searched stars
\newcommand{\NknownRecovered}{30}           % PC detections matched by the Section 5.3 test, incl. via promoted CTOIs
\newcommand{\FracKnownRecovered}{16\%}      % 30 / 190 = 15.8%

\newcommand{\TICdeepDepth}{683}

\newcommand{\TICdeepShare}{3.6\%}
\newcommand{\TICdeepDip}{2.5\%}             % 683 ppt x 0.036
\newcommand{\TICdeepNFPP}{0.93}
\newcommand{\TICbigRp}{12}
\newcommand{\TICbigFPP}{0.9995}
\newcommand{\TIClowShare}{3.5\%}            % TIC 401320705 S25, 1 - f
\newcommand{\TIClowShareAmp}{28}            % UNCONFIRMED: 1/0.035 = 28.6; Batch 11 item 5
\newcommand{\NPoorlyVetted}{1}              % TIC 401320705 S25
\newcommand{\LNNemin}{0.40}
\newcommand{\LPemin}{0.18}
\newcommand{\KPlninetyeightSectors}{six}
\newcommand{\PaperICrowdMag}{15}

\newcommand{\PaperINewPC}{24}
\newcommand{\NPaperIOnTargets}{23}
\newcommand{\PaperIminP}{1.25}
\newcommand{\PaperImaxP}{6.84}
\newcommand{\PaperIminRP}{1.26}
\newcommand{\PaperImaxRP}{5.31}

\newcommand{\KcurveMstar}{0.33}             % M_sun; median host mass of the 262 planet candidates
\newcommand{\NHZtoi}{8}                     % 11-10 render, TOIs by TFOPWG disposition (PC, APC, CP, KP)
\newcommand{\NHZctoi}{4}                    % CTOIs in the optimistic HZ, as drawn

\newcommand{\PLDorder}{3}
\newcommand{\PLDnPCA}{3}
\newcommand{\PLDvarThr}{95\%}
\newcommand{\PLDmaxCombos}{5{,}000}
\newcommand{\PLDdiagReg}{\ensuremath{10^{-8}}}
\newcommand{\PLDgpWindow}{0.5}
\newcommand{\PLDgateNMAD}{15}
\newcommand{\PLDgateWindow}{1.0}
\newcommand{\PLDdipSigma}{4}
\newcommand{\PLDdipGrow}{1}
\newcommand{\PLDdipIter}{3}
\newcommand{\PLDmaskCap}{50\%}

\newcommand{\PCsToFollowUp}{13}        % new PCs, MES >= 8, above Kempton TSM threshold
\newcommand{\TSMs}{90}                 % Kempton TSM threshold, 1.5 to 10 R_earth
\newcommand{\ESMs}{7.5}                % Kempton ESM threshold, terrestrial planets only
\newcommand{\TSMsTerr}{10}             % Kempton TSM threshold, R_P < 1.5 R_earth
\newcommand{\RpTerr}{1.5}              % R_earth, Kempton terrestrial bin edge
\newcommand{\MpTerr}{2.9}              % M_earth, CK17 mass at 1.5 R_earth (2.86)
\newcommand{\TSMRmax}{10}              % R_earth, TSM undefined at and above
\newcommand{\CKRmax}{14.3}             % R_earth, no CK17 mass above
\newcommand{\NFollowNoMass}{5}         % of the 262, manuscript_counts e
\newcommand{\NFollowTSMclean}{10}      % of \PCsToFollowUp, not grazing, manuscript_counts c
\newcommand{\NFollowTSMsmall}{8}       % of those, R_P < 8, manuscript_counts c
\newcommand{\NFollowTSMlarge}{2}       % of those, R_P > 8, both P < 1 d, manuscript_counts c
\newcommand{\FollowRpLo}{2.5}          % R_earth, 2.488, manuscript_counts c
\newcommand{\FollowRpHi}{5.6}          % R_earth, 5.622, manuscript_counts c
\newcommand{\FollowRpMinHigh}{2.1}     % R_earth, 2.0799, manuscript_counts b
\newcommand{\NFollowKmass}{24}         % of the 25 with an inferred mass
\newcommand{\FollowKmin}{2.9}          % m/s, 2.876
\newcommand{\FollowKmed}{10.2}         % m/s, 10.231
\newcommand{\FollowJmed}{11.2}         % 11.168
\newcommand{\NFollowNbr}{14}           % of the 25, nearby star not ruled out, manuscript_counts a
\newcommand{\NFollowTerrLowMES}{3}     % new, R_P < 1.5, TSM > 10, all MES < 8, manuscript_counts d

\submitjournal{The Astronomical Journal}
\begin{document}

\title{\NEMESIS: Exoplanet Tra\textcolor{mycolor}{N}sit Surv\textcolor{mycolor}{E}y of Nearby \textcolor{mycolor}{M}-dwarfs in T\textcolor{mycolor}{ES}S FF\textcolor{mycolor}{IS} II\\ -- \NFavored{} Transiting Planet Candidates from the TESS Primary Mission}

% affiliation list
% ---- Affiliations -------------------------------------------------------------
\newcommand{\CCA}{Center for Computational Astrophysics, Flatiron Institute, 162 Fifth Avenue, New York, NY 10010, USA}
\newcommand{\AMNH}{Department of Astrophysics, American Museum of Natural History, 200 Central Park West, New York, NY 10024, USA}
\newcommand{\Vanderbilt}{Department of Physics and Astronomy, Vanderbilt University, Nashville, TN 37235, USA}
\newcommand{\Fisk}{Department of Physics, Fisk University, 1000 17th Avenue North, Nashville, TN 37208, USA}
\newcommand{\GeorgeMason}{George Mason University, 4400 University Drive, Fairfax, VA 22030, USA}
\newcommand{\UCostaRica}{Escuela de Física, Universidad de Costa Rica, 11501-2060 San José, Costa Rica}
\newcommand{\ASU}{Arizona State University School of Earth and Space Exploration, P.O. Box 871404, Tempe, AZ 85287, USA}
\newcommand{\Barnard}{Department of Physics and Astronomy, Barnard College of Columbia University, New York, NY 10027, USA}
\newcommand{\Columbia}{Astronomy Department, Columbia University, New York, 10027, New York, USA.}
\newcommand{\NYU}{Center for Cosmology and Particle Physics, Department of Physics, New York University, New York, NY
10003, USA}
\newcommand{\SRMP}{Science Research Mentoring Program, American Museum of Natural History, 200 Central Park West, New York, NY 10024, USA}

\newcommand{\Princeton}{Princeton University, 1 Nassau Hall, Princeton, NJ 08544-0070, USA} % is this right?
\newcommand{\BU}{Boston University, 233 Bay State Road, Boston, MA 02215}
\newcommand{\Hunter}{Hunter College, The City University of New York, 695 Park Avenue, New York, NY 10065, USA}
\newcommand{\Villanova}{Villanova University, 800 Lancaster Ave, Villanova, PA 19085}
\newcommand{\STScI}{Space Telescope Science Institute, 3700 San Martin Drive, Baltimore, MD 21218, USA} % is this right?

%author list
% it me, i wrote dis
\author[orcid=0000-0002-2457-7889,gname=Dax,sname=Feliz]{Dax L. Feliz}
\email[show]{dfeliz@flatironinstitute.org}
\affiliation{\CCA}
\affiliation{\AMNH}
\correspondingauthor{Dax L. Feliz }

% Summer students
%consented
\author[orcid=0009-0009-8668-3060,gname=Madeline,sname=Maldonado Gutierrez]{Madeline Maldonado Gutierrez}
\email{mjm2418@alum.barnard.edu} % TODO: confirm email
\affiliation{\Barnard}

%consented
\author[orcid=0000-0002-9947-7348,gname=Maliyah,sname=Adams]{Maliyah Adams}
% \altaffiliation{Wyoming Stargazing, Teton Science Schools, Jackson, WY 83002, USA}
\affiliation{\ASU}
\email{maliyahxadams@icloud.com}

% SRMP students
%consented
\author[orcid=0009-0009-1161-8942,gname=Shreeya,sname=KC]{Shreeya KC}
\email{shreeyakc07@gmail.com} % TODO: confirm email
\affiliation{\SRMP}
\affiliation{\Princeton}

%consented
\author[orcid=0009-0003-2071-3577,gname=Kylor,sname=Ghai]{Kylor G. Ghai}
\email{kghai@bu.edu} % TODO: confirm email
\affiliation{\SRMP}
\affiliation{\BU}

% consented 
\author[orcid=0009-0009-5931-9298,gname=Thamim,sname=Chowdhury]{Thamim Chowdhury}
\email{THAMIM.CHOWDHURY13@login.cuny.edu} % TODO: confirm email
\affiliation{\Hunter}
\affiliation{\SRMP}

% consented
\author[orcid=0009-0009-6542-9098,gname=August,sname=Fischer]{August Fischer}
\email{august.fischer@villanova.edu} % TODO: confirm email
\affiliation{\SRMP}
\affiliation{\Villanova}

% consented
\author[orcid=0009-0009-4119-5792,gname=Jashcelyn Sescon,sname=Canada]{Jashcelyn S. Canada}
\email{jashcelyncanada@gmail.com}
\affiliation{\SRMP}
\affiliation{\BU}

% NOT consented but will be given credit up until conset required for submission — no ORCID yet, drop optional arg
\author[gname=Donovan,sname=Bradley]{Donovan Bradley}
\email{dmbradley1023@gmail.com} % TODO: confirm email if reinstated
\affiliation{\SRMP}

% NEMESIS I team

% graduated, now HS teacher — consented, AMNH and BDNYC
\author[orcid=0000-0002-5000-9316,gname=Mary,sname=Jimenez]{Mary Jimenez}
\email{mjimenez1@amnh.org}
\affiliation{\AMNH}

% Vandy group (NEMESIS I) — consented
\author[orcid=0000-0001-9729-9413,gname=Samantha,sname=Bianco]{Samantha N. Bianco}
\email{sbianco@stsci.edu} % TODO: confirm email
\affiliation{\STScI}

% Summer Student (NEMESIS I) — consented, no longer in academia
\author[orcid=0000-0002-4788-5147,gname=Bryan,sname=Villarreal Alvarado]{Bryan Villarreal Alvarado}
\email{bryanjva@gmail.com} % 
\affiliation{\UCostaRica}

% consented 
\author[orcid=0000-0002-3481-9052,gname=Keivan,sname=Stassun]{Keivan G.\ Stassun}
\email{keivan.stassun@vanderbilt.edu}
\affiliation{\Fisk}
\affiliation{\Vanderbilt}

% consented
\author[orcid=0000-0002-8864-1667,gname=Peter,sname=Plavchan]{Peter Plavchan}
\email{pplavchan@gmu.edu} % TODO: confirm email if reinstated
\affiliation{\GeorgeMason}

% consented
\author[orcid=0000-0003-2781-3207,gname=Kevin,sname=Collins]{Kevin I. Collins}
\email{kcollins@gmu.edu} % TODO: confirm email
\affiliation{\GeorgeMason}

% consented
\author[orcid=0000-0003-4540-5661,gname=Ruth,sname=Angus]{Ruth Angus}
\email{rangus@amnh.org}
\affiliation{\AMNH}
% \affiliation{\Columbia}

%NOT consented
% \author[orcid=0000-0003-2866-9403,gname=David,sname=Hogg]{David Hogg}
% \email{david.hogg@nyu.edu}
% \affiliation{\CCA}
% \affiliation{\NYU}

\begin{abstract}
\noindent
We present NEMESIS~II, a blind transit search of nearby M-dwarfs in TESS Primary Mission full-frame images (Sectors 1--26), including stars lacking 2-minute observations. Using the open-source \NEMESIS pipeline, we extracted, detrended, and searched \NStarSectors{} single-sector light curves of \NSearchedTICs{} M-dwarfs within \DistMax{}~pc, yielding \TCEs{} threshold crossing events. Automated \leovetter{} vetting, applied identically to the survey and its injection-recovery and orbit-scrambling controls, retains \PCs{} detections corresponding to \PCsUnique{} distinct signals with converged transit fits and periods of \minP{}--\maxP{}~d. Of these, \NCatalogNew{} lie on stars without TOI, CTOI, or NASA Exoplanet Archive transiting-planet entries. Orbit-scrambling predicts \NexpFakePC{} false alarms among the \PCs{} detections, all but \NexpFakeStrong{} of them among the \NPCweakbin{} below Multiple Event Statistic \MESbinWeakHi{}. Assuming an astrophysical signal, \texttt{TRICERATOPS} favors a planetary origin for \NFavored{} of \NTriEvaluated{} candidates (\NNewFavored{} new), which we classify as planet candidates; it also excludes resolved-neighbor eclipses for \NLikelyPlanet{}. Known planets and TOI candidates survive vetting at \KPratio{}~$\pm$~\KPratioErr{} of the injection-predicted rate. Correcting for each light curve's sensitivity, false alarms, and astrophysical false positives, we infer $\IntOcc{} \pm \IntOccErr{}$ planets per star for $P\in[\OccMinP{},\OccMaxP{}]$~d and $R_P\in[\OccMinRP{},\OccMaxRP{}]$~\RE{}. This is a lower limit, because our injections recover few or no planets near and below 1~\RE{} and assume central transits. Light curves, catalogs, vetting reports, and injection-recovery results will be released as a High Level Science Product at MAST upon publication. 
% \NoteToSelf{Add one sentence on CTOI submission once the rule is decided.} 
Documentation for the NEMESIS pipeline is available at \NEMESISDocs.
\end{abstract}

% TODO: Add CTOI submission statement after finalizing the rule.


\NewPageAfterKeywords

% \vspace*{10\baselineskip}
\section{Introduction} \label{sec:intro}
NASA's Transiting Exoplanet Survey Satellite (TESS; \citealt{Ricker:2015}) launched in April 2018 as the first nearly all-sky, space-based search for transiting exoplanets. Over its two-year Primary Mission, TESS divided the sky into 26 sectors and observed each one for 27.4 days through a $\sim 24^{\circ} \times 96^{\circ}$ field of view, covering roughly 73\% of the sky in the process. Every sector produced two kinds of photometry. Around 20,000 pre-selected stars were observed as small ``postage stamp'' cutouts at a 2-minute cadence, while the entire field was recorded as full-frame images (FFIs) at a 30-minute cadence. Searches of the Primary Mission data produced 2,241 transiting planet candidates, cataloged as TESS Objects of Interest (TOIs; \citealt{Guerrero:2021}). Two official pipelines generate the light curves behind most of those detections: the Science Processing Operations Center pipeline (SPOC; \citealt{Jenkins:2016}), which handles the 2-minute targets, and the Quick-Look Pipeline (QLP; \citealt{Huang:2020, Kunimoto:2022}), developed at the TESS Science Office at MIT to search the FFIs. The distinction between the two data products matters for what follows: the 2-minute target list is a curated subset, while the FFIs record every star that happened to fall on a detector.

\noindent
M dwarfs are the natural place to spend that FFI coverage. They make up roughly 70\% of the stars in the Milky Way \citep{Bochanski:2010}, so a volume-limited sample of the solar neighborhood is mostly made of them. Because they are small, a planet of a given size blocks a larger fraction of the stellar disk, and the transit it produces is correspondingly deeper and easier to recover from noisy photometry. Because they are cool, their habitable zones sit at small orbital separations, so a planet in that region completes several orbits within the 27-day span of a single TESS sector. The difficulty is that these same stars are faint at optical wavelengths, and faintness is precisely the property that keeps most of them off the 2-minute target lists. The nearby M dwarfs we most want to study are, for the most part, recorded only in the FFIs.

\noindent
\textit{Kepler} established the statistical foundation for M-dwarf planet demographics, showing that small, close-in planets are common around low-mass stars and producing the first occurrence rates and the first look at radius-period structure for M-dwarf hosts \citep[e.g.,][and references therein]{Dressing:2015}. Those measurements came from a fixed field of view and a target list weighted toward distant field stars. TESS has since extended this work to a nearer and brighter population \citep{Kunimoto:2022, Eschen:2024, Kunimoto:2025}, and separate efforts have measured the occurrence of giant planets around low-mass stars in the same data \citep{Gan:2023, Bryant:2023}. The picture for nearby M dwarfs is therefore no longer empty, but it is assembled from surveys with different target lists, different photometry, and different vetting, and it is correspondingly difficult to combine. What limits a demographic measurement in this regime is rarely the number of detections. It is knowing the selection function well enough to divide by it. An occurrence rate needs two numbers that are properties of the specific catalog being counted: the fraction of real planets the search would have found, and the fraction of the catalog that is not real. Both depend on the photometry, the search, and the vetting that produced that particular catalog, so neither can be borrowed from another survey.

\noindent
Alternative analyses of FFI transit surveys of low-mass stars by other pipelines include \textsf{ORION}  \citep{Cloutier:2019}, \textsf{DIAMANTE} and its autoregressive DTARPS search (\citealt{Montalto:2020}, \citealt{Montalto:2022}, \citealt{Melton:2024a}, \citealt{Melton:2024b}), QLP's Faint Star search and its M-dwarf extension \citep{Kunimoto:2022, Eschen:2024}, the CNN-based FFI search of \citet{Olmschenk:2021}, giant-planet occurrence and discovery surveys around low-mass stars in TESS FFIs \citep{Gan:2023, Bryant:2023}, and the TESS--Gaia Light Curve GEMS survey \citep{Han:2023, Glusman:2025}. Most recently, \citet{Lafarga:2026} applied the \textsf{RAVEN} pipeline to TESS-SPOC FFI light curves from Sectors 1--55, performing a uniform BLS search across $\sim2.26$ million main-sequence stars well-characterized by Gaia. \textsf{RAVEN} ranks candidates with machine-learning classifiers trained on simulated transits and false positives, and statistically validated 118 planets. \NEMESIS complements this effort by targeting a distance-limited sample of nearby M dwarfs ($d \leq 100$ pc) with custom aperture photometry optimized for faint, red targets, and by providing injection-recovery-based completeness characterization necessary for occurrence rate estimates in the M-dwarf regime.

\noindent
In \citet{Feliz:2021}, hereafter Paper~I, we introduced \NEMESIS as a dedicated framework for extracting and detrending time-series photometry from the 30-minute FFIs and for running a blind, single-sector transit search tailored to nearby M dwarfs. That first survey covered Sectors 1--5 and turned up \PaperINewPC{} new planet candidates, with orbital periods from \PaperIminP{} to \PaperImaxP{}~days and radii from \PaperIminRP{} to \PaperImaxRP{}~\RE. It showed that custom FFI photometry and detrending can pull transit signals out of a regime where cadence, instrumental systematics, and stellar variability would otherwise limit sensitivity. We built \NEMESIS to meet three practical requirements that any FFI catalog intended for demographics has to satisfy. The first is reach: we want the planets around faint or poorly cataloged stars that never make the short-cadence target lists. The second is uniformity: the search and the vetting have to be applied identically to every star, so that the same procedure can be run on injected signals and the completeness measured honestly. The third is fitness for purpose: the light curves and the diagnostics have to be built for M-dwarf variability and for transits lasting a few hours, not for the average field star. In this work we extend the survey to the full Primary Mission baseline, Sectors 1--26. The longer baseline gives us a uniform search across a substantially larger sample, and it lets us
connect what we detect to the demographics of short-period ($P \lesssim 10$~d) planets around nearby M dwarfs.

\noindent
The pipeline has changed since Paper~I in several places that matter at survey scale, and we describe each change where the relevant step is introduced in Section~\ref{subsec:tess}. We now (i) replace the blanket exclusion of every cadence carrying a nonzero TESS QUALITY flag with a narrower bitmask that removes only cadences flagged for spacecraft, pointing, scattered-light, or calibration problems (Section~\ref{subsubsec:SAP}); (ii) tighten the contamination search radius from 3 to 2 TESS pixels to reduce false neighbor associations (Section~\ref{subsubsec:SAP}); (iii) introduce a quaternion engineering telemetry mask to suppress pointing jitter (Section~\ref{subsubsec:masks}); and (iv) add a pixel excision step for resolved neighbor contaminants (Section~\ref{subsubsec:SAP}). We also revised the PLD detrending (Section~\ref{subsubsec:PLD}): the design matrix now carries one reduced block per product order from the second up to order \PLDorder{}, each keeping an adaptive rather than fixed number of singular vectors; cadences dominated by stellar variability are down-weighted in the fit; and cadences with significant flux decrements are iteratively excluded from it.

\noindent
The largest change since Paper~I is in the vetting. Every disposition in this work comes from a single automated rule built on \leovetter{} \citep{Kunimoto:2025}, and no candidate is promoted, demoted, or removed by visual inspection. Manual inspection introduces selection effects that injection-recovery and scramble analyses cannot reproduce, so a catalog vetted by eye cannot be characterized by the completeness and effectiveness measurements that occurrence rates depend on \citep{Kunimoto:2025, Bashi:2026}. 

\noindent
The rest of this paper is organized as follows. Section~\ref{sec:obs} describes how we selected our target stars and how we extract and detrend their photometry from the FFIs. Section~\ref{sec:transit_search} covers the transit search and the automated vetting chain that turns a detection into a disposition. Section~\ref{sec:results} presents our results: the candidate catalog itself, the transit fits and false positive probabilities behind it, and the two control analyses, injection-recovery and the scramble test, that tell us how complete that catalog is and how often noise survives our vetting. Section~\ref{sec:discussion} discusses what the candidates imply for short-period planet demographics around nearby M dwarfs, reports our occurrence rates, compares our detections against the TOI and CTOI catalogs, and sets out the limitations of a
single-sector search. Section~\ref{sec:conclusion} summarizes our conclusions and the work we intend to do next.
\setcounter{footnote}{0}

\section{Observations and Data Reduction} \label{sec:obs}

\subsection{Target Selection}\label{sec:target_selection}
\begin{figure*}
    \centering
    \includegraphics[width=1\textwidth,scale=0.5]{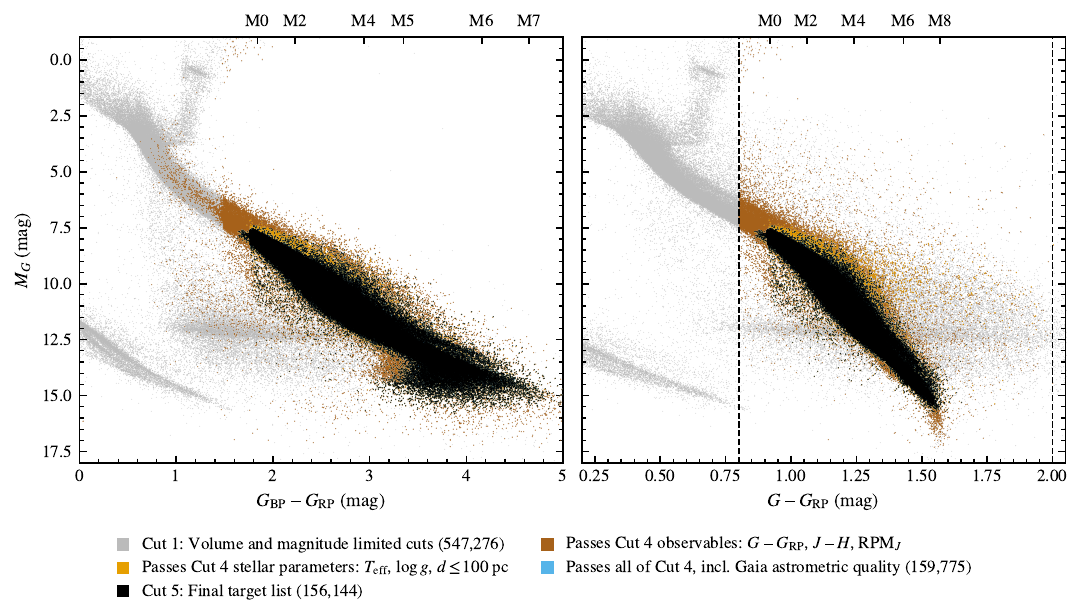}
    \caption{Color-magnitude diagrams of the \NEMESIS~II target selection, showing absolute Gaia $G$ magnitude against $G_{\rm BP} - G_{\rm RP}$ (left) and $G - G_{\rm RP}$ (right). Each color shows the stars surviving a cut of Table~\ref{tab:selection_cuts}, from all TIC sources with $T \le 18$ within 150~pc, before phantom removal (Cut~1, grey), to the final target list (black). The dashed line in the right panel marks the color cut $G - G_{\rm RP} = 0.8$. Spectral types on the top axes follow \citet{Pecaut:2013}.  }
    \label{fig:selection_cut}
\end{figure*}

What we want from this section is a clean, distance-limited list of nearby M-dwarfs that TESS actually observed. We build it in four steps: a deliberately broad query of the TESS Input Catalog, a crossmatch to Gaia DR3 for better astrometry and distances, a set of photometric and kinematic cuts that keep M-dwarfs while rejecting the giants, white dwarfs, and blended solutions that resemble them, and a coverage check that tells us which Primary Mission sectors put each star on silicon. Table~\ref{tab:selection_cuts} gives each cut and the number of targets surviving it, and Figure~\ref{fig:selection_cut} shows the properties of the final sample.

\subsubsection{Target Selection from the TIC}\label{sec:target_selection_TIC}

We began with the TESS Input Catalog (TIC Version 8; \citealt{Stassun:2019}), queried through the CasJobs service at the Barbara A. Mikulski Archive for Space Telescopes (MAST)\footnote{\url{http://mastweb.stsci.edu/mcasjobs/}}, and pulled every source within 150 pc and brighter than $T = 18$. We set this first cut deliberately loose. Our intended volume limit is 100 pc, but TIC distances are not the distances we ultimately trust, and we would rather discard a star later on a Gaia parallax than lose it here on a TIC value. The northern and southern queries were exported separately, and we concatenated them into a single input table. The TIC carries a population of spurious entries known as phantoms: duplicated or split records that do not correspond to a real star \citep{Paegert:2021}. We removed every source flagged with a disposition of ARTIFACT, DUPLICATE, or SPLIT. We then kept only those targets carrying an associated Gaia DR2 source identifier, since that identifier is what carries each target into Gaia DR3 in the next step. Table~\ref{tab:selection_cuts} gives the number of targets surviving each cut.

\subsubsection{M-dwarf filtering from TIC-GAIA DR3 cross matches}\label{sec:target_selection_GAIA}

We crossmatched the TIC against Gaia DR3 \citep{Gaia:2016, Gaia:2023} to get astrometry and photometry good enough to select on. TIC entries carry a Gaia DR2 source identifier \citep{Gaia:2018} rather than a DR3 one, so we first mapped DR2 identifiers onto DR3 using the \texttt{gaiaedr3.dr2\_neighbourhood} table, then retrieved the corresponding DR3 entries from \texttt{gaiadr3.gaia\_source} through asynchronous ADQL queries. From those entries we took positions, parallaxes, proper motions, and magnitudes.

Crowded fields make this mapping ambiguous more often than one would like, so we imposed two consistency requirements before accepting a match. For each TIC source we took the first DR3 source returned for its DR2 identifier by the \texttt{gaiaedr3.dr2\_neighbourhood} table, propagated its position to the TIC epoch (J2000) with its proper motion, and accepted it if it lay within 21\arcsec{} of the TIC position and agreed with the TIC $G$ magnitude to within 2.5\,mag. Sources without an accepted counterpart remain in the Cut 3 count of Table~\ref{tab:selection_cuts} but carry no Gaia DR3 values, and the cuts of Section~\ref{sec:target_selection_cuts} remove all 22{,}479 of them.

Distances follow from the Gaia DR3 parallaxes. A few measured parallaxes are negative, for which direct inversion gives an unphysical distance, so we evaluated the distance posterior
\begin{equation}
p(d \,|\, \varpi) \propto
\mathcal{N}\!\left(\varpi \,\middle|\, d^{-1}, \sigma_\varpi\right)\,\Pi(d)
\label{eq:distance_posterior}
\end{equation}
where $\mathcal{N}$ is a Gaussian likelihood in parallax and $\Pi(d)$ is a uniform prior truncated to $0 < d < 200$~pc. We normalized the posterior numerically on a fixed distance grid, adopted the maximum a posteriori distance, and took uncertainty bounds from the posterior cumulative distribution function at a 99.7\% credible interval. These are the distances we carry forward, and it is on these that we finally restrict the sample to $d \le 100$~pc. Because the prior is uniform, the maximum a posteriori distance equals $1/\varpi$ whenever that value lies inside the prior range, so for every retained star the adopted distance agrees with the inverted parallax to within the 0.2\,pc grid spacing; the posterior's role is to place negative and very small parallaxes at the 200\,pc boundary and to set the uncertainty bounds. 

\begin{deluxetable*}{llr}
\tablecaption{Nearby M-dwarf target selection: cumulative cuts and retention.
\label{tab:selection_cuts}}
\tablewidth{0pt}
\tablehead{
  \colhead{Cut} &
  \colhead{Criterion} &
  \colhead{N targets}
}
\startdata
Cut 1: Volume and magnitude limited cuts
  & $T_{\mathrm{mag}} \leq 18$ \& $d_{\mathrm{TIC}} \leq 150$ pc
  & 547,276 \\
Cut 2: Removal of TIC phantoms
  & Exclude disposition $\in$ \{SPLIT, ARTIFACT, DUPLICATE\}
  & 520,860 \\
Cut 3: TIC to Gaia crossmatch
  & \shortstack[l]{Require Gaia DR2 source ID\\
    Crossmatch DR2$\rightarrow$DR3 via \texttt{gaiaedr3.dr2\_neighbourhood}\\
    Query \texttt{gaiadr3.gaia\_source}}
  & 519,441 \\
Cut 4: Photometric / quality cuts
  & \shortstack[l]{$0.8 \leq G - G_{\mathrm{RP}} \leq 2$
    \& $d_{\mathrm{Bayesian}} \leq 100$ pc \& RUWE $\leq 1.4$\\
    $4.543 \leq \log g_{\mathrm{TIC}} \leq 5.585$
    \& $2300$ K $\leq T_{\mathrm{eff,TIC}} \leq 4000$ K\\
    $0.3 \leq J{-}H \leq 0.8$ \& $7 \leq \mathrm{RPM_{J}} \leq 25$\\
    \& \texttt{ipd\_frac\_multi\_peak} $\leq 2$ \& \texttt{visibility\_periods\_used} $\geq 1$\\
    \& \texttt{non\_single\_star} $= 0$}
  & 159,775 \\
\textbf{Cut 5: Final target list}
  & \shortstack[l]{Collect TESS observations
    \& calculate flux contamination ratios\\
    \& remove duplicate IDs}
  & \textbf{156,144} \\
\enddata
\end{deluxetable*}

\subsubsection{Photometric and Astrometric Selection}\label{sec:target_selection_cuts}

To identify cool, dwarf-like stars while minimizing contamination from red giants and white dwarfs, we applied a set of photometric and astrometric selection criteria using Gaia DR3 and the TESS Input Catalog (TIC). We first selected stars with Gaia optical colors consistent with late-type dwarfs by requiring $0.8 \le G - G_{\rm RP} \le 2$. To mitigate blended or otherwise problematic Gaia solutions that can bias colors and astrometry, we additionally required RUWE $\le 1.4$, \texttt{ipd\_frac\_multi\_peak} $\le 2$, \texttt{visibility\_periods\_used} $\ge 1$, and \texttt{non\_single\_star} $= 0$. We then imposed TIC-based stellar-parameter and near-infrared color cuts, retaining only stars with $2300 \le T_{\rm eff} \le 4000$\,K and $0.3 \le J - H \le 0.8$, and with surface gravities consistent with dwarfs, $4.543 \le \log g \le 5.585$. These bounds are the surface gravities of two model dwarfs, one with $M_\star = 0.7\,M_\odot$ and $R_\star = 0.7\,R_\odot$ and one with $M_\star = 0.08\,M_\odot$ and $R_\star = 0.08\,R_\odot$, widened by 0.05\,dex on each side to allow for errors in the TIC values.

We further applied a reduced proper-motion cut in the $J$ band to exclude objects with atypical kinematics and to reduce contamination from sources that occupy distinct loci in color--magnitude space. We define the $J$-band reduced proper motion as
\begin{equation}
    \mathrm{RPM_{J} = J + 5 \log_{10}\!\bigg(\sqrt{\mu_{\delta}^{2} + \mu_{\alpha}^{2} \cos{\delta}}\bigg) + 5}
\end{equation}
where $\mu_{\alpha}$ and $\mu_{\delta}$ are the proper motions in right ascension and declination, respectively, and $\delta$ is the declination. We retained targets with $7 \le \mathrm{RPM_{J}} \le 25$, which removes the most extreme outliers while preserving the bulk of the dwarf locus. The cumulative impact of these criteria is summarized in Table~\ref{tab:selection_cuts}, and the resulting distributions in color--magnitude and color space are shown in Figure~\ref{fig:selection_cut}.

The spectral types annotated on the top axes of both panels in Figure~\ref{fig:selection_cut} are taken from \citet{Pecaut:2013}\footnote{An updated table from \citet{Pecaut:2013} is maintained by E. Mamajek at \url{http://www.pas.rochester.edu/~emamajek/EEM_dwarf_UBVIJHK_colors_Teff.txt}}.

\subsubsection{Applying \textsf{TESS-POINT} to Enforce Observational Coverage Cuts}\label{sec:tesspoint}

To determine which Primary Mission sectors provide usable coverage for each target, we used TESS-POINT \citep{Burke:2020} to evaluate the TESS pointing geometry at the target sky position across Sectors 1–26. For each target–sector pair, TESS-POINT predicts whether the source falls on silicon and returns the corresponding camera/CCD as well as an expected detector position. We used these predictions to (i) build the list of sectors searched for each star and (ii) flag “edge-warning” configurations where the expected target position lies too close to a CCD edge, chip gap, or other boundary where fixed-size TESSCut cutouts can yield partial apertures, unstable background estimation, or intermittent loss of flux. We flag a target-sector pair when the predicted position lies within 6 pixels of the edge of the CCD's science pixel area; 2{,}270 of the \NObsStarSectors{} observable star-sectors defined below are flagged. The flag does not change the target list. Instead, the light-curve stage of \NEMESIS repeats the same test for each star-sector it processes and skips any that it flags, so a flagged star-sector yields no light curve, and the 845 targets whose every Primary Mission sector is flagged are not searched (Section~\ref{sec:results}). At this stage we also recorded sector-level metadata used downstream in the photometric extraction, including the observed sector baseline (number of consecutive sectors) and crowding/contamination diagnostics (the flux contamination ratio within 2 TESS pixels of the target; Section~\ref{subsubsec:SAP} for the contamination correction applied to the Simple Aperture Photometry). 

\NEMESIS II searches single-sector light curves, so a star observed in more than one Primary Mission sector yields one independent light curve and one independent transit search per sector. Throughout this work we refer to the resulting unit of analysis as a \texttt{star-sector}: a single (TIC ID, Sector) pair. Counts of light curves, TCEs, and PC detections are counts of star-sectors unless stated otherwise; reconciliation of multi-sector detections of the same physical object is described in Section~\ref{sec:reconciliation}. The \NObsTargets{} stars surviving the coverage cuts correspond to \NObsStarSectors{} observable star-sectors across Sectors 1--26. We retained only targets that fall on silicon in at least one Primary Mission sector. Of these \NObsTargets{} targets, 89{,}090 (77\%) were on no SPOC 2-minute target list in any Primary Mission sector, so for most of our sample the Primary Mission recorded the star only in the 30-minute FFIs. 

Stellar radii and masses are taken from the TIC \citep{Stassun:2019}, and every searched star-sector has TIC values for both. For the cool dwarfs that make up our sample, TIC~v8 adopts the values of its Cool Dwarf List, which derives $M_\star$ from the $M_{K_S}$ relation of \citet{Mann:2019} and $R_\star$ from the $M_{K_S}$ relation of \citet{Mann:2015}, both evaluated from 2MASS \citep{Skrutskie:2006} $K_S$ and the Gaia DR2 parallax over $\MannMKsMin \le M_{K_S} \le \MannMKsMax$. Recomputing both relations from the TIC $K_S$ magnitudes and distances, we find that \NStarSecMannConsistent{} of the \NStarSectors{} searched star-sectors (\FracMannConsistent{}\%) reproduce the TIC radius and mass to within \MannMatchTol\%; the remainder carry TIC values that depart from these relations by more than that. No stellar parameters were imputed.

\begin{figure*}
    \centering
    \includegraphics[width=1\textwidth,scale=0.5]{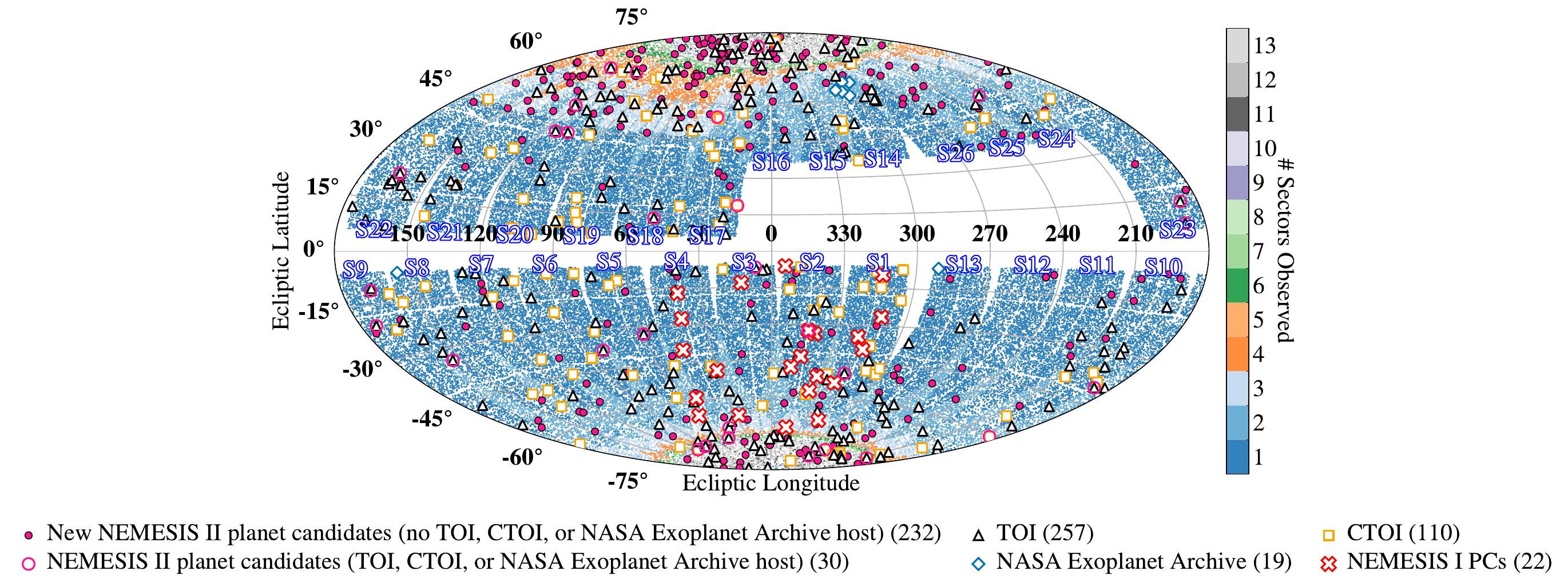}
    \caption{Sky positions of the \NObsTargets{} target stars with Primary Mission coverage, in ecliptic coordinates, colored by the number of Primary Mission sectors in which each was observed. Filled circles mark the stars that host our \NNewFavored{} new planet candidates (one star, TIC~230092943, hosts two), and open circles the stars that host our other planet candidates, each of which has a TOI, CTOI, or transiting NASA Exoplanet Archive entry (Section~\ref{sec:candidates}). The other markers show catalog entries on stars in our target list, from the lists of 2026 October 12 (Section~\ref{sec:comparison}), each placed at its host star: TOIs with a TFOPWG disposition of PC, APC, CP, or KP (triangles), CTOIs with a planetary or blank disposition that have not become TOIs (squares), the Paper I candidates we submitted as CTOIs, under the same rule (crosses), and transiting NASA Exoplanet Archive planets on stars with no TOI or CTOI (diamonds). One of the \NPaperIOnTargets{} Paper I candidates on our target list has since become TOI-7384.01 and is drawn as a TOI. Legend counts give objects, not stars.} 
\label{fig:target_star_locations}
\end{figure*}

\begin{figure}%[h!]
    \centering
    \includegraphics[width=1\columnwidth,scale=0.5]{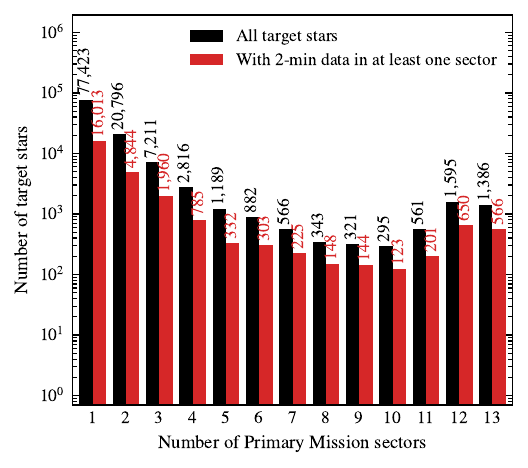}
    \caption{Number of target stars observed in a given number of Primary Mission sectors, after the on-silicon coverage cut of Section~\ref{sec:tesspoint}, before the edge test. Black bars count all target stars; red bars count the subset observed at 2-minute cadence in at least one sector. Because \NEMESIS~II searches each sector separately, a star contributes one independent search per sector (Section~\ref{sec:reconciliation}). }
    \label{fig:multi_sector_target_stars}
\end{figure}

% \clearpage %REMOVE AFTER DRAFTING
\subsection{Overview of \NEMESIS Processing of TESS Photometry} \label{subsec:tess}
The extraction described below follows Paper~I in outline; each change listed in Section~\ref{sec:intro} is described in the subsection where the relevant step appears.

%UPDATED FLOWCHART
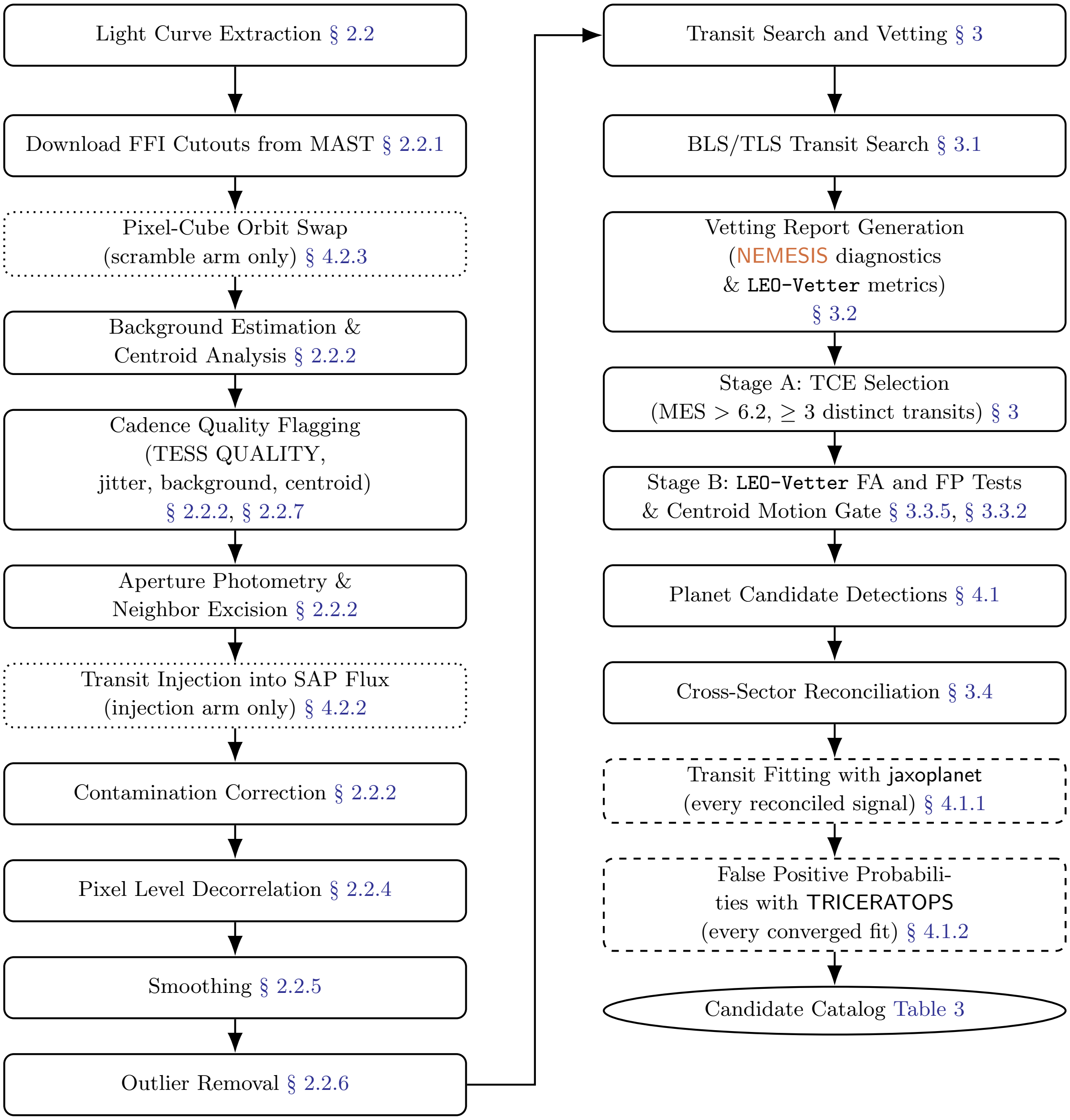
\begin{figure*}[t]
\centering
\begin{tikzpicture}[node distance=5mm,
    box/.append style={text width=6.4cm, align=center},
    lane/.style={box, draw, thick, minimum width=6.8cm, align=center},
    armbox/.style={box, dotted, thick},
    charbox/.style={box, dashed}]

  % Lane headers
  \node[lane] (lc) {Light Curve Extraction \autoref{subsec:tess}};
  \node[lane, right=2.0cm of lc] (ts) {Transit Search and Vetting \autoref{sec:transit_search}};

  % LC lane (top to bottom)
  \node[box, below=7mm of lc] (dl) {Download FFI Cutouts from MAST \autoref{subsubsec:FFI}};
  \node[armbox, below=of dl] (arm) {Pixel-Cube Orbit Swap\\
    (scramble arm only) \autoref{sec:scramble}};
  \node[box, below=of arm] (bkg) {Background Estimation \&\\
    Centroid Analysis \autoref{subsubsec:SAP}};
  \node[box, below=of bkg] (flag) {Cadence Quality Flagging\\
    (TESS QUALITY,\\ jitter, background, centroid)\\
    \autoref{subsubsec:SAP}, \autoref{subsubsec:masking}};
  \node[box, below=of flag] (sap) {Aperture Photometry \& Neighbor Excision \autoref{subsubsec:SAP}};
  \node[armbox, below=of sap] (inj) {Transit Injection into SAP Flux\\
    (injection arm only) \autoref{sec:detection_efficiency}};
  \node[box, below=of inj] (cc) {Contamination Correction \autoref{subsubsec:SAP}};
  \node[box, below=of cc] (pld) {Pixel Level Decorrelation \autoref{subsubsec:PLD}};
  \node[box, below=of pld] (smooth) {Smoothing \autoref{subsubsec:smoothing}};
  \node[box, below=of smooth] (outlier) {Outlier Removal \autoref{subsubsec:outlier_removal}};

  % Handoff: unflagged cadences only
  \draw[arrow] (outlier.east) -- ++(1.0,0) |- (ts.west);

  % TS lane (top to bottom)
  \node[box, below=7mm of ts] (scan) {BLS/TLS Transit Search \autoref{subsec:BLS_TLS_search}};
  \node[box, below=of scan] (report) {Vetting Report Generation\\
    (\NEMESIS diagnostics \& \leovetter{} metrics)\\
    \autoref{sec:Vetting_reports}};
  \node[box, below=of report] (stagea) {Stage A: TCE Selection\\
    (MES $>$ \MESthresh{}, $\geq$ \NTransitMin{} distinct transits) \autoref{sec:transit_search}};
  \node[box, below=of stagea] (stageb) {Stage B: \leovetter{} FA and FP Tests\\
    \& Centroid Motion Gate \autoref{sec:leovetter}, \autoref{subsect:centroid_motion}};
  \node[box, below=of stageb] (pcdet) {Planet Candidate Detections \autoref{sec:candidates}};
  \node[box, below=of pcdet] (recon) {Cross-Sector Reconciliation \autoref{sec:reconciliation}};
  \node[charbox, below=of recon] (jax) {Transit Fitting with \textsf{jaxoplanet}\\
    (every reconciled signal) \autoref{sec:jaxoplanet}};
  \node[charbox, below=of jax] (tri) {False Positive Probabilities with \textsf{TRICERATOPS}\\
    (every converged fit) \autoref{sec:triceratops}};
  \node[ellipse, draw, thick, below=of tri, minimum width=6.8cm, align=center] (cat)
    {Candidate Catalog \autoref{tab:candidates}};

  % Arrows (order must match node placement above)
  \foreach \a/\b in {lc/dl, dl/arm, arm/bkg, bkg/flag, flag/sap, sap/inj, inj/cc, cc/pld,
                     pld/smooth, smooth/outlier,
                     ts/scan, scan/report, report/stagea, stagea/stageb, stageb/pcdet,
                     pcdet/recon, recon/jax, jax/tri, tri/cat}
    \draw[arrow] (\a) -- (\b);
\end{tikzpicture}
\caption{Schematic overview of the \NEMESIS~II workflow. The left branch summarizes light-curve extraction from TESSCut FFI cutouts (Section~\ref{subsec:tess}). Only cadences with no quality flag set are passed to the transit search. The right branch summarizes the
transit search, the two-stage disposition rule, and candidate characterization (Section~\ref{sec:transit_search}). The two dotted boxes mark the only steps at which the scramble and injection-recovery arms differ from the blind survey (Section~\ref{sec:injection_analysis}). All three arms share every other step, including the Stage~A and Stage~B labeling rule. Dashed boxes mark characterization steps, which assign transit parameters and false positive probabilities but do not change any disposition. The overall structure follows Paper~I. The changes adopted in this work are described in Sections~\ref{subsec:tess} and \ref{sec:transit_search}.}
\label{fig:flowchart}
\end{figure*}

\subsubsection{Acquisition and Processing of TESS Full Frame Images}\label{subsubsec:FFI}

For each target, the \texttt{gethdu} routine within \NEMESIS queries the TIC with a 21\arcsec{} cone search through \textsf{Lightkurve} \citep{LightKurve:2018}, takes the stellar radius and mass from the TIC through \textsf{Astroquery} \citep{Astroquery:2019} and quadratic limb-darkening coefficients from \citet{Claret:2017}, as in Paper~I, and downloads an FFI cutout for each Primary Mission sector from the MAST \textsf{TESSCut} service \citep{TESScut:2019}. It rejects images with anomalous median photon counts or flagged by the quality bitmask of Section~\ref{subsubsec:SAP}, and records the camera, CCD, and reference pixel of each cutout for the steps that follow.

\subsubsection{Background Subtraction, Centroid Analysis, Deblending, and Simple Aperture Photometry}\label{subsubsec:SAP}
We process each TESSCut FFI cutout to extract Simple Aperture Photometry (SAP) for each target star, incorporating detailed steps for background estimation, aperture mask optimization, and flux deblending. To begin, we apply a TESS QUALITY bitmask that excludes cadences flagged for spacecraft anomalies such as momentum dumps, for scattered light, or for bad calibration.

Following background extraction and centroid analysis, we assign each cadence a quality bitmask built from four flags: jitter (flag~1; bit value 2$^0$ = 1), anomalous background flux (flag~2; 2$^1$ = 2), centroid excursion (flag~3; 2$^2$ = 4), and sigma-clipped outlier (flag~4; 2$^3$ = 8). Cadences carrying flags 1, 2, or 4 (bit values 1, 2, or 8 set; the flux-corrupting subset) are excluded from normalization and deblending; cadences carrying only flag~3 keep their flux but, like every flagged cadence, are excluded from the transit search (Section~\ref{subsubsec:masking}). A cadence is a centroid excursion when the radial offset of its centroid from the median position exceeds the median offset by more than $5\,\sigma_{\rm MAD}$, where $\sigma_{\rm MAD} = 1.4826 \times {\rm MAD}$ throughout. The TESS QUALITY bitmask removes cadences with bits 1 to 6, 8, 10, 13, or 15 set (attitude tweak, safe mode, coarse point, Earth point, Argabrightening, reaction wheel desaturation, manual exclude, impulsive outlier, scattered light, and bad calibration; \citealt{Tenenbaum:2018}), the set the TESS Archive Manual\footnote{\url{https://outerspace.stsci.edu/display/TESS/2.0+-+Data+Product+Overview}} identifies as likely to mark lesser-quality cadences.

We define the background aperture as all cutout pixels whose median-image flux does not exceed the image median by more than $10^{-3}\,\sigma_{\rm MAD}$, approximately the fainter half of the cutout. In each cadence, the sky level is the median of these pixels after iterative $3\,\sigma_{\rm MAD}$ clipping (at most five iterations), floored at zero, and is subtracted from each aperture pixel. We flag a cadence for anomalous background when, after removal of a 0.5-day biweight trend from the background time series, its residual lies more than $5\,\sigma_{\rm MAD}$ above the median residual and also exceeds 5\% of the trend level; these cadences receive a quality flag but are not immediately discarded from the light curve. Next, we identify the optimal aperture mask for the target star by performing a centroid analysis of the median brightness image, using a bivariate quadratic function to approximate the core of the Point Spread Function (PSF). A similar centroiding approach is implemented in the Eleanor FFI pipeline \citep{Feinstein:2019}; further methodological details are given in \citet{Vakili:2016}.  Pixels brighter than 7.5 standard deviations above the median cutout flux are included in the aperture mask. This threshold was chosen to produce compact, source-centered apertures. If the measured centroid position deviates by more than 1.5 pixels from the expected reference position, a circular aperture of fixed radius is placed at the reference pixel instead, preventing the threshold mask from locking onto a brighter neighboring source. 

The background-subtracted Simple Aperture Photometry (SAP) flux is calculated by summing the flux within the target aperture and subtracting the background contribution:

\begin{equation}
F_{\rm SAP}(t) =
\sum_{k=1}^{N_{\rm TGT}} F_k(t)
- N_{\rm TGT} F_{\rm BKG}(t)
\label{eq:sap}
\end{equation}

where F$_k$(t) is the flux of the (k)-th pixel in the target aperture, N$_{\rm TGT}$ is the number of pixels in the target aperture, and F$_{\rm BKG}$(t) is the median background flux per pixel, estimated using the sigma-clipping procedure described above. The SAP flux is then normalized to a baseline of unity.  

To account for flux from nearby stars that may contaminate the aperture, we query the TESS Input Catalog (TIC) for all sources within 2 TESS pixels ($\sim 42$\arcsec{}) of the target and record their TESS magnitudes. Sources with $\Delta T_{\rm mag} \geq 5$ relative to the target are discarded as negligible contaminants ($\lesssim 1\%$ of the target flux). Each remaining neighbor is then handled according to its separation from the target. For sources resolved at the TESS pixel scale (separation $\geq 2$ pixels), we excise the aperture pixels lying within $1.5$ pixels of the neighbor centroid, never removing the target's own reference pixel; if this excision would reduce the aperture below two pixels, the original aperture is retained. Neighbors within 2 pixels cannot be separated from the target by editing around them, so for these we trim the aperture to pixels within 1.5 pixels of the target's reference pixel, again keeping at least two pixels; the light they still contribute is accounted for by the flux contamination correction of Equation~\ref{eq:contamination}. The only model-based treatment of neighbor flux is the one inside \texttt{TRICERATOPS} (\citealt{Giacalone:2021}; Section~\ref{sec:triceratops}). Relative to Paper~I, the search radius has been reduced from 3 to 2 TESS pixels to avoid false associations with unrelated background sources at larger separations. 

For every target we compute the flux contamination ratio as:
\begin{equation}
f = 1 - \frac{10^{-T_{\rm mag,target}/2.5}}{\sum_{i}^{N_{\rm stars\,in\,aperture}} 10^{-T_{\rm mag,i}/2.5}}
\label{eq:contamination}
\end{equation}
where the sum runs over all catalog sources within the search cone and $T_{\rm mag}$ is the TESS magnitude of the target and of each source $i$; the instrumental zero point cancels in the ratio. We also propagate a $1\sigma$ uncertainty on $f$ from the TIC magnitude errors. We subtract $f$ from the normalized SAP flux and renormalize the result to unity, which divides the transit depth by $(1-f)$, and we scale the flux uncertainties by the same factor after adding the propagated uncertainty on $f$ in quadrature. The transit search, vetting, and the transit-model fits of Section~\ref{sec:jaxoplanet} therefore operate on dilution-corrected flux. 

\subsubsection{Additional Masking of Timestamps Using TESS Quaternions}\label{subsubsec:masks}\label{subsubsec:masking}
To suppress pointing-induced systematics, we developed our own implementation of the quaternion-based approach of the QLP \citep{Kunimoto:2022}. The TESS engineering quaternions, available through MAST\footnote{\url{https://archive.stsci.edu/missions/tess/engineering/}}, describe the spacecraft attitude at 2-second cadence for each camera. For the camera that observed each target, we bin the quaternion components $Q_1$, $Q_2$, and $Q_3$ to the 30-minute light-curve cadence with cadence-midpoint bin edges, so that no samples are assigned across the gaps at momentum dumps and data downlinks, and we require at least 225 samples per bin (25\% of the expected 900). From the standard deviation of each component within a bin we compute the jitter $J = \sqrt{\sigma_{Q1}^2 + \sigma_{Q2}^2 + \sigma_{Q3}^2}$, and we mask cadences whose jitter exceeds the per-sector median by more than $5\,\sigma_{\rm MAD}$, where $\sigma_{\rm MAD} = 1.4826 \times {\rm MAD}$, or whose jitter is undefined. Masked cadences carry the jitter flag (bit value 1) and are excluded from deblending. Because the mask is recomputed from each sector's own telemetry, it replaces the hand-set jitter intervals of the TESS Data Release Notes that Paper~I used. Unlike Paper~I, we excise no additional window around momentum dumps, which the TESS QUALITY bitmask already removes (Section~\ref{subsubsec:SAP}); the quaternion mask also flags the elevated jitter that accompanies most dumps. For display on the vetting reports (Section~\ref{sec:Vetting_reports}) we also store a 36-feature summary of the quaternions in each bin, filling bins with fewer than 225 samples by interpolation; it does not enter the mask.

\subsubsection{Pixel Level Decorrelation}\label{subsubsec:PLD}
As in Paper~I, we build the PLD model from the aperture pixel fluxes, each normalized by the total aperture flux \citep{Deming:2015, Luger:2016, Luger:2018}. To capture higher-order pixel responses, we form, for each product order from the second up to order \PLDorder{}, every product of that many first-order components, allowing repeated factors. When an order yields more than \PLDmaxCombos{} products, which occurs only for the largest apertures, we keep a fixed random subset of \PLDmaxCombos{} of them. Because the products within an order are highly redundant, we replace each order's block with its leading left singular vectors, computed without subtracting the column means. We retain the smallest number of vectors whose squared singular values reach \PLDvarThr{} of the block total, up to a maximum of \PLDnPCA{}; the maximum order and this cap equal the order and number of terms adopted in Paper~I.
% After verification (Next step 2): In practice this retains \PLDnKeep{} vector per order for most targets.
The design matrix is the concatenation of a constant column, the first-order components, and the retained vectors of each higher order. As in Paper~I, the constant and first-order blocks enter unreduced, and we solve for the coefficients by weighted linear least squares, adding \PLDdiagReg{} to the diagonal of the normal matrix for numerical stability; we apply no other regularization. 

The fit draws on every cadence with finite flux, including those that carry a NEMESIS quality flag; flagged cadences are removed afterward, before the transit search (Section~\ref{subsubsec:masking}). We fit the model in stages. The first fit weights each cadence by the inverse variance of its flux uncertainty, $\sigma_{\rm flux}^{-2}(t)$. To keep the fit from absorbing stellar variability, we smooth the first-fit residuals with a biweight filter \citep{Tukey:1977, Wotan} of width \PLDgpWindow{}~d and define $\sigma_{\rm stellar}(t)$ as the absolute departure of the smoothed residuals from their median. The second fit uses weights $[\sigma_{\rm flux}^{2}(t) + \sigma_{\rm stellar}^{2}(t)]^{-1}$, so cadences at which the star departs from its typical level count for less. This reweights independent cadences rather than modeling correlated noise, and it is simpler than the Gaussian-process treatment of stellar variability in EVEREST \citep{Luger:2016, Luger:2018}. The second fit also gives zero weight to cadences that lie more than \PLDgateNMAD{} median absolute deviations below the median of the SAP flux after removal of a \PLDgateWindow{}-day biweight trend; upward excursions are not gated. Finally, so that transits do not bias the fit, we flag cadences whose residuals from the current fit fall more than \PLDdipSigma{} standard deviations below the residual median, with the standard deviation estimated from the normalized median absolute deviation. We extend each flagged cadence by \PLDdipGrow{} cadence on either side, give the flagged cadences zero weight, and refit. We repeat this step up to \PLDdipIter{} times. We stop early if no new cadences are flagged, and we stop and keep the previous fit if fewer than \PLDmaskCap{} of cadences would retain nonzero weight. 

We remove the PLD model additively, as in Paper~I, subtracting the model and restoring its median before re-normalizing to unit median, so that flare amplitudes and transit depths are not rescaled by the systematics model. We divide the flux uncertainties by the same normalization factor. Cadences that received zero weight are neither removed nor replaced: they keep their corrected flux, because the model at each cadence is computed from that cadence's own pixel data. Figure~\ref{fig:cdpp} compares the CDPP of every searched star-sector before and after PLD.

\subsubsection{Smoothing}\label{subsubsec:smoothing}
After PLD, we remove long-term stellar variability such as rotational modulation with a time-windowed biweight filter \citep{Tukey:1977} implemented in \textsf{W\={o}tan} \citep{Wotan}, as in Paper~I.

To determine the optimal window size for the smoothing filter, we compute the theoretical longest transit duration for a circular orbit of an Earth-sized planet with an orbital period equal to half the light-curve time span (about 13.5 d for a single sector), assuming a semi-major axis aligned with the observer's line of sight (impact parameter $b$ = 0, inclination $i$=90\degrees).  Using the stellar radius and mass obtained from the TESS Input Catalog (TIC), we calculate the transit duration $\mathrm{T_{dur}}$ following the formula:

\begin{equation} \label{eq:T_dur}
\small
    \mathrm{T_{dur} = \frac{P}{\pi}\arcsin\bigg( (R_{Star}+R_{Planet})(\frac{4\pi^2}{GM_{Star}P^2})^{1/3}\bigg)}
\end{equation} 

where P is the orbital period, R$_\mathrm{star}$ and R$_\mathrm{planet}$ are the stellar and planetary radii, and M$_\mathrm{star}$ is the stellar mass. For each target star, the smoothing window is set to three times this calculated transit duration, ensuring that longer-term trends are removed without affecting the transit signal.

\textsf{W\={o}tan} splits the light curve into segments at large time gaps, using its default break tolerance, and detrends each segment separately. The original light curve is then divided by the smoothed flux to produce a detrended light curve. 

\subsubsection{Outlier Removal}\label{subsubsec:outlier_removal}

Finally, we remove outliers using contiguous, non-overlapping windows spanning five times the $b = 0$ transit duration of an Earth-sized planet at the maximum search period. Within each window we flag cadences whose PLD-corrected flux deviates from the window median by more than twice the median absolute deviation (MAD), the same criterion as Paper~I. Isolated flagged cadences on either side are removed. Runs of two or more consecutive low cadences are kept as potential transits, while runs of two or more consecutive high cadences, such as flare sequences, are removed. The flagged cadences are dropped from the detrended light curve. 

To assess the quality of our processed light curves, we compute the Combined Differential Photometric Precision (CDPP) in parts per million per square root of hour (ppm hr$^{-1/2}$), a standard metric that quantifies photometric precision on short timescales \citep{Christiansen:2012}. We compare our results to the expected photometric noise floor as derived in \citet{Bouma:2019}, which serves as an approximation of the preflight theoretical precision predictions for TESS from \citet{Sullivan:2015}. 

\begin{figure*}%[h!]
    \centering    \includegraphics[width=1\textwidth]{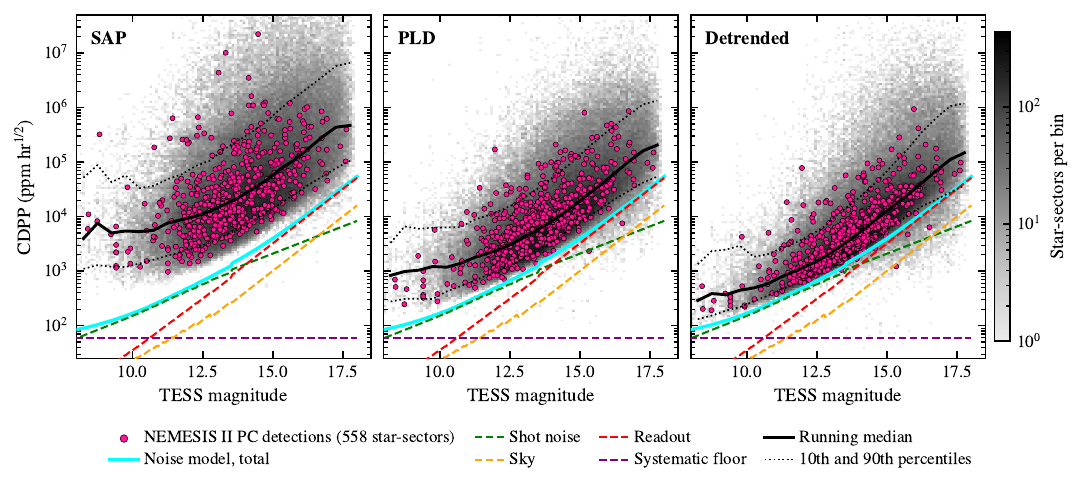}
    \caption{CDPP against TESS magnitude for every searched star-sector, at three stages of processing: SAP (Section~\ref{subsubsec:SAP}), PLD-corrected (Section~\ref{subsubsec:PLD}), and smoothed and outlier-removed. Pink points mark the hosts of the \PCs{} PC detections, and the black line in each panel is the running median. The cyan line is the noise model of \citet{Bouma:2019}, and the dashed lines are its components. }
    \label{fig:cdpp}
    \vspace*{1cm}
\end{figure*}

In Figure \ref{fig:cdpp}, we present individual contributions to the total photometric noise, including source shot noise, sky background noise from unresolved stars, read noise, and a systematic noise floor, each represented by dashed lines. 

\subsubsection{Default Masking 
Configuration}\label{subsubsec:masking} 
Before the transit search we combine two masking layers: the TESS QUALITY bitmask of Section~\ref{subsubsec:SAP}, which removes momentum dumps, and the quaternion jitter mask of Section~\ref{subsubsec:masks}, which independently flags the elevated jitter that accompanies most dumps. The two layers are partially redundant by design and provide a mutual cross-check on cadence quality. Only cadences carrying no NEMESIS quality flag (jitter, background, centroid excursion, or outlier) are passed to the transit search. The vetting reports apply the same filter to their light-curve and transit panels but show flagged cadences on the processing-summary page (Section~\ref{sec:Vetting_reports}). 

\section{Transit Detection and Vetting} \label{sec:transit_search}

We search all detrended \NEMESIS light curves for periodic transit-like signals using the Box-fitting Least Squares algorithm (BLS; \citealt{Kovacs:2002}) followed by the Transit Least Squares algorithm (TLS; \citealt{Hippke:2019}). We run both algorithms on every light curve. TLS searches the same period range as BLS with a transit-shaped kernel rather than a box, and we use the TLS best-fit ephemeris for all downstream vetting and candidate characterization. Every search result passes through a two-stage disposition chain. Stage~A promotes a result to a threshold crossing event (TCE) when the Multiple Event Statistic (MES) reported by \leovetter{} exceeds $\MESthresh$, its distributed detection floor, and TLS reports at least \NTransitMin{} distinct transits within the sector. The MES is the value of the phase-folded signal-to-noise time series at zero phase, the folded transit depth divided by the pink noise on the transit timescale \citep[][Eqs.~1--3]{Kunimoto:2025}; because the pink-noise estimator combines white and correlated components \citep{Pont:2006}, it suits 30-minute FFI photometry of M dwarfs, where the correlated term is not negligible. We use the value \leovetter{} reports without modification and call it the MES throughout, reserving SNR for the TLS statistic that we report but do not use for classification. Unlike in Paper~I, we do not gate on the TLS signal detection efficiency (SDE), because its value depends on the trial period grid and its usual threshold was calibrated on white noise alone \citep{Hippke:2019}; however, we do report it for every detection. Requiring \NTransitMin{} transits suppresses single-epoch systematics and stellar flares. Stage~B then labels each TCE a planet candidate (PC), an astrophysical false positive (FP), or a false alarm (FA) with \leovetter{}, which provides 13 flux-level false alarm tests and 4 astrophysical false positive tests, of which we apply 11 and 4 (Section~\ref{sec:leovetter}). Because \leovetter{} was developed on TESS FFI light curves of M dwarfs and uses pass-fail thresholds rather than a trained classifier, we apply it to our light curves without retraining.

The search configuration and the two-stage disposition rule described in this section are applied identically to the blind survey, to the transit injection-recovery experiment, and to the scramble test. The reason this identity is required, rather than merely convenient, is set out in Section~\ref{sec:injection_analysis}.

\subsection{BLS and TLS Transit Searches}\label{subsec:BLS_TLS_search}
Both BLS and TLS are run on every detrended light curve. Trial periods span $P \in [0.5,\ T_\mathrm{span}/N_\mathrm{tr}]$ days, where $T_\mathrm{span}$ is the baseline of the light curve and $N_\mathrm{tr}=\NTransitMin$ is the minimum number of transits required for a detection. Requiring three transits caps the maximum searchable period at one third of the baseline, or $\sim 9$ days for a typical single Primary Mission sector after quality masking. The period grid is constructed with the TLS \texttt{period\_grid} function, which sets optimal frequency sampling from the stellar mass and radius following \citet{Ofir:2014}, with an oversampling factor of $9$. Trial durations are constructed with the TLS \texttt{duration\_grid} function using a logarithmic step of $1.05$, and durations shorter than $N_\mathrm{tr}$ cadences are discarded as unresolvable at 30-minute sampling. Only cadences with \texttt{QUALITY\_FLAG}$=0$ are passed to either search; cadences flagged for spacecraft jitter, anomalous background, centroid excursion, or sigma-clipped outliers are excluded (Section~\ref{subsubsec:masking}), although they remain visible on the vetting reports.

BLS is run first, using the Astropy implementation \citep{Astropy:2018}. We then run TLS \citep{Hippke:2019} on the same light curve over the same period range, which replaces the box with a limb-darkened, transit-shaped matched filter. Unlike Paper~I, we do not gate TLS on the BLS detection significance, because running TLS on every star-sector removes a search-order dependence from the completeness map and costs only compute. The one exception is a light curve whose BLS SDE spectrum has no finite, non-negative peak, which TLS skips. If the BLS or TLS power spectrum returns all NaN values, which indicates non-Gaussian flux uncertainties on a high-CDPP light curve, we repeat the search once with the per-point uncertainty replaced by the point-to-point RMS of the flux. We record star-sectors that fail on the second attempt as search failures and exclude them from the count of searched star-sectors in Section~\ref{sec:results}. In the blind survey, \NnoBLS{} light curves produced no BLS output, and for the remaining \NTLSskip{} the BLS search completed but TLS produced no output; together they account for all \NFailSearch{} star-sectors that have a light curve but no completed search (Table~\ref{tab:disposition}).

The TLS best-fit period, epoch, and duration are used for all downstream vetting, injection recovery, and catalog reporting. BLS best-fit parameters are retained on disk and are used as an independent cross-check of the injection-recovery completeness map (Section~\ref{sec:detection_efficiency}).

\subsection{NEMESIS Vetting Reports}
\label{sec:Vetting_reports}

For every completed transit search, \NEMESIS generates a six-page vetting report that consolidates the photometric, centroid, and diagnostic information needed for candidate review into a single PDF. Reports are produced unconditionally, before any detection threshold is applied, and one is generated for each of the BLS and TLS searches on every star-sector. This ordering is required rather than incidental: the \leovetter{} flux-level metrics, including the MES on which the Stage A criterion of Section~\ref{sec:transit_search} is defined, are computed during generation of the report's sixth page and appended to the best-fit table as
\texttt{LEOVetter\_*} columns. The TCE criterion is therefore evaluated on the report's own output, not upstream of it. Across the survey this produced a TLS report for each of the \NStarSectors{} searched star-sectors.

\begin{figure*}
\includegraphics[width=\textwidth]{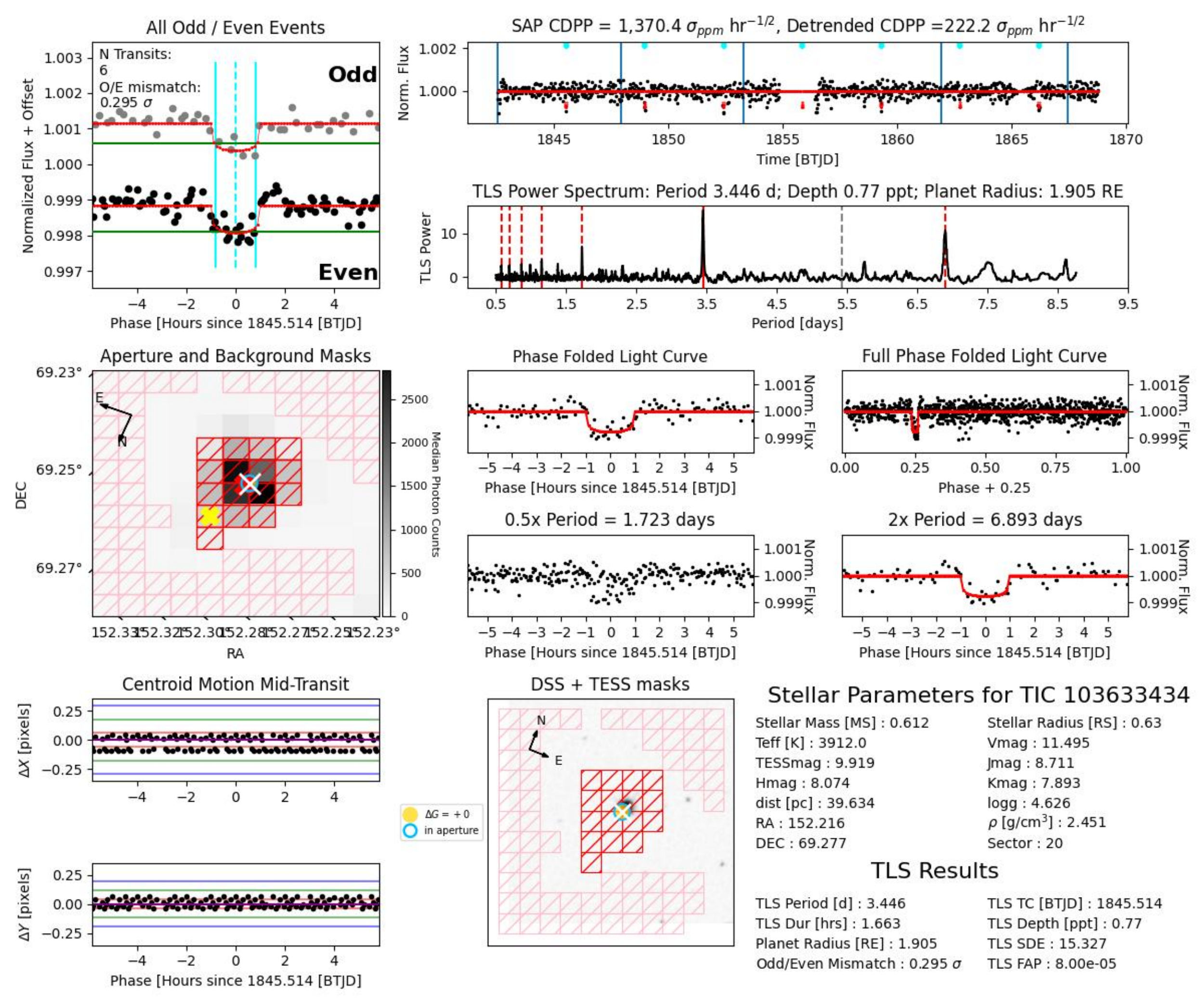}
\centering
\caption{Page~1 of the seven-page \NEMESIS vetting report for TOI-1235.01 (TIC~103633434) in Sector~20, which summarizes the TLS search. TOI-1235.01 is the confirmed planet TOI-1235~b \citep{Cloutier:2020b, Bluhm:2020}, which our blind search recovered in Sectors 20 and 21; we show a confirmed planet so that each page of the report can be read against an independent confirmation. Top: odd and even transits folded at the TLS period, with the number of transits and the odd-even depth mismatch; the detrended light curve with the TLS model and the predicted transit times, titled with the SAP and detrended CDPP; and the TLS power spectrum, with the best period and its aliases marked. Middle: the median FFI cutout with the photometric aperture (red hatching), the background pixels (pink hatching), and nearby Gaia sources labeled by their $G$-band magnitude difference from the target; and the light curve folded at the TLS period around transit and over the full orbit, and at half and twice that period. Bottom: column and row centroid offsets around mid-transit; a Digitized Sky Survey image of the field with the same masks; and the TIC stellar parameters and TLS results. The stellar density there is labeled in g~cm$^{-3}$ but printed in solar units: 2.451~$\rho_\odot$ is 3.455~g~cm$^{-3}$. Pages 2 to 7 are shown in Appendix~\ref{apndx:vetting_report}; page~7, the \texttt{TRICERATOPS} analysis, is added only to the reports of catalog candidates (Section~\ref{sec:Vetting_reports}).}
\label{fig:vetting_report_example}
\end{figure*}

Page~1 summarizes the transit search and is shown in Figure~\ref{fig:vetting_report_example}. Page~2 follows the light curve through each processing stage of Section~\ref{subsec:tess}, together with the background, centroid, and quaternion diagnostics behind the cadence flags. Page~3 compares Lomb-Scargle periodograms of the SAP, PLD-corrected, and detrended flux (Section~\ref{subsect:LS}) and shows odd and even transits folded at half, one, and two times the best-fit period. Page~4 shows the centroid motion and transit symmetry tests of Section~\ref{sec:internal_diagnostics}, with a summary of the \NEMESIS diagnostics and their verdicts. Page~5 shows individual transits, and page~6 the \leovetter{} modshift analysis and every flux-level false alarm and false positive metric (Section~\ref{sec:leovetter}). For each of the \NCatalog{} candidates in our catalog, all of which we evaluate with \texttt{TRICERATOPS} (\citealt{Giacalone:2021}; Section~\ref{sec:triceratops}), we append a seventh page to the report of its adopted sector. It shows the nearby sources that \texttt{TRICERATOPS} considers on the image with the \NEMESIS aperture, the transit parameters we supply, the FPP and NFPP across trials with the probability of each scenario, the resulting label, and the fit of each scenario from one trial, which is absent for the one candidate whose target cannot produce the observed depth (Section~\ref{sec:triceratops}). Appendix~\ref{apndx:vetting_report} shows pages 2 to 7 for the star-sector of Figure~\ref{fig:vetting_report_example}.

Report generation is wrapped so that a failure in any individual diagnostic does not abort the remaining pages or discard the search result. Each of the \leovetter{} flux-level stages is driven independently under its own exception handler, so metrics from stages that completed before a failure are retained and written, and the stages that did not complete are recorded as missing rather than as passing. Star-sectors for which the metric set is
incomplete are identified by a provenance column in the best-fit table and are reported separately in Section~\ref{sec:candidates} rather than being
silently absorbed into the false alarm count.

The same reports are produced by the transit injection-recovery and scramble arms (Section~\ref{sec:injection_analysis}), using the identical code path. This is the mechanism by which the three arms carry the same \texttt{LEOVetter\_*} columns and can therefore be labeled by the same rule. 

In Paper~I, each TCE was dispositioned by eye on these reports. Here no disposition depends on them, and we used them only for post-hoc inspection of the final candidates; they remain in the released data products as visual aids for readers and follow-up observers. Every label reported in this work follows deterministically from the criteria of Section~\ref{sec:transit_search} and Section~\ref{sec:leovetter}, applied identically to every star-sector, and no candidate was promoted, demoted, or removed on the basis of visual inspection. This change is deliberate. Manual inspection introduces selection effects that cannot be reproduced in the injection-recovery and scramble analyses, and therefore cannot be characterized by the completeness and effectiveness measurements that the occurrence rates in Section~\ref{sec:occurrence} depend on \citep{Kunimoto:2025, Bashi:2026}.

% AJ VERSION (POST-PUBLICATION)
% We release the complete set of reports with the \NEMESIS light curves as a MAST High Level Science Product, and we attach the reports of candidates submitted as Community TESS Objects of Interest to their ExoFOP-TESS entries.
% ARXIV VERSION (PRE-PUBLICATION)
Upon publication, we will release the TLS report of every searched star-sector with the \NEMESIS light curves as a MAST High Level Science Product, including the \texttt{TRICERATOPS} page of every catalog candidate, and we will attach the reports of candidates submitted as Community TESS Objects of Interest to their ExoFOP-TESS entries.

% ---- Vetting report example figure (placeholder) ----

\subsection{Internal Vetting Diagnostics}
\label{sec:internal_diagnostics}

Alongside the \leovetter{} metrics, \NEMESIS computes its own diagnostics while generating each vetting report: Lomb-Scargle periodograms (Section~\ref{subsect:LS}), the TLS odd-even depth comparison, a centroid motion test (Section~\ref{subsect:centroid_motion}), a test for left-right asymmetry of the folded transit (Section~\ref{sec:symmetry}), and a comparison of the photometric noise inside and outside the transit windows (Section~\ref{sec:intransit_noise}). 
%
%POST-PUBLICATION version (AJ)
% All are written to the best-fit table of every star-sector, which we release with the light curves, and shown on the reports. 
%PRE-PUBLICATION version (arxiv)
All are written to the best-fit table of every star-sector, which we will release with the light curves, and shown on the reports. 
Only the centroid motion test gates, because it takes the place of the \leovetter{} pixel-level offset test; the others either overlap an active \leovetter{} test or have not been measured in the injection and scramble arms (Section~\ref{sec:injection_analysis}), and we report them only. Publishing a catalog with no off-target screen would not be defensible, because blended eclipsing binaries are the dominant astrophysical false positives for TESS \citep{Sullivan:2015, Guerrero:2021} and the 21\arcsec{} TESS pixels make blending the default condition for our few-pixel apertures.

\subsubsection{Lomb-Scargle Analysis}\label{subsect:LS}
We compute Lomb-Scargle periodograms \citep{Lomb:1976, Scargle:1982} of the SAP, PLD-corrected, and detrended flux with \textsf{Astropy}, over a frequency grid bounded by the minimum and maximum search periods, and show them on page 3 of each vetting report. Comparing the three stages shows how PLD (Section~\ref{subsubsec:PLD}) and smoothing (Section~\ref{subsubsec:smoothing}) change the periodic content of a light curve, and whether a strong stellar or instrumental periodicity may lie near the period that BLS or TLS reports. For each flux stage we write the peak period, its power signal-to-noise, its false alarm probability, a harmonic flag, and an estimated rotation period to the best-fit table of each star-sector. These values support interpretation of individual candidates and do not enter the disposition chain.

\subsubsection{Centroid Motion Test}\label{subsect:centroid_motion}
As an additional diagnostic for off-target transit signals, we compute a centroid motion metric for each TCE and display it on Page~4 of the \NEMESIS vetting report (Section~\ref{sec:Vetting_reports}).
For each cadence we track the photocenter of the target star as a two-dimensional position$(\hat{x}_t, \hat{y}_t)$ in image column and row coordinates, estimated using the bivariate quadratic centroiding method described in Section~\ref{subsubsec:SAP}. We define the centroid displacement series as:
\begin{equation}
    \Delta x_t = \hat{x}_t - \widetilde{x}, \quad \Delta y_t = \hat{y}_t - \widetilde{y}
\end{equation}

where $\widetilde{x}$ and $\widetilde{y}$ are the median centroid positions over the full sector, so that the origin of the displacement coordinate system coincides with the time-averaged photocenter of the target.

We test whether the centroid is systematically offset during the transit window relative to its out-of-transit (OOT) position. We classify as in-transit every cadence within $\CentroidHalfWindow\,\times T_{14}$ of any predicted mid-transit time in the sector, a window of total width $\NCentroidWin\,\times T_{14}$ around each transit, where $T_{14}$ is the TLS transit duration, and we classify all remaining cadences as OOT. We compute a normalized mean-shift score separately for each axis,

\begin{equation}
z_x = \frac{|\overline{\Delta x}_{\mathrm{in}} - \overline{\Delta x}_{\mathrm{out}}|}
{\sigma_{x,\mathrm{out}} / \sqrt{N_{\mathrm{in}}}}
~,\qquad
z_y = \frac{|\overline{\Delta y}_{\mathrm{in}} - \overline{\Delta y}_{\mathrm{out}}|}
{\sigma_{y,\mathrm{out}} / \sqrt{N_{\mathrm{in}}}}
\label{eq:centroid_zscore}
\end{equation}
\noindent
where over-bars denote means over the respective cadence subsets,
$\sigma_{x,\mathrm{out}}$ and $\sigma_{y,\mathrm{out}}$ are the standard deviations of the OOT centroid displacements, and $N_{\mathrm{in}}$ is the number of in-transit cadences. The reported centroid motion score is the mean of the two axis scores, $\bar{z} = (z_x + z_y)/2$. If $\bar{z} > 3$, the TCE is flagged as potentially off-target. If the in-transit window contains no finite cadences (e.g., due to data gaps at $T_0$), the test result is recorded as indeterminate rather than a failure. The score in Equation~(\ref{eq:centroid_zscore}) is the ratio of the observed mean centroid offset during transit to the uncertainty on that mean estimated from the OOT scatter. Using the mean of the in-transit centroid positions, rather than the minimum or maximum, makes the statistic robust to individual outlier cadences; this is particularly important at 30-minute FFI cadence, where a typical transit window contains only a few cadences. 

We record the in-transit mean-shift score and its verdict in the best-fit table as \texttt{centroid\_motion\_score} and \texttt{centroid\_motion\_flag}. Unlike the other diagnostics of Section~\ref{sec:internal_diagnostics}, this test is a gate, and it is the only off-target screen in the survey. We apply it in the false positive family, after the \leovetter{} flux-level tests and before final classification. Because false alarm tests take precedence (Section~\ref{sec:leovetter}), the gate can change the label only of a TCE that has already survived flux-level vetting, which is the order in which \citet{Kunimoto:2025} recommend applying pixel-level tests. Of the \TCEs{} TCEs in the blind survey, \NCentroidFlag{} are flagged as off-target and \NCentroidIndet{} return an indeterminate result. Most of the flagged TCEs also fail a flux-level test, so the effect of the gate on our catalog is better measured among the \NFluxPC{} TCEs that pass every \leovetter{} flux-level test, of which it alone moves \NCentroidOnlyPC{} into the false positive category. The 1, 3, and 5 sigma circles on the $\Delta X$ versus $\Delta Y$ panel of the vetting report are there for visual reference, and we do not use them for gating.

\subsubsection{Transit Symmetry Test}
\label{sec:symmetry}
As an additional screen for eclipsing binaries, we test the phase-folded, detrended transit for left-right asymmetry. We split the folded profile at mid-transit and fit depth and fractional duration independently to each half, obtaining $(\delta_L, \sigma_{\delta_L}, q_L, \sigma_{q_L})$ and $(\delta_R, \sigma_{\delta_R}, q_R, \sigma_{q_R})$, and form
\begin{equation}
S_\delta = \frac{|\delta_L - \delta_R|}{\sqrt{\sigma_{\delta_L}^2 + \sigma_{\delta_R}^2}},
\qquad
S_q = \frac{|q_L - q_R|}{\sqrt{\sigma_{q_L}^2 + \sigma_{q_R}^2}} 
\label{eq:symmetry}
\end{equation}
A profile is recorded as asymmetric if $S_\delta > 3$ or $S_q > 5$, and the test returns \emph{passed} where either uncertainty is undefined. We measured this test as a candidate gate in the false alarm family on the full blind sample and did not enable it. Of the \NSymFlagged{} TCEs it flags, \NSymRedundant{} already fail an active false alarm test, so it would change only \NSymUnique{} dispositions, and it would remove \NSymPClost{} of the \PCs{} PC detections, far more than the few tenths of a percent that a three-sigma test should reject among genuine transits. The likely cause is per-half depth uncertainties that are biased low at 30-minute cadence, where a 1.5-hour transit leaves one or two cadences on each side. We report $S_\delta$, $S_q$, and the verdict in the best-fit table, as \texttt{symmetry\_depth\_SNR}, \texttt{symmetry\_qtran\_SNR}, and \texttt{symmetry\_flag}, and on page 4 of the vetting report, but no disposition depends on them. \leovetter{}'s \texttt{asymmetric} test remains active and fires on \NfAsym{} of the \TCEs{} TCEs.

\subsubsection{In-Transit and Out-of-Transit Noise Consistency}\label{sec:intransit_noise}
A genuine transit adds scatter to the cadences it occupies, because the flux there dips below the baseline, whereas a detection assembled from noise need not. We therefore compare the photometric precision of the detrended flux inside and outside the predicted transit windows. We classify as in-transit every cadence within $T_{14}$ of a predicted mid-transit time, and we compute the CDPP separately for the in-transit cadences, the out-of-transit cadences, and the full light curve, giving $\mathrm{CDPP}_{\rm in}$, $\mathrm{CDPP}_{\rm out}$, and $\mathrm{CDPP}_{\rm LC}$. We record the normalized difference
\begin{equation}
\mathcal{R} = \frac{\left|\mathrm{CDPP}_{\rm in} - \mathrm{CDPP}_{\rm out}\right|}{\mathrm{CDPP}_{\rm LC}}
\label{eq:cdpp_ratio}
\end{equation}
and flag the TCE when $\mathcal{R} < \CDPPratioMin$, that is, when the detected signal adds no measurable scatter within its own transit windows. $\mathcal{R}$, both CDPP values, and the flag are recorded in the best-fit table as \texttt{CDPP\_ratio}, \texttt{CDPP\_in\_transit}, \texttt{CDPP\_out\_of\_transit}, and \texttt{CDPP\_flag}, and a flagged TCE is marked in red on page 4 of the vetting report. Like the other diagnostics of Section~\ref{sec:internal_diagnostics} other than the centroid test, this comparison is reported and does not gate.

\subsubsection{Automated Vetting with \leovetter{}}\label{sec:leovetter}
We perform Stage B disposition with the Lazy-Exoplanets-Operations Vetter (\leovetter{}) Python package \citep{Kunimoto:2025}. \leovetter{} applies its tests in two families. False alarm tests identify signals that are inconsistent with a transiting astrophysical source, and false positive tests identify signals that are astrophysical but not planetary. We disposition a TCE that fails no active test in either family as a planet candidate.

We use \leovetter{}'s distributed default thresholds throughout and make no per-target adjustments. We do not apply three of its tests, and we apply a fourth test, \texttt{secondary}, with the phase condition described at the end of this section. The first is the pixel-level \texttt{offset} test, which requires difference imaging that \NEMESIS does not produce, and which we replace with the centroid motion test of Section~\ref{subsect:centroid_motion}. The second is \texttt{invalid\_transits}, which reapplies the MES floor and the three-transit requirement to the MES and transit count that \leovetter{} recomputes after discarding events it judges unusable. We exclude it because Stage A already imposes both criteria, using the full MES and the TLS transit count, and a second application to a culled count would reject signals that Stage A admitted for the same reason. The third is \texttt{bad\_fit}, which fails a TCE when the \leovetter{} least-squares transit fit fails, or when it does not improve on a straight line by \BfAICmarginLow{} in the Akaike Information Criterion statistic (AIC) for $N_{\rm tr} \le \BfNtrSplit{}$, or by \BfAICmarginHigh{} otherwise. At 30-minute cadence the fit window holds few cadences, so a fixed AIC margin acts as a second detection floor well above the Stage~A criterion. Among recovered injected planets with $N_{\rm tr} \le \BfNtrSplit{}$ and MES below \BfLowMESedge{}, it rejects \BfInjLowRej{} of \BfInjLowN{} (\BfInjLowPct{}). Across all recovered injected TCEs it fails \BfInjAllPct{}, against \KunBfInjPct{} in \citet{Kunimoto:2025}. Applied to our catalog, it would remove \BfBlindPC{} of the \PCs{} PC detections and \BfKnownCP{} of the \KPnCPPC{} confirmed or known planets that our rule retains (Section~\ref{sec:comparison}). It does reject \BfLeakCaught{} of the \NScrLeak{} scrambled TCEs that pass every false-alarm test. However, \BfLeakLowMES{} of those lie below MES of \BfLowMESmax{}, where, for $N_{\rm tr} \le \BfNtrSplit{}$, it rejects at least \BfInjLowRateMin{} of injected planets as well. Its gain there is the gain of raising the MES floor, which we characterize explicitly through $R_{\rm FA}$ as a function of MES (Section~\ref{sec:scramble}). We therefore exclude it. Because we measure vetting completeness and effectiveness with the same exclusions in place (Section~\ref{sec:injection_analysis}), the effect of each exclusion on our catalog is carried by those measurements rather than left unquantified.  Table~\ref{tab:leovetter} lists every active test.

When a TCE fails tests in both families, we count it as a false alarm. We follow \citet{Kunimoto:2025} in this false-alarm-precedence convention because it is required to reproduce their published false alarm reliability, and the alternative, counting such objects as false positives, does not reproduce it. The choice matters for our catalog, since \NFAandFP{} of the \FAs{} false alarms in the blind survey fail tests in both families, against \NFAonly{} that fail false alarm tests alone. Completeness, effectiveness, and reliability are defined on disjoint categories, so we fix the convention before computing any of them and apply it identically in all three analysis arms  (Section~\ref{sec:injection_analysis}). We give the per-test failure counts for the blind survey in Table~\ref{tab:leovetter}. They sum to more than the number of failing TCEs, because a single TCE frequently fails several tests.

\begin{deluxetable*}{llcc}[h!]
\tabletypesize{\small}
\tablecaption{\leovetter{} vetting tests applied to \NEMESIS~II threshold crossing events. Counts are the number of TCEs in the blind survey failing each test and sum to more than the number of failing TCEs because a single TCE frequently fails several. Thresholds marked ``default'' are the values distributed with \leovetter{} \citep{Kunimoto:2025} and were not modified.\label{tab:leovetter}}
\tablehead{
\colhead{Test} & \colhead{Criterion} & \colhead{Threshold} &
\colhead{$N_{\rm fail}$}
}
\startdata
\cutinhead{False alarm tests}
\texttt{weak}\tablenotemark{a} & MES below the detection floor & default & \NfWeak \\
\texttt{non\_unique} & Primary event not unique against red noise, tertiary, and positive events & default & \NfNonUnique \\
\texttt{chases} & Individual transits not distinct from nearby features, for $N_{\rm tr} \le 5$ & default & \NfChases \\
\texttt{bad\_shape} & Flux increases during the event & default & \NfBadShape \\
\texttt{single\_event} & MES dominated by one transit, for $N_{\rm tr} \le 10$ & default & \NfSingleEvent \\
\texttt{dmm} & Mean and median individual transit depths inconsistent & default & \NfDMM \\
\texttt{sinusoidal} & Signal better described by a sinusoid, for $P < 10$~d & default & \NfSinusoidal \\
\texttt{unphysical\_duration} & $a/R_\star$ too small, or duration too long for a circular orbit & default & \NfUnphysDur \\
\texttt{asymmetric} & Left and right transit halves differ in duration & default & \NfAsym \\
\texttt{chi} & Individual transit SNRs inconsistent with a common depth & default & \NfChi \\
\texttt{data\_gapped} & At least half of the transits fall near data gaps & default & \NfDataGap \\
\cutinhead{False positive tests}
\texttt{odd\_even} & Odd and even transit depths or epochs inconsistent & default & \NfOddEven \\
\texttt{vshaped} & $b + R_P/R_\star$ exceeds the grazing limit & default & \NfVshaped \\
\texttt{large} & Implied planet radius exceeds the planetary regime & default & \NfLarge \\
\texttt{secondary}\tablenotemark{c} & Significant secondary eclipse at a phase a binary can produce & default & \NfSecondary \\
\texttt{centroid\_offset}\tablenotemark{b} & In-transit photocenter shift $\bar{z} > 3$ & $3\sigma$ & \NCentroidFlag \\
\cutinhead{Tests not applied}
\texttt{offset} & Pixel-level centroid offset from difference imaging & \nodata & \nodata \\
\texttt{invalid\_transits} & Too few valid transits, or MES too low, after culling bad events & \nodata & \nodata \\
\texttt{bad\_fit} & Transit model fit fails or is not preferred over a straight line & \nodata & \nodata
\enddata
\tablenotetext{a}{\texttt{weak} cannot fire on a TCE, because Stage A already
requires the MES to exceed the same floor. It is listed for completeness.}
\tablenotetext{b}{NEMESIS in-transit centroid motion test
(Section~\ref{subsect:centroid_motion}), substituted for \leovetter{}'s pixel-level
\texttt{offset} test, which requires difference imaging that \NEMESIS does not
produce.}
\tablenotetext{c}{Applied with the phase condition of Equation~\ref{eq:emin}, so the test fails a TCE only when $e_{\min}\le\EMAX{}$ and the distributed criteria are also met. Without the condition it fails \NsecFired{} TCEs.}
\end{deluxetable*}

We apply one false positive test, \texttt{secondary}, with a modification. \leovetter{} searches the phase-folded light curve for the strongest event away from the transit and fails the TCE when that event is significant \citep{Kunimoto:2025}. The test assumes that the only astrophysical signal in the light curve is the one being vetted, and in our data two things break that assumption. A second transiting planet close to a low-order period ratio with the first places its transits at nearly the same phase of the first planet's fold, and a single outlying cadence can do the same for a short transit, because at 30-minute cadence one exposure spans roughly half of a one-hour transit. Neither mechanism is tied to phase 0.5, which is where tidal circularization places the secondary eclipse of a short-period binary. We therefore overrule \texttt{secondary} when the phase of the event is one that an eclipsing binary cannot plausibly produce.

For a primary eclipse at phase $\phi_{\rm pri}$, the secondary eclipse is displaced from $\phi_{\rm pri}+1/2$ by $(2/\pi)\,e\cos\omega$ to first order in eccentricity \citep{Winn:2010}. An event at phase $\phi_{\rm sec}$ therefore requires an eccentricity of at least
\begin{equation}
e_{\min} = \frac{\pi}{2}\left|\phi_{\rm sec}-\phi_{\rm pri}-\frac{1}{2}\right|
\label{eq:emin}
\end{equation}
and this is a strict lower bound, because the exact relation never displaces the secondary by more than $(2/\pi)\,e$. We take $\phi_{\rm sec}$ and $\phi_{\rm pri}$ from the \leovetter{} model-shift analysis. A TCE fails \texttt{secondary} only if it meets the distributed \leovetter{} criteria and $e_{\min}\le\EMAX{}$, and a TCE whose secondary phase is undefined keeps the default behavior.

\begin{figure}%[ht!]
\centering
\includegraphics[width=\columnwidth,keepaspectratio]{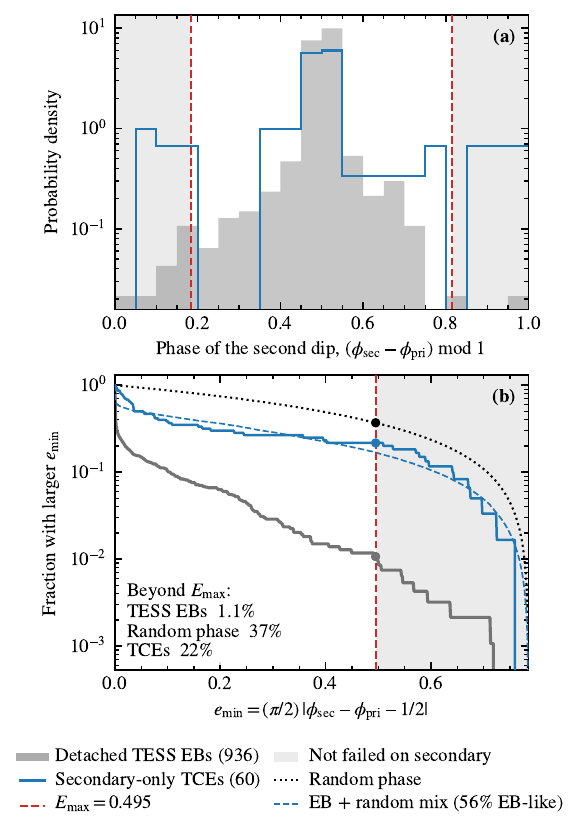}
\caption{Calibration of the phase condition on the \texttt{secondary} test. Grey shows the \NEBcal{} detached eclipsing binaries of the TESS Eclipsing Binary Catalog with periods in our search range and a measured secondary eclipse, and blue shows the \NsecOnly{} blind TCEs that fail the distributed \texttt{secondary} test and no other test. (a) Phase of the strongest secondary event relative to the primary. The dashed lines enclose phases within $2E_{\max}/\pi=\EMAXphaseHalf{}$ of 0.5, where \texttt{secondary} can still fail a TCE; in the shaded regions it cannot. (b) Fraction of each sample with $e_{\min}$ (Equation~\ref{eq:emin}) larger than a given value. The dashed line marks $E_{\max}=\EMAX{}$, the \EMAXpct{}th percentile of the binary distribution, and the fractions of each sample beyond it are given in the panel. The dotted line is the expectation for an event at a random phase, for which $e_{\min}$ is uniform on $[0,\pi/4]$, and the thin dashed line is the mixture of binary-like and random-phase events from which we estimate the number of binaries the condition admits (Section~\ref{sec:leovetter}).}
\label{fig:emax}
\end{figure}

We set $E_{\max}$ from real eclipsing binaries rather than from our own detections, so that the threshold does not depend on which planets it recovers. From the TESS Eclipsing Binary Catalog \citep{Prsa:2022} we took the \NEBcal{} detached systems (morphology coefficient below \EBmorphMax{}) with periods inside our search range and a measured secondary eclipse, computed $e_{\min}$ from their fitted eclipse positions with Equation~\ref{eq:emin}, and adopted the \EMAXpct{}th percentile of that distribution, $E_{\max}=\EMAX{}$ (Figure~\ref{fig:emax}). We fixed the percentile before computing $E_{\max}$. By construction, the condition excuses about \EMAXleakPct{}\% of real binaries of this kind. Among the \NsecOnly{} blind TCEs that fail the distributed \texttt{secondary} test and no other test, \NsecOnlyHalf{} place the event within \SecHalfWidth{} of phase 0.5, where a random phase would put about \SecUniformPct{}\% of them, which is the concentration that eclipsing binaries produce. Applied to this population, the calibrated condition admits an expected \NEBadmitted{} eclipsing binaries. The polyfit eclipse models of the same catalog give $E_{\max}=\EMAXpf{}$, and dropping the morphology cut gives \EMAXall{}, so the value we adopt is the most conservative of the three.

The condition changes the label of \NsecExempt{} blind TCEs from false positive to planet candidate. One of them is GJ~436~b \citep{Butler:2004,Gillon:2007}, which our survey detects in a single sector and which would otherwise be absent from our catalog, and another is TOI-270~c \citep{Gunther:2019} in Sector~5 (Section~\ref{sec:limitations}). A third, TIC~159344139, is too deep to be a planet on this star, and \texttt{TRICERATOPS} attributes it to a nearby eclipsing binary (Section~\ref{sec:limitations}). Because the condition acts only within the false positive family, it cannot move a TCE into or out of the false alarm category, so $E$ and $R_{\rm FA}$ (Section~\ref{sec:scramble}) are unchanged by it. Its effect on completeness is carried by $C(P,R_P)$, since the injection arm is labeled by the same rule. The condition cannot help when a stray event falls near phase 0.5, and we describe that residual loss in Section~\ref{sec:limitations}. 

\subsection{Cross-Sector Candidate Reconciliation}
\label{sec:reconciliation}
We run the \NEMESIS~II transit search independently on each star-sector and keep only the strongest TLS peak, so a star-sector contributes at most one TCE. The same physical signal can therefore appear more than once only across sectors, or on neighboring stars whose apertures overlap. We reconcile the first case below. For the second, which occurs for \NClosePairs{} pairs of candidates separated by less than one TESS pixel, we report both members, and \texttt{TRICERATOPS} labels all four likely nearby false positives (Section~\ref{sec:triceratops}). A TIC observed in several Primary Mission sectors can therefore carry several planet candidate detections for three distinct reasons. The first is a repeated detection of the same periodic signal in more than one sector. The second is a multi-planet system in which different planets dominate the search in different sectors. The third is a period alias driven by gapped coverage or by the window function. Reporting each detection as an independent candidate would inflate the candidate list, so we reconcile them with a lightweight step that works on the per-sector ephemerides alone. We deliberately stop short of a joint multi-sector fit, because that would take the detection stage beyond the single-sector design that the rest of this work characterizes, and we defer a global joint analysis to future work.

We apply reconciliation to the planet candidates of Section~\ref{sec:candidates} rather than to the full TCE list, and it changes the candidate table only. It does not change any occurrence rate. In Section~\ref{sec:occurrence} we count the effective number of stars as a sum over searched star-sectors, since a star observed in two sectors gave us two chances to detect its planet. For the same reason, the numerator of each rate is a count of star-sector detections, and substituting the reconciled count would place a deduplicated numerator over a star-sector denominator and bias every rate low.

For every TIC that carries two or more planet candidates, we evaluate each pair $(i, j)$ with a period-ratio test and an epoch-consistency test, and we merge matching pairs transitively, so that if $i \sim j$ and $j \sim k$ then $i$, $j$, and $k$ form one cluster. We report each cluster as a single candidate, and we report distinct clusters on the same TIC as distinct candidates, flagged for targeted multi-sector follow-up.

\runin{Period-ratio test} Let $P_i$ and $P_j$ be the recovered periods of two candidates on the same TIC. For every coprime pair of integers $(m, n)$ with $1 \le m, n \le N_{\max}$ and $N_{\max} = \ReconNMax$, we form the candidate common period $P_\star = \tfrac{1}{2}\left(P_i/m + P_j/n\right)$ and the fractional residual
\begin{equation}
\delta_P(m, n) = \frac{\left| P_i/m - P_j/n \right|}{P_\star}
\label{SLOT:eq10}
\end{equation}
and we adopt the pair that minimizes it. The two candidates are consistent with a common true period if \begin{equation}
\min_{m,n}\, \delta_P(m, n) < \tau_P 
\label{SLOT:eq11}
\end{equation}
with $\tau_P = \ReconTauP$ held fixed. This formulation handles the common 2:1 and 3:1 aliases produced by gapped coverage, as well as the less common case in which two sectors recover different aliases of a shared true period. We restrict the grid to coprime pairs because a pair such as $(2, 2)$ only reproduces the residual of $(1, 1)$. We select on the fractional residual, the quantity that Equation~\ref{SLOT:eq11} tests, because the absolute residual shrinks whenever both periods are divided by larger integers and would favor large $(m, n)$ for no physical reason.

\runin{Period uncertainty} The epoch-consistency test needs an uncertainty on $P_\star$. We do not take it directly from TLS, because the formal period uncertainty that TLS reports is roughly an order of magnitude smaller than a single-sector search can actually resolve (Section~\ref{sec:detection_efficiency}). For each candidate we instead adopt
\begin{equation}
\sigma_P = \max\left(\sigma_{P,{\rm TLS}},\ \frac{T_{14}}{N_{\rm ep} - 1}\right)
\label{eq:sigmaP_eff}
\end{equation}
where $N_{\rm ep}$ is the number of distinct transit epochs that contain in-transit data. The second term is the period error at which the ephemeris drifts by one transit duration across the observed epochs, which is the same scale that defines a recovered injection in  Equation~\ref{eq:coherence}. Where neither term is available, we fall back to $\ReconSigPFloor\,P$. The uncertainty on the common period is then $\sigma_{P_\star} = \tfrac{1}{2}\sqrt{(\sigma_{P,i}/m)^2 + (\sigma_{P,j}/n)^2}$.

\runin{Epoch-consistency test} A small period ratio alone cannot distinguish two aliases of one planet from two distinct planets near a low-order mean-motion resonance, so we also require the epochs to agree at the candidate period $P_\star$. For transit midtimes $T_{0,i}$
and $T_{0,j}$, we define the wrapped epoch difference as
\begin{equation}
\begin{split}
\Delta T_0 ={}&
\left[(T_{0,j}-T_{0,i})+\frac{P_\star}{2}\right]
\bmod P_\star -\frac{P_\star}{2}
\end{split}
\label{SLOT:eq12}
\end{equation}
which lies in $[-P_\star/2,P_\star/2)$. We accept a pair when $|\Delta T_0|<\tau_{T_0}$, where
\begin{equation}
\begin{split}
\tau_{T_0} = \max\Bigg\{&
f_T T_{14,\mathrm{ref}},\\
&\left[
\sigma_{T_0,i}^2+\sigma_{T_0,j}^2
+(N_{\mathrm{cyc}}\sigma_{P_\star})^2
\right]^{1/2}
\Bigg\}
\end{split}
\label{SLOT:eq13}
\end{equation}
Here, $T_{14,\mathrm{ref}}=\min(T_{14,i},T_{14,j})$, $N_{\mathrm{cyc}}=|T_{0,j}-T_{0,i}|/P_\star$, and $f_T=\ReconFT$ by default. Increasing $f_T$ allows greater deviations due to transit timing variations or ephemeris drift, while decreasing it imposes a stricter epoch-consistency criterion.

The last term under the square root carries the period uncertainty across the $N_{\rm cyc}$ orbits that separate the two epochs. We include it because Primary Mission observations of the same star can lie up to $\sim$\ReconMaxGap{}~d apart. Over such a gap, an ephemeris that is well determined within one sector predicts the epoch in the other with an uncertainty that grows linearly with the number of cycles. Without this term, the test would reject genuine recoveries of short-period signals in widely separated sectors. TLS reports no epoch uncertainty, so we set $\sigma_{T_0,i} = \sigma_{T_0,j} = 0$, and the statistical term reduces to the propagated period uncertainty. When $\tau_{T_0} \ge P_\star/2$, the folded offset cannot exceed the tolerance and the pair passes the epoch test automatically. Such a link rests on the period test alone, and we report the number of these links in Section~\ref{sec:candidates}.

\runin{Single-transit and low-coverage candidates} A single-transit detection leaves the period essentially unconstrained within one sector. We tag any candidate with $N_{\rm ep} < \NMinEpochs$ with a single-transit flag and exclude it from the period-ratio test. We keep its epoch, however, for a consistency check against multi-transit candidates on the same target. If the ephemeris of a multi-transit candidate predicts a transit within $\tau_{T_0}$ of the single event (Equation~\ref{SLOT:eq13}), we take this as corroborating evidence and cluster the two. We never link two single-transit candidates, since neither constrains a period. 

\runin{Multi-planet systems} When two candidates on the same TIC fail both tests at every admissible $(m, n)$, we retain them as distinct planet candidates, which recovers multi-planet systems in which different members dominate the search in different sectors. Because each star-sector search keeps only its strongest peak, however, our survey cannot recover a planet that is subdominant in every sector in which it is observed. Our injection experiment (Section~\ref{sec:injection_analysis}) places one planet in each light curve, so it does not measure this loss, and our sensitivity map describes the dominant signal in a light curve.

\runin{Ephemeris adoption and provenance} Within a cluster, we adopt the ephemeris $(P, T_0, T_{14})$ of the sector with the largest number of in-transit cadences, and we break ties with the TLS SDE. We retain every contributing sector, its individually recovered $(P_i, T_{0,i})$, and its period ratio $(m{:}n)$ relative to the adopted period as provenance columns in the candidate table (Table~\ref{tab:candidates}). Reconciliation uses period and timing only, and we do not compare depths or radii when forming clusters.

\runin{Support from non-candidate detections} Because we reconcile planet candidates only, a cluster can hide the case in which the same ephemeris was recovered in another sector as a TCE that vetting labeled a false alarm or false positive there. For each cluster we therefore record the number of sectors in which a TCE of any label matches its ephemeris under the same two tests. This count tells us about the vetting rather than about the candidate, and we never use it to promote or demote a candidate.

\runin{Chance coincidences} The period test accepts unrelated periods at a rate set by $\tau_P$ and by the density of the $(m, n)$ grid. An analytic estimate of that rate is unreliable here, because the periods that TESS searches recover pile up near sector-length and momentum-dump aliases instead of spreading uniformly. We therefore measure it directly. We permute the TIC labels among the reconciled rows \ReconNullTrials{} times, which keeps the number of candidates per star but destroys any physical association between them, and we rerun both tests on each permutation. The mean number of links formed per permutation is the number of chance links we expect in the real run, and we compare the two in Section~\ref{sec:candidates}.

\runin{Tolerance choices} 
We set $\tau_P$ (Equation~\ref{SLOT:eq11}) comfortably above the period precision of a single-sector search, $T_{14}/T_{\rm span} \sim \SearchPeriodRes$ (Section~\ref{sec:detection_efficiency}), but small enough that unrelated periodicities in the same system do not match, and we fixed both tolerances before processing the full sample, in order to preserve the blindness of the survey.

\runin{A merge by hand} The two detections on TIC~454291823 (Sectors~11 and 12) pass the period test but not the epoch test. Two planets on separate orbits cannot be stable with periods this close, so we merge the pair into one signal and adopt the ephemeris that the adoption rule of this section selects. The merge changes no disposition and, because both periods lie below one day, no occurrence rate.

\section{Planet Candidates and Transit Detection Sensitivity}
\label{sec:results}

Of the \NObsTargets{} targets with usable Primary Mission coverage
(Section~\ref{sec:tesspoint}), \NEMESIS produced a light curve and a completed transit search for \NSearchedTICs{} unique TIC entries across \NStarSectors{} star-sectors, a coverage fraction of \CoverageFrac{} of the observable star-sector list. The shortfall is accounted for by \NFailCutout{} star-sectors for which no light curve was produced, 2{,}270 of them skipped by the edge test of Section~\ref{sec:tesspoint}, and \NFailSearch{} star-sectors on which the transit search or the vetting-report chain did not run to completion (Section~\ref{subsec:BLS_TLS_search}). The occurrence rates of Section~\ref{sec:occurrence} are normalized to the \NMetricCompleteSS{} of the searched star-sectors that have a complete \leovetter{} metric set (Section~\ref{sec:candidates}), not to the target list of Table~\ref{tab:selection_cuts}. Because the survey is single-sector, PC detections are counted in star-sectors; Section~\ref{sec:reconciliation} combines them into signals.

\subsection{Vetted Planet Candidates}
\label{sec:candidates}

Applying the Stage~A criterion of Section~\ref{sec:transit_search} to the \NStarSectors{} searched star-sectors yields \TCEs{} threshold crossing events. Stage~B disposition (Section~\ref{sec:leovetter}) classifies \PCs{} of these as PC detections (\PCfrac{}), \FPs{} as astrophysical false positives, and \FAs{} as false alarms. Because a TCE failing tests in both families is counted as a false alarm, the three categories are disjoint and sum to the TCE total by construction. A further \NIndeterminate{} star-sectors returned an incomplete \leovetter{} metric set and are identified by the provenance column described in Section~\ref{sec:Vetting_reports}; these are excluded from the TCE count above and from every completeness, effectiveness, and reliability quantity in Section~\ref{sec:injection_analysis}, rather than being absorbed into the false alarm total. Per-test failure counts are given in Table~\ref{tab:leovetter}, and the full funnel from target list to candidate catalog is given in Table~\ref{tab:disposition}.

We call these \PCs{} PC detections, because each is one star-sector. After the cross-sector reconciliation of Section~\ref{sec:reconciliation} they correspond to \PCsUnique{} distinct signals on \NCatalogHosts{} stars, once we drop the \NUnconverged{} signals whose transit fits do not converge (Section~\ref{sec:jaxoplanet}). Of these signals, \PCsMultiSector{} were independently recovered in more than one Primary Mission sector, and \PCsMultiCluster{} stars host more than one distinct signal. Apart from one pair that we merge by hand (Section~\ref{sec:reconciliation}), we formed these clusters from \ReconNLinks{} matched pairs, against \ReconNullMean{} pairs expected by chance from the permutation test of Section~\ref{sec:reconciliation}, and \ReconPeriodOnlyLinks{} of the matched pairs rests on the period test alone. Independent recovery of the same ephemeris in a second sector is not used as a vetting criterion, because it is unavailable for the majority of the sample and applying it would make the selection function depend on sky position; it is nonetheless reported in Table~\ref{tab:candidates}, since it is the strongest internal consistency check available for a single-sector detection. 

Two distinct notions of reliability apply to this catalog and they are not interchangeable. The scramble test of Section~\ref{sec:scramble} measures $R_{\rm FA}$, a population-level statement about what fraction of the surviving catalog is expected to be instrumental or stellar noise rather than a transit. The \texttt{TRICERATOPS} analysis of Section~\ref{sec:triceratops} measures the false positive probability (FPP), a per-candidate statement about whether a signal that is genuinely astrophysical is planetary rather than stellar. A catalog can have high $R_{\rm FA}$ and still contain eclipsing binaries, and an individual candidate can have low FPP while belonging to a catalog with poor $R_{\rm FA}$. We report both.

\begin{figure}%[ht!]
\centering
\includegraphics[width=\columnwidth,keepaspectratio]{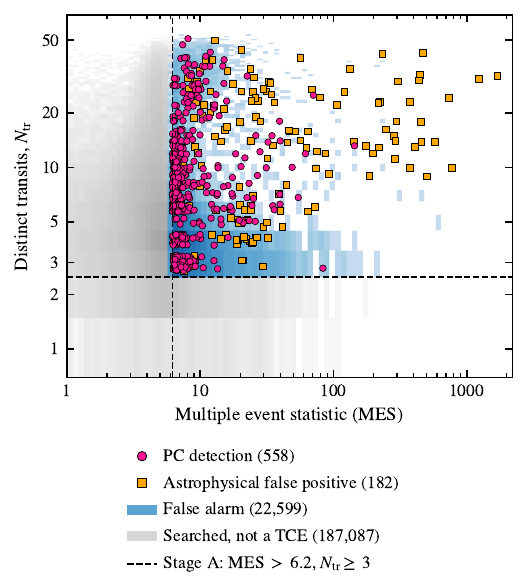}
\caption{
The \NEMESIS~II threshold crossing events in the plane of multiple event statistic and number of observed transits. All \TCEs{} TCEs are shown, colored by their Stage~B label (Section~\ref{sec:leovetter}): PC detection, astrophysical false positive, or false alarm. The Stage~A selection boundaries ($\mathrm{MES} >$~\MESthresh{} and at least \NTransitMin{} distinct transits) are marked. Grey shading shows the searched star-sectors without a TCE that have a positive MES and at least one transit, the population the legend counts; those with MES below 1 fall left of the plotted range. This is the same plane in which the false alarm effectiveness and reliability of Section~\ref{sec:scramble} are reported (Figure~\ref{fig:reliability}), so the reliability applicable to an individual detection can be read off by locating it here.}
\label{fig:catalog_plane}
\end{figure}

The \NCatalog{} reconciled signals are the candidates of our catalog (Table~\ref{tab:candidates}); each has a converged limb-darkened transit fit (Section~\ref{sec:jaxoplanet}), and we evaluate each with \texttt{TRICERATOPS} (Section~\ref{sec:triceratops}).
Of these candidates, \NCatalogNew{}, on \NCatalogNewHosts{} stars, lie on stars with no TOI, CTOI, or transiting planet in the NASA Exoplanet Archive \citep{Akeson:2013, NEA:PSCompPars}\footnote{Accessed on 2026 October 12, returning 6{,}445 rows, of which we use the 4{,}811 transiting planets.}, and we report them here as new candidates. The remaining \NCatalogKnown{} lie on stars that already host at least one of these. Whether a host is known is decided by TIC alone, and a TOI or CTOI of any disposition, false positives included, makes it known. Only transiting Archive planets count, so one new candidate, TIC~256419669 (Sector~24, $P = 3.03$~d), orbits Gl~49, which hosts the non-transiting planet Gl~49~b found by radial velocities ($P = 13.85$~d; \citealt{Perger:2019}). A candidate is identified with a particular TOI or CTOI only when its period agrees with the catalog period within $\tau_P$ at an integer ratio of up to three and its transit times agree with the catalog ephemeris, the test of Section~\ref{sec:comparison}. Where the known-planet audit of Section~\ref{sec:comparison} matches one of them to a cataloged signal, we list the archival period in the machine-readable version of Table~\ref{tab:candidates} (column \texttt{P\_arch}; Table~\ref{tab:candidates_cols}), as an independent check on the \NEMESIS ephemeris. Reliability and occurrence rates count PC detections rather than candidates, because the injection and scramble arms reproduce the vetting rule but neither the refit nor \texttt{TRICERATOPS}; the occurrence rates use the \texttt{TRICERATOPS} probabilities only as weights (Section~\ref{sec:occurrence}). The distribution of all \TCEs{} TCEs in the plane of MES and number of observed transits, colored by disposition, is shown in Figure~\ref{fig:catalog_plane}. The same plane is used in Section~\ref{sec:scramble} to report catalog reliability, so the two figures can be read against each other.

\begin{deluxetable*}{rccccccl}
\digitalasset
\tabletypesize{\footnotesize}
\tablecaption{NEMESIS~II candidate catalog.\label{tab:candidates}}
\tablehead{
\colhead{TIC} & \colhead{Sector} & \colhead{$P$} & \colhead{$T_0$} &
\colhead{$R_P$} & \colhead{MES} & \colhead{FPP} & \colhead{Label} \\
\colhead{} & \colhead{} & \colhead{(d)} & \colhead{(BTJD)} &
\colhead{(\RE)} & \colhead{} & \colhead{} & \colhead{}
}
\colnumbers
\startdata
7322727  & 6  & $1.90662^{+0.00062}_{-0.00059}$ & $1468.4947^{+0.0032}_{-0.0032}$ & $3.55^{+0.33}_{-0.25}$  & 9.0  & 0.084 & Inconclusive (nearby star not ruled out) \\
7586683  & 4  & $3.89912^{+0.00360}_{-0.00426}$ & $1411.2330^{+0.0163}_{-0.0143}$ & $3.07^{+23.21}_{-0.81}$ & 6.5  & 0.054 & Inconclusive (nearby star not ruled out) \\
8347680  & 24 & $5.13766^{+0.00405}_{-0.00318}$ & $1955.8294^{+0.0076}_{-0.0093}$ & $2.64^{+6.75}_{-0.33}$  & 7.4  & 0.240 & Likely nearby false positive \\
8415512  & 25 & $3.55497^{+0.00143}_{-0.00138}$ & $1987.0256^{+0.0058}_{-0.0056}$ & $9.54^{+14.93}_{-5.36}$ & 6.6  & 0.494 & Likely planet \\
9038546  & 24 & $3.28212^{+0.00267}_{-0.00292}$ & $1956.4049^{+0.0130}_{-0.0118}$ & $2.29^{+0.34}_{-0.35}$  & 6.7  & 0.044 & Likely planet \\
9472926  & 25 & $4.25383^{+0.00537}_{-0.00551}$ & $1986.3237^{+0.0128}_{-0.0109}$ & $2.48^{+0.39}_{-0.41}$  & 7.8  & 0.017 & Likely planet \\
9590211  & 26 & $4.92074^{+0.00261}_{-0.00222}$ & $2012.6438^{+0.0069}_{-0.0087}$ & $9.47^{+3.89}_{-1.90}$  & 10.3 & 1.000 & Likely nearby false positive \\
11895422 & 22 & $2.96672^{+0.00155}_{-0.00208}$ & $1902.8242^{+0.0118}_{-0.0074}$ & $2.62^{+1.42}_{-0.37}$  & 6.8  & 0.049 & Likely planet \\
11940899 & 21 & $3.75792^{+0.00117}_{-0.00130}$ & $1872.1300^{+0.0044}_{-0.0040}$ & $7.87^{+5.28}_{-2.91}$  & 6.4  & 0.587 & False positive favored \\
18269311 & 26 & $3.43277^{+0.00147}_{-0.00123}$ & $2011.3702^{+0.0044}_{-0.0051}$ & $4.72^{+7.27}_{-0.48}$  & 7.0  & 0.300 & Likely planet \\
\enddata
\tablecomments{Table~\ref{tab:candidates} is published in its entirety in machine-readable format; \NCatalog{} candidates are listed there and a portion is shown here for guidance regarding form and content. Columns shown are the subset most useful in print; the complete column set of the machine-readable version is described in Table~\ref{tab:candidates_cols}. Each row gives one catalog candidate, and every quantity in it comes from the sector whose ephemeris we adopted in Section~\ref{sec:reconciliation}. We give each uncertainty as the distance from the posterior median to the 16th or 84th percentile of the posterior (Section~\ref{sec:jaxoplanet}). Candidates with $R_P > \TriRadiusCeiling{}$~\RE{} are flagged in the machine-readable table; their FPP is reported but is less informative (Section~\ref{sec:triceratops}). Candidates whose posterior median impact parameter places the transit in grazing geometry, $b > 1 - R_P/R_\star$, are flagged, because their radii are degenerate with $b$ (Section~\ref{sec:jaxoplanet}); for TIC~266307162 (Sector~10), the TLS depth gives $R_P = \GrazExRpTLS$~\RE{} against the refit value of \GrazExRpFit~\RE{}. For TIC~9590211 (Sector~26), the target cannot produce the observed depth under the \texttt{TRICERATOPS} dilution model, so we set its FPP and NFPP to 1 rather than compute them (Table~\ref{tab:triceratops}).} 
\end{deluxetable*}

% \end{deluxetable*}
\begin{deluxetable*}{lll}
\tabletypesize{\scriptsize}
\tablecaption{Column description for the machine-readable version of
Table~\ref{tab:candidates}.\label{tab:candidates_cols}}
\tablehead{\colhead{Column} & \colhead{Unit} & \colhead{Description}}
\startdata
\cutinhead{Identification}
\texttt{TIC}, \texttt{Sector} & \nodata & TESS Input Catalog identifier and adopted sector \\
  \texttt{RA}, \texttt{Dec} & deg & TIC position (ICRS, epoch J2000) \\
\texttt{Sectors\_rec} & \nodata & All sectors with an independent detection of this ephemeris \\
\texttt{Sectors\_P, Sectors\_T0} & d, BTJD & Period and epoch recovered in each sector of \texttt{Sectors\_rec} \\
\texttt{Sectors\_ratio} & \nodata & Period ratio $m{:}n$ of each sector's detection relative to \texttt{P} \\
\texttt{N\_TCE\_support} & \nodata & Sectors in which a TCE of any label matches this ephemeris (Section~\ref{sec:reconciliation}) \\
\cutinhead{Stellar parameters}
\texttt{Tmag}, \texttt{Teff} & mag, K & TESS magnitude and stellar effective temperature \\
\texttt{Rstar}, \texttt{Mstar} & \RS, \MS & Stellar radius and mass \\
\cutinhead{Transit model, Section~\ref{sec:jaxoplanet}}
\texttt{P}, \texttt{T0}\tablenotemark{a} & d, BTJD & Orbital period and mid-transit epoch \\
  \texttt{T14}\tablenotemark{a} & h & Total transit duration \\
\texttt{RpRs}, \texttt{b}\tablenotemark{a} & \nodata & Planet-to-star radius ratio and impact parameter \\
\texttt{Rp}\tablenotemark{a} & \RE & Planet radius from \texttt{RpRs} and \Rstar \\
\texttt{aRs} & \nodata & Scaled semi-major axis from $P$, \Mstar, and \Rstar \\
\texttt{Sinc} & $S_\oplus$ & Incident stellar flux (Equation~\ref{eq:insolation}) \\
\texttt{rhat\_max} & \nodata & Largest rank-normalized split-$\hat{r}$ across fitted parameters \\
\texttt{Fit\_flag} & \nodata & 0 converged, 1 prior-limited $R_P/R_\star$ \\
\texttt{Grazing} & \nodata & 1 if the posterior median $b > 1 - R_P/R_\star$ (Section~\ref{sec:jaxoplanet}) \\
\cutinhead{Detection statistics}
\texttt{MES}, \texttt{Ntransit}, \texttt{SDE} & \nodata & \leovetter{} MES; TLS transit count in the adopted sector and SDE \\ \texttt{P\_TLS}, \texttt{Rp\_TLS} & d, \RE & TLS period and TLS-depth radius of the adopted detection \\
\cutinhead{False positive probabilities, Section~\ref{sec:triceratops}}
\texttt{FPP}, \texttt{NFPP}\tablenotemark{b} & \nodata & \texttt{TRICERATOPS} FPP and NFPP, medians across trials \\
\texttt{TRI\_label} & \nodata & \texttt{TRICERATOPS} label (Table~\ref{tab:triceratops}) \\
\texttt{Planet\_cand} & \nodata & 1 if a planet candidate (Section~\ref{sec:triceratops}) \\
\texttt{New} & \nodata & 1 if the host has no TOI, CTOI, or transiting NASA Exoplanet Archive planet \\
\texttt{Poorly\_vetted} & \nodata & 1 if poorly vetted (Section~\ref{sec:triceratops}) \\
\texttt{Large\_flag} & \nodata & 1 if $R_P > \TriRadiusCeiling$~\RE, where FPP is less informative \\
\cutinhead{Contamination and cross-match}
\texttt{FCR} & \nodata & Flux contamination ratio of the photometric aperture \\
\texttt{TOI}, \texttt{CTOI} & \nodata & TOI and CTOI matched to this signal (Section~\ref{sec:comparison}) \\
\texttt{TESS\_EB} & \nodata & 1 if the host appears in the TESS Eclipsing Binary Catalog \citep{Prsa:2022} \\
\texttt{P\_arch} & d & Period of the matched TOI, else of the matched CTOI \\
\texttt{C\_PRp} & \nodata & Sensitivity $C(P,R_P)$ at this candidate's grid location
\enddata
\tablenotetext{a}{\scriptsize Posterior median; columns suffixed \texttt{\_lo} and \texttt{\_hi} give the offsets to the 16th and 84th percentiles (days for \texttt{T0}). Intervals on \texttt{Rp} exclude the uncertainty on \Rstar.}
\tablenotetext{b}{\scriptsize Columns suffixed \texttt{\_lo} and \texttt{\_hi} give the 16th and 84th percentiles across trials.}
\tablecomments{\scriptsize \texttt{Sectors\_rec} lists the Primary Mission sectors that independently recovered this ephemeris (Section~\ref{sec:reconciliation}); it is not a vetting criterion. \texttt{C\_PRp} is the Sensitivity of the injection-grid cell (Section~\ref{sec:detection_efficiency}) containing the adopted detection's TLS period and TLS-depth radius (\texttt{P\_TLS} and \texttt{Rp\_TLS}), the quantities the occurrence rates use (Section~\ref{sec:occurrence}), and is empty outside the grid. \texttt{New} is set by TIC identifier alone, for catalog entries of any disposition; \texttt{TOI}, \texttt{CTOI}, and \texttt{P\_arch} come from the period and transit-time test of Section~\ref{sec:comparison}, list a CTOI since promoted to a TOI as that TOI, and are empty when no cataloged signal matches, even where the star hosts one. \texttt{FPP} and \texttt{NFPP} are set to 1, not computed, for the one candidate whose target cannot produce the observed depth, and are empty for the one labeled inconclusive from negative flux (Table~\ref{tab:triceratops}). \texttt{TESS\_EB} is a flag only and changes no label: the two new planet candidates on stars in that catalog, TIC~311188656 (Sector~22) and TIC~408474216 (Sector~14), remain planet candidates.}
\end{deluxetable*}

\begin{figure*}[p]
\centering
\includegraphics[width=\textwidth,height=0.76\textheight,keepaspectratio]{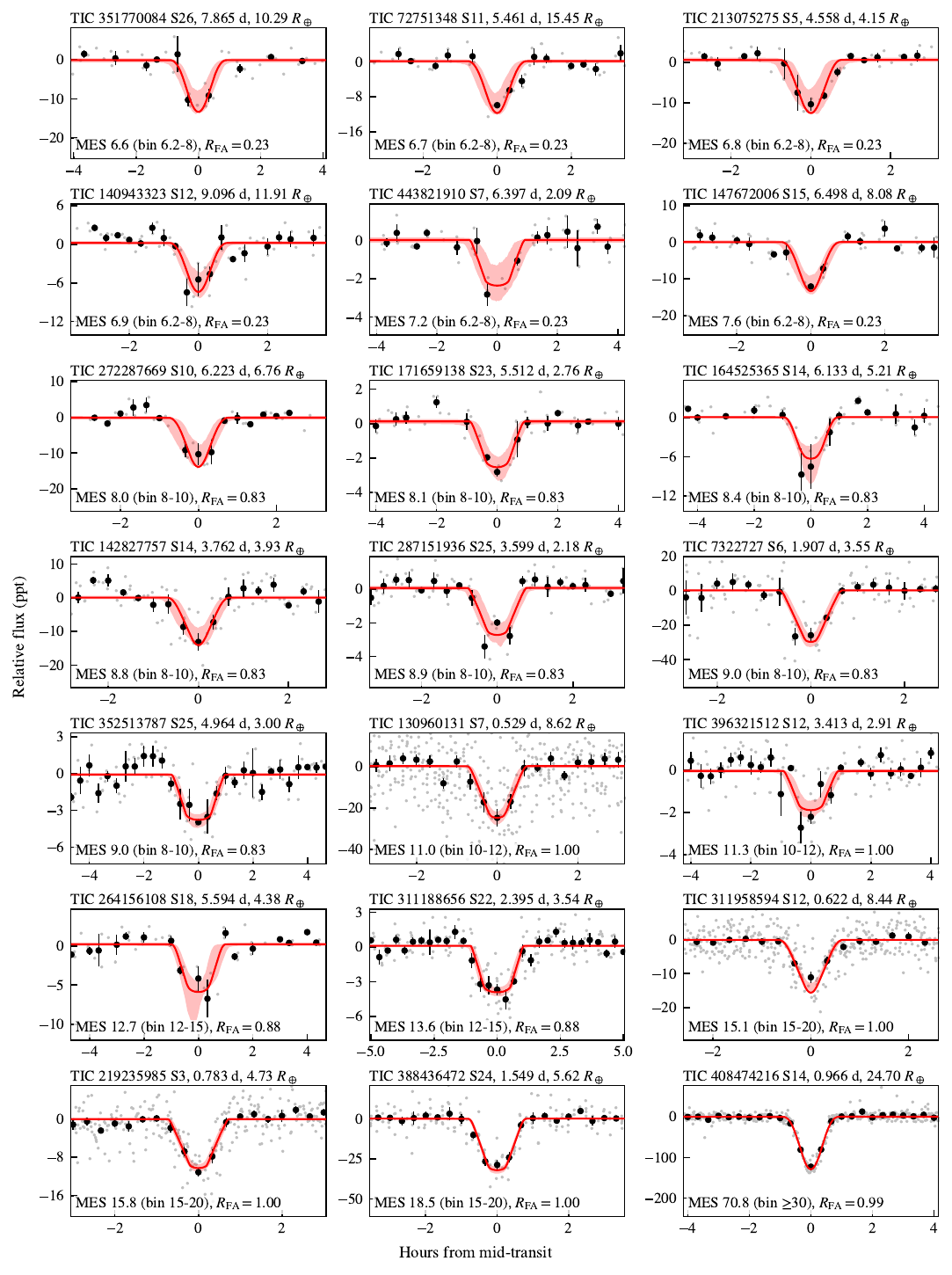}
\caption{Phase-folded \NEMESIS detrended light curves of \NGalleryPanels{} new planet candidates (Section~\ref{sec:triceratops}), ordered by MES from left to right and top to bottom. Each MES bin of Figure~\ref{fig:reliability} receives up to \NGalleryPerBin{} panels, and a slot that a bin cannot fill passes to the bins with candidates to spare, spread evenly between them and nearest first, so the bins show between none (MES 20 to 30, which holds no new planet candidate) and seven panels. Where a bin holds more candidates than panels, we show those whose folded transits stand out most clearly from the out-of-transit scatter, and a bin with no more candidates than panels shows all of them. We then removed three of the selected panels, two with sparse phase coverage and one with strong out-of-transit variation, and filled their slots by the same rule; this affects only the figure, and these are illustrative examples, not typical ones. We do not filter the panels on the large-radius, grazing, or prior-limited flags of Table~\ref{tab:candidates_cols}. Grey points are 30-minute cadences of the adopted sector against hours from mid-transit, black points are the same data in 20-minute phase bins, and the red line and pink band are the posterior median model and its 16th to 84th percentile range from Section~\ref{sec:jaxoplanet}, integrated over the exposure as in the fit. Each panel gives the TIC, sector, period, and planet radius, and beneath it the MES, its bin, and the reliability against false alarms $R_{\rm FA}$ of that bin. TIC~311188656 (Sector~22) and TIC~408474216 (Sector~14) are listed in the TESS Eclipsing Binary Catalog \citep{Prsa:2022}; neither has a TOI, CTOI, or NASA Exoplanet Archive entry, and both pass our vetting, so we keep them as planet candidates. TIC~408474216, at 24.7~\RE{}, also lies far above \TriRadiusCeiling{}~\RE{}, where FPP is less informative (Section~\ref{sec:triceratops}). The vetting report of every catalog candidate, whose first page shows its folded light curve, and its transit-fit posterior samples will be released with the High Level Science Product (Section~\ref{sec:Vetting_reports}).
}
\label{fig:gallery}
\end{figure*}

\subsubsection{Transit Parameter Estimation with \texttt{jaxoplanet}}
\label{sec:jaxoplanet}
\leovetter{} performs a least-squares transit fit as part of its metric computation, but its authors caution that these fits are intended for vetting rather than for publication: the impact parameter is capped at unity to aid convergence, which biases the radii of grazing transits low \citep{Kunimoto:2025}. Following Paper~I, we therefore refit every signal independently before reporting parameters. We model each candidate with the \texttt{jaxoplanet} package \citep{jaxoplanet}, a \texttt{JAX} re-implementation of \texttt{exoplanet} \citep{exoplanet:exoplanet} that evaluates the analytic quadratically limb-darkened light curve of \citet{exoplanet:agol19} for a circular orbit. Paper~I used \texttt{exoplanet} directly. We changed package for sampling speed across the larger candidate list, and we changed the model in the ways described below. We fit each reconciled signal once, using the \NEMESIS detrended flux of the sector whose ephemeris we adopted in Section~\ref{sec:reconciliation}, with flagged cadences removed (Section~\ref{subsubsec:masking}). The occurrence rates of Section~\ref{sec:occurrence} do not use these fits. Table~\ref{tab:candidates} reports one fit for each candidate in our catalog.

The free parameters are the mid-transit time $T_0$, the period $P$, the transit duration $T_{14}$, the radius ratio $R_P/R_\star$, the impact parameter $b$, two limb-darkening coefficients, a flux offset, and a jitter term that we add in quadrature to the flux uncertainties. We center the priors on the TLS solution. $T_0$ is normal with mean at the TLS epoch and width equal to the TLS duration. $\ln P$ is normal with width equal to the fractional TLS period uncertainty, clipped to $[\SigLnPmin, \SigLnPmax]$, or \SigLnPdefault{} where TLS reports no uncertainty. $\ln T_{14}$ is normal about the TLS duration with width \DurLogSigma. $R_P/R_\star$ is uniform on $[0.01, 1]$, and $b$ is uniform on $[0.01, 1 + R_P/R_\star]$, which, unlike the \leovetter{} fit, admits grazing geometries. Grazing posterior medians often reflect the prior rather than the data: the joint prior on $R_P/R_\star$ and $b$ places more than half of its mass in grazing configurations, and the 30-minute light curves constrain $b$ weakly, so grazing medians are more common below MES of \MESbinWeakHi{} than above it, and far more common among candidates labeled false positive favored than among the planet candidates. At 30-minute cadence a larger planet at high impact parameter produces a transit of nearly the same shape as a smaller one at low impact parameter, so an upper limit on $R_P/R_\star$ near the search value would truncate the posterior; a limit at unity does not. For \NCeilFinal{} candidates the posterior nonetheless reaches this limit, so the upper uncertainty on $R_P$ is set by the prior, and we flag them in Table~\ref{tab:candidates}. We place normal priors of width \LDsigma{}, truncated to the interval $[0, 1]$, on the limb-darkening parameters $q_1$ and $q_2$ of \citet{exoplanet:kipping13}, centered on the transform of the \citet{Claret:2017} TESS-band quadratic coefficients for each host, the values also supplied to \leovetter{} and to the injections. Paper~I left the coefficients free. We center them on tabulated values
instead, because at 30-minute cadence the ingress is unresolved and free
coefficients would trade off against $R_P/R_\star$ and $b$ without being
constrained by the data.

We use a Student-t likelihood with $\nu = \StudentNu$ degrees of freedom rather than a normal one, taking as its scale the reported flux uncertainty added in quadrature to the jitter term. At 30-minute cadence a single outlying exposure spans a large fraction of a short transit. Under a normal likelihood the jitter term grows to accommodate such a cadence and inflates the uncertainty on every other cadence, whereas the heavier tails of the Student-t absorb it locally. We fix $\nu$ rather than sampling it, so that every candidate is fit with the same noise model. For $\nu = \StudentNu$ the variance is twice the square of the scale, so the fitted jitter is not directly comparable to a Gaussian standard deviation.

At 30-minute cadence the ingress is not resolved, so the light curve alone constrains $b$ poorly. We therefore add a log-normal prior on the stellar density, with a fractional width of \RhoFracErr\%, centered on the density implied by the TIC stellar mass and radius. For a circular orbit, $P$, $T_{14}$, $b$, and $R_P/R_\star$ together fix the stellar density, so this prior lets the transit duration determine $b$.

Because the 30-minute exposure is a substantial fraction of a short-period M-dwarf transit, evaluating the model at cadence midpoints alone would smear ingress and egress and bias the fitted duration and impact parameter. For every model evaluation we therefore average \NOversample{} subsamples across each exposure.

We fit only the cadences within \FitWindowDur{} transit durations of a predicted mid-transit time, and we fall back to the full light curve when fewer than \FitWindowMinCad{} cadences remain. We start each fit at the maximum a posteriori solution and sample the posterior with the No-U-Turn Sampler \citep{Hoffman:2014} as implemented in \texttt{numpyro} \citep{numpyro:2019}, running \NChains{} chains of \NTune{} tuning steps and \NDraws{} posterior draws each, with a target acceptance probability of \TargetAccept{}. We accept a fit as converged when the rank-normalized split-$\hat{r}$ \citep{Gelman:1992, Vehtari:2021} satisfies $\hat{r} \le \RhatMax$ for every parameter, the bulk and tail effective sample sizes of \citet{Vehtari:2021} are both at least \ESSmin, and the sampler reports no divergent transitions. The fits of \NUnconverged{} signals do not pass this test. We drop these signals: without a converged fit, their transit parameters, and therefore the inputs to the false positive analysis of Section~\ref{sec:triceratops}, are not constrained. Their detections remain among the \PCs{} PC detections and in the reliability analysis of Section~\ref{sec:scramble}, because the vetting rule shared by the blind, injection, and scramble analyses does not involve the refit; having no FPP, they do not enter the occurrence rates (Section~\ref{sec:occurrence}). They are TIC~67772635 (Sector~24), TIC~271725674 (Sector~6), TIC~340638718 (Sector~10), TIC~371857332 (Sector~23), and TIC~375090800 (Sector~3).

We derive planet radii from $R_P/R_\star$ and the TIC stellar radius, and the semi-major axis from $P$ and the TIC stellar mass through Kepler's third law. We adopt the TIC stellar parameters as fixed values, so the intervals we report for $R_P$ reflect the transit-fit posterior alone and do not include the uncertainty on $R_\star$. We list $a/R_\star$ and $S_{\rm inc}$ without intervals: both follow from $P$ and stellar parameters that we hold fixed ($M_\star$ and $R_\star$ for $a/R_\star$, and also $T_{\rm eff}$ for $S_{\rm inc}$), so their posterior widths would reflect only the period posterior. We treat the photometric noise as independent between cadences, apart from the jitter term, and we do not model time-correlated noise, so where such noise is present the uncertainties in Table~\ref{tab:candidates} are underestimated. We defer a joint multi-sector fit, and the extension of the survey itself to multi-sector light curves, to a future paper.

\begin{figure*}[ht!]
\centering
\includegraphics[width=\textwidth]{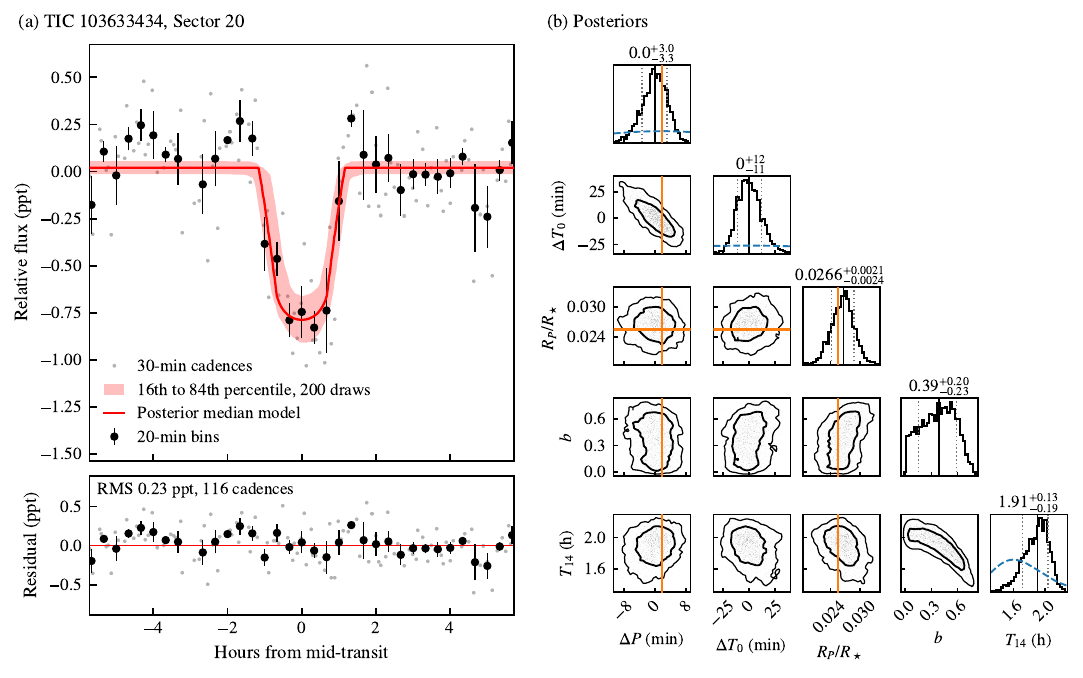}
\caption{The transit fit of Section~\ref{sec:jaxoplanet} for TOI-1235.01 (TIC~103633434, Sector~20). (a) The \NEMESIS detrended light curve folded on the posterior median ephemeris: 30-minute cadences (grey), the same data in 20-minute phase bins (black), the posterior median model integrated over the 30-minute exposure (red), and the 16th to 84th percentile range of 200 posterior draws (pink), with residuals below (RMS 0.23~ppt). (b) Posteriors of the period and mid-transit time, as offsets from their medians in minutes, $R_P/R_\star$, $b$, and $T_{14}$, with 68\% and 95\% contours. Dashed blue curves show the priors on period, epoch, and $T_{14}$, and orange lines mark the TLS period and the $R_P/R_\star$ implied by the TLS depth. The data constrain the ephemeris, $T_{14}$, and $R_P/R_\star$; the stellar-density prior sets the density and most of the constraint on $b$. Limb darkening and the jitter term are fitted but not shown. Marginals and contours use the 2{,}400 saved posterior draws (every fifth draw); their medians agree with those of Table~\ref{tab:candidates}, computed from all 12{,}000 draws, to within $0.05\sigma$.}
\label{fig:jaxoplanet_example}
\end{figure*}

Figure~\ref{fig:jaxoplanet_example} shows one fit in detail, for TOI-1235.01 (TIC~103633434, Sector~20), the confirmed planet of Figure~\ref{fig:vetting_report_example}. Our search recovered it in Sectors 20 and 21, with MES of 12.76 and 7.75. TESS also observed it in Sector~14, where the light curve was about 75\% noisier than in Sector~20 and the strongest TLS peak, at 2.54~d with MES 4.98, fell below our detection threshold. The fit converges well within our criteria, with $\hat{r} = 1.00$ for every parameter, bulk and tail effective sample sizes of 2{,}508 and 2{,}615, and no divergent transitions. The data constrain the ephemeris, $T_{14}$, and $R_P/R_\star$: the posteriors on period and epoch are 0.17 and 0.12 times as wide as their priors, and the posterior on $T_{14}$, half as wide as its prior, moves from the TLS value of 1.66~h to $1.90^{+0.14}_{-0.19}$~h. The priors set the rest. The stellar density, $3.50^{+0.51}_{-0.44}$~g~cm$^{-3}$, is nearly as wide as its TIC-based prior, the limb-darkening coefficients return their prior values, and $b = 0.39^{+0.20}_{-0.24}$ is weakly constrained, as expected when the 30-minute cadence leaves ingress and egress unresolved. We find $R_P = 1.82^{+0.15}_{-0.16}$~\RE{}, within $1\sigma$ of both the $1.738^{+0.087}_{-0.076}$~\RE{} of \citet{Cloutier:2020b}, who adopt the same stellar radius, 0.630~\RS{}, so that the comparison tests our transit fit rather than the stellar parameters, and the $1.69 \pm 0.08$~\RE{} of \citet{Bluhm:2020}.

Figure~\ref{fig:gallery} shows the phase-folded light curve and fitted model of \NGalleryPanels{} of our new planet candidates; the vetting report of every catalog candidate will be released with the High Level Science Product (Section~\ref{sec:Vetting_reports}). We spread the printed examples across the MES bins of Figure~\ref{fig:reliability} rather than drawing them from the strongest signals, because \NPCweakbin{} of the \PCs{} detections lie below MES of \MESbinWeakHi{}. Where a bin holds more candidates than panels we show the clearest, and we replaced three panels with sparse phase coverage or strong out-of-transit variation (Figure~\ref{fig:gallery}), so the panels show what a detection can look like at each MES rather than a typical one.

\subsubsection{False Positive Probabilities with \texttt{TRICERATOPS}}
\label{sec:triceratops}

Vetting establishes that a signal is transit-like and is not an obvious eclipsing binary. It does not establish that the signal is planetary. To quantify the remaining astrophysical false positive risk for individual candidates we use \texttt{TRICERATOPS} \citep{Giacalone:2021}, which returns the false positive probability (FPP) that a signal assumed to be astrophysical is something other than a planet on the target, and the nearby false positive probability (NFPP) that it arises on a resolved neighbor.

We evaluate every candidate in the catalog, including those whose fitted radius ratio is limited by its prior. For each, we supply \texttt{TRICERATOPS} \citep[version~\TriVersion{};][]{Giacalone:2020} with the \NEMESIS detrended light curve of the adopted sector, the TIC stellar parameters, and the actual pixel mask used for the \NEMESIS photometry of that star-sector (Section~\ref{subsubsec:SAP}). Supplying the true aperture rather than a nominal one matters, because both the dilution of the observed depth and the set of neighbors able to produce that depth follow from which pixels were summed, and \NEMESIS apertures are constructed per target. \NEMESIS removes dilution during photometry, subtracting the flux contamination ratio $f$ and renormalizing by $1-f$ (Section~\ref{subsubsec:SAP}), whereas \texttt{TRICERATOPS} expects a light curve normalized to the total flux in the aperture and applies its own dilution model to every candidate host. We therefore multiply the departure of each light curve from unity, its uncertainty, and the input transit depth by $1-f$, which exactly reverses the \NEMESIS correction, so that dilution enters the calculation once. In these units a transit of depth $\delta$ in the deblended light curve has depth $(1-f)\,\delta$, and \texttt{TRICERATOPS} attributes $(1-f)\,\delta/X$ to the target, where $X$ is its own estimate of the fraction of the aperture flux that the target contributes; this equals $\delta$ only where $X = 1-f$. We fold the light curve on the posterior median period, mid-transit time, and duration of Section~\ref{sec:jaxoplanet} and keep the unbinned cadences within \TriWindowDur{} transit durations of mid-transit, or within $\pm\TriWindowPhase$ in phase where that window would exceed the period. A wider window admits isolated outlying cadences far from transit, which can drive the likelihood of every scenario below the range of double-precision arithmetic; at \TriWindowWide{} durations this left two candidates with an undefined FPP. We assign every cadence a single uncertainty, $1.4826$ times the median absolute deviation of the cadences more than \TriOOTdur{} transit durations from mid-transit. The input depth, which determines which sources in the field could produce the signal, is the fractional TLS depth of the adopted detection (Section~\ref{sec:transit_search}), converted in the same way. It is a median \TriDepthShallow{} shallower than the deepest point of the refit model (\TriDepthShallowOK{} where the radius ratio is not limited by its prior). A shallower depth admits fainter sources as possible hosts, so this choice makes the dispositions more conservative: compared with the model depth, it adds at least one possible host for \NTriDepthMoreHosts{} candidates and removes at least one for \NTriDepthFewerHosts{}.

Stellar masses, radii, effective temperatures, and parallaxes are the TIC values (with Gaia DR2 parallaxes), identical to those adopted by \NEMESIS, and neighbors without TIC stellar properties take the default values that \texttt{TRICERATOPS} assigns. Background populations come from TRILEGAL \citep{Girardi:2005}, queried at each target's coordinates. We set the exposure time of the model light curves to the cadence of the light curve being evaluated, \TriExptime{} minutes for the Primary Mission FFIs, because the \texttt{TRICERATOPS} default of two minutes compares sharp model transits with smeared data. We have no high-resolution imaging for these targets, so we supply no contrast curve, and the probabilities we report are conditioned only on catalog-resolved and statistically modeled companions.

Because the scenario likelihoods are evaluated by Monte Carlo sampling, repeated runs on identical inputs return different probabilities. We run \NTriTrials{} independent trials per candidate, each with \TriMCsamples{} samples, and report the median FPP and NFPP across trials together with their 16th and 84th percentiles. Within each trial we sum the probability of each scenario type over all stars, take the median of these sums across trials, and report as the top scenario the type with the largest summed median, together with the star that contributes most to it.

We evaluate every catalog candidate regardless of radius. Above approximately \TriRadiusCeiling{}~\RE, giant planets, brown dwarfs, and the lowest-mass stars occupy overlapping radii, so transit depth alone no longer separates the planetary and stellar scenarios and a low FPP is correspondingly less informative. We therefore report \texttt{TRICERATOPS} results for the \NCatalogLarge{} candidates above this radius but flag them in Table~\ref{tab:candidates} as requiring spectroscopic follow-up before their FPP is interpreted. We discuss two of them in Section~\ref{sec:limitations}.

\citet[][Section~4.2]{Giacalone:2021} caution against validating planets with \texttt{TRICERATOPS} from 30-minute photometry alone, while its likely-planet and likely-nearby-false-positive labels remain informative at that cadence. Because \NEMESIS~II uses only 30-minute data, we claim no candidate as a validated planet. Nonetheless, \NValidated{} candidates, all labeled likely planets, meet the validation thresholds of that work ($\mathrm{FPP} < 0.015$ and $\mathrm{NFPP} < \NFPPmax$) at both the median and the 84th percentile across trials. All but one, TOI-5486.01, are new. One of the new ones, TIC~408474216, has $R_P = 24.7$~\RE{}, far above \TriRadiusCeiling{}~\RE{}, and its host is listed in the TESS Eclipsing Binary Catalog \citep{Prsa:2022}; the other four are the most promising targets for validation with 2-minute or 10-minute photometry. We assign each candidate the first label in Table~\ref{tab:triceratops} whose criterion it meets, using the median and the 16th and 84th percentiles across trials. A likely planet requires $\mathrm{FPP} < \FPPmax$ and $\mathrm{NFPP} < \NFPPmax$ at the 84th percentile, so that a favorable minority of trials cannot earn the label, and a false positive is favored when the 16th percentile of FPP exceeds \FPPmax{}, the mirror of that condition. A likely nearby false positive requires a median NFPP above \NFPPnfp{}. Two outcomes are set before the trials are summarized. When the \texttt{TRICERATOPS} dilution model implies that the target would need more than its entire flux to produce the observed depth, stock \texttt{TRICERATOPS} evaluates a neighbor in the target's place; we instead do not run it, set $\mathrm{FPP} = \mathrm{NFPP} = 1$, and label the candidate a likely nearby false positive. We also count the cadences below zero flux in the converted light curve. In aperture units the flux is all the light the aperture recorded, which no star, eclipse, or blend can drive below zero, so such a cadence marks a data-quality problem rather than an astrophysical signal. We flag any candidate with one as poorly vetted, without removing data, and label it inconclusive when in addition no trial returns a finite probability. \NPoorlyVetted{} of the \NTriEvaluated{} candidates is flagged, TIC~401320705 (Sector~25): its target supplies about \TIClowShare{} of the aperture flux, so deblending magnifies its noise by a factor of about \TIClowShareAmp{}, and every trial for it is undefined.

\begin{deluxetable*}{cllr}
\tabletypesize{\footnotesize}
\tablecaption{\texttt{TRICERATOPS} classification of the \NTriEvaluated{} evaluated candidates.\label{tab:triceratops}}
\tablehead{\colhead{Order} & \colhead{Label} & \colhead{Criterion} & \colhead{$N$}}
\startdata
1 & Inconclusive (evidence of negative flux) & No finite trial; a cadence below zero flux & \NTriNegFlux \\
2 & Likely nearby false positive              & Target cannot produce the depth; $\mathrm{FPP}=\mathrm{NFPP}=1$ set    & \NTriNoDip \\
3 & Likely nearby false positive              & $\mathrm{NFPP}_{50} > \NFPPnfp$                                         & \NLikelyNFP \\
4 & False positive favored                    & $\mathrm{FPP}_{16} > \FPPmax$                                           & \NFPfavored \\
5 & Likely planet & $\mathrm{FPP}_{84} < \FPPmax$ and $\mathrm{NFPP}_{84} < \NFPPmax$ & \NLikelyPlanet \\
6 & Inconclusive (nearby star not ruled out) & $\mathrm{FPP}_{84} < \FPPmax$ and $\mathrm{NFPP}_{84} \geq \NFPPmax$    & \NTriNearbyOpen \\
7 & Inconclusive (trials split)               & None of the above                                                       & \NTriSplit \\
\enddata
\tablecomments{Each candidate receives the first label whose criterion it meets. Subscripts give the percentile across the \NTriTrials{} trials. The rule set also defines two outcomes that no candidate meets: no finite trial without negative flux, and no source in the field able to produce the depth. No validated tier is assigned at 30-minute cadence \citep[][Section~4.2]{Giacalone:2021}.}
\end{deluxetable*}

\begin{figure}[ht!]
\centering
% \includegraphics[width=0.5\textwidth]{example-image-a}
% Final: 
\includegraphics[width=\columnwidth]{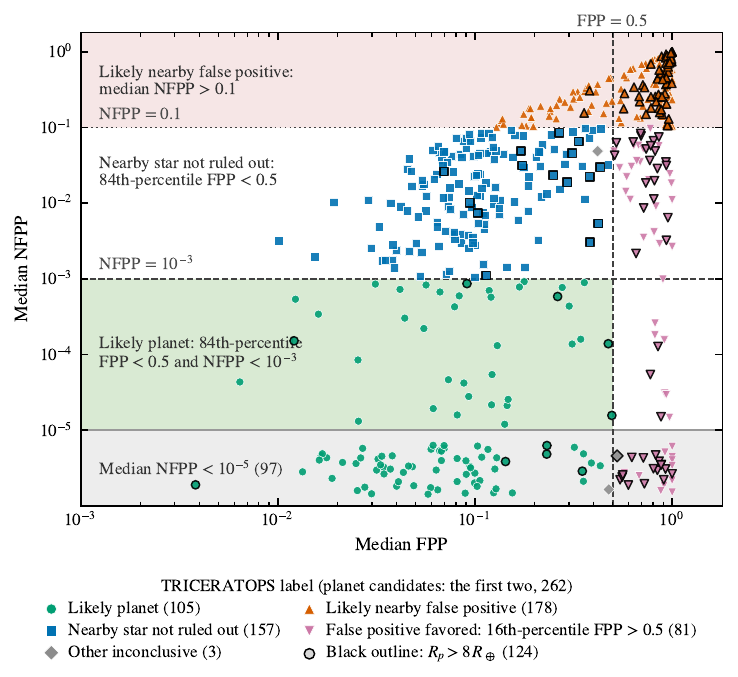}
\caption{
False positive probabilities of \NTriPlotted{} of the \NTriEvaluated{} candidates evaluated with \texttt{TRICERATOPS} (Section~\ref{sec:triceratops}); the one with no finite trial, which has $R_P > 8$~\RE{}, is omitted. Each point is the median FPP and NFPP across \NTriTrials{} trials. Medians of NFPP below $\TriPlotFloor$, including the \NNFPPzero{} candidates with $\mathrm{NFPP} = 0$, are spread at random heights across the grey strip below that value so that they overlap less, and the candidate whose target cannot produce the observed depth is drawn at $(1, 1)$. Dashed lines mark $\mathrm{FPP} = \FPPmax$ and $\mathrm{NFPP} = \NFPPmax$, and the dotted line marks $\mathrm{NFPP} = \NFPPnfp$. Marker shape and color give the label of Table~\ref{tab:triceratops}: likely planet (circles), inconclusive with a nearby star not ruled out (squares), other inconclusive (diamonds; the trials-split label), likely nearby false positive (upward triangles), and false positive favored (downward triangles). The first two form our planet candidates. A likely planet needs the 84th percentiles of FPP and NFPP, not only the medians, to fall in the shaded likely-planet region, so a candidate whose medians lie there can still be inconclusive. Black outlines mark candidates with $R_P > \TriRadiusCeiling$~\RE{}, for which FPP is less informative; the omitted candidate is also above this radius.
}
\label{fig:triceratops}
\end{figure}

These labels characterize individual candidates, and we do not use them to remove any candidate from the catalog. Of the \NTriEvaluated{} candidates, \NLikelyPlanet{} are likely planets, \NLikelyNFPall{} are likely nearby false positives, \NFPfavored{} favor a false positive, and \NTriUncertain{} are inconclusive (Table~\ref{tab:triceratops}). \NLargeLP{} of the likely planets have $R_P > \TriRadiusCeiling$~\RE, where planets, brown dwarfs, and the smallest stars overlap in radius and \texttt{TRICERATOPS}, which has no brown-dwarf scenario, cannot tell them apart (Section~\ref{sec:limitations}). Three of the inconclusive candidates have trials that straddle the FPP threshold in both directions. At \TriMCsamples{} samples the trial-to-trial scatter is small compared with these thresholds: across the computed candidates, the 16th and 84th percentiles of FPP differ by a median of \TriScatter{}. Figure~\ref{fig:triceratops} places the evaluated candidates in the plane of FPP and NFPP: the likely planets sit at low FPP (median \TriMedFPPPlanet{}) and the likely nearby false positives at high values of both (medians \TriMedFPPNFP{} and \TriMedNFPPNFP{}).

We report as planet candidates the \NFavored{} candidates for which \texttt{TRICERATOPS} favors a planetary origin: those labeled likely planets or inconclusive with a nearby star not ruled out (Table~\ref{tab:triceratops}). \NNewFavored{} of them lie on stars with no TOI, CTOI, or transiting planet in the NASA Exoplanet Archive \citep{Akeson:2013}. Here and below, ``planet candidate'' refers to this set, while a ``PC detection'' is a star-sector that carries the vetting label ``PC'' of Section~\ref{sec:transit_search}; the other candidates stay in the catalog with their labels. Both groups have $\mathrm{FPP} < \FPPmax$ at the 84th percentile across trials, and they differ in whether an eclipse on a resolved neighbor is excluded, which holds for \NNewFavoredClear{} of the new planet candidates. The other \NNewFavoredNbr{} need seeing-limited photometry of their neighbors or high-resolution imaging first; TOI-3785~b, a confirmed planet on a known host, carries the same label (below). Because \texttt{TRICERATOPS} assumes that a signal is astrophysical, it does not screen false alarms: \NNewFavoredLowMES{} of the new planet candidates have MES below \MESbinWeakHi{}, where PC detections have $R_{\rm FA} \approx \RFAweakbin$ (Section~\ref{sec:scramble}), and the \NNewFavoredHighMES{} at or above it are the strongest targets for follow-up (Section~\ref{sec:followup}). \NNewFavoredLarge{} have $R_P > \TriRadiusCeiling$~\RE{}, where FPP is less informative. The occurrence rates of Section~\ref{sec:occurrence} do not select on these labels; the rate that corrects for astrophysical false positives weights every PC detection by its FPP instead.

As a check on these labels, we compare them with the dispositions of the TESS Follow-up Observing Program Working Group (TFOPWG) for the \NTriTOI{} evaluated candidates that match a TOI in period and transit timing (Section~\ref{sec:comparison}), taking each disposition from the TFOPWG column of ExoFOP-TESS, or from the TOI catalog where ExoFOP gives none. Of the \NTriTOIknown{} confirmed or known planets among them, \NTriTOIknownLP{} are likely planets, including TOI-270~c \citep{Gunther:2019}, GJ~436~b \citep{Butler:2004, Gillon:2007}, and L~98-59~c \citep{Kostov:2019a}, and \NTriTOIknownNbr{} are inconclusive with a nearby star not ruled out. Of the rest, one has split trials, \NTriTOIknownNFP{} are likely nearby false positives (among them TOI-2407.01), and \NTriTOIknownFPF{} are false positive favored: TOI-5205.01 and TOI-6894.01, whose refit radii exceed \TriRadiusCeiling~\RE, where FPP is less informative. Of the three TFOPWG false positives among the matches, TOI-5268.01 and TOI-5695.01 are false positive favored and TOI-497.01 is a likely nearby false positive. TOI-3785.01 shows what the inconclusive label means. It is the confirmed planet TOI-3785~b \citep{Powers:2023}, and our fitted radius of \TOIpubRpNEM{}~\RE{} agrees with their $\TOIpubRp \pm \TOIpubRpErr$~\RE{}. Its FPP is low, but its NFPP does not exclude an eclipse on a resolved neighbor, so it is labeled inconclusive: the label records what our data cannot exclude, not where the signal lies.

\subsection{Characterizing Completeness and Reliability}
\label{sec:injection_analysis}

The candidate list of Section~\ref{sec:candidates} is not interpretable on its own. Two numbers are needed to turn it into a measurement: the fraction of planets present in the data that the pipeline would have found and kept, and the fraction of the surviving list that is not a planet at all. We measure both empirically, by running the same pipeline on two modified versions of the same light curves.

In the injection arm (Sections~\ref{sec:repstars} and
\ref{sec:detection_efficiency}), synthetic transits of known period and radius are added to the background-subtracted aperture flux of a representative subsample of survey stars, before contamination correction, PLD correction, and detrending. Because the injected planet is known, this arm measures the pipeline's sensitivity,
\begin{equation}
C(P, R_P) = \frac{N(\mathrm{recovered\ and\ classified\ PC})}{N(\mathrm{injected})}
\label{eq:sensitivity}
\end{equation}
on a grid in orbital period and planet radius. $C(P, R_P)$ sets the level of the completeness correction that the occurrence rates of Section~\ref{sec:occurrence} divide by (Equation~\ref{eq:pdet}).

$C$, $E$, and $R_{\rm FA}$ are meaningful only if the three arms of this analysis are comparable. For this reason, the three arms share one code path: the scramble arm modifies the pixel cube before SAP extraction, the injection arm modifies the SAP flux immediately after it, and from there contamination correction, PLD, detrending, outlier rejection, the BLS and TLS searches, vetting report generation, and the Stage~A and Stage~B rule are the same code (Figure~\ref{fig:flowchart}). The one exception is that the injection arm applies Stage~A as a flag rather than a filter, so that injections that never became TCEs remain in the completeness denominator.

Four quantities appear in what follows and are easily conflated, because three of them are called some form of completeness. Their definitions and denominators are collected in Table~\ref{tab:quantities} and we use the names in that table consistently thereafter.

\begin{deluxetable*}{lll}
\tabletypesize{\small}
\tablecaption{Completeness-like quantities used in this work. All are
distinct and none may be substituted for another.\label{tab:quantities}}
\tablehead{
\colhead{Quantity} & \colhead{Numerator} & \colhead{Denominator}
}
\startdata
Recoverability         & ephemeris recovered (Eq.~\ref{eq:coherence}) & all injections placed \\
Vetter-Completeness    & recovered and classified PC                  & recovered injections \\
Sensitivity $C(P,R_P)$ & recovered and classified PC                  & all injections placed \\
$C_{\rm vet}$          & recovered injected TCEs classified PC        & recovered injected TCEs
\enddata
\tablecomments{Cell by cell, Recoverability $\times$ Vetter-Completeness
$=$ Sensitivity, which we use as an internal consistency check on the maps.
Survey completeness, shown in Figure~\ref{fig:completeness} for reference, is a fifth quantity: the product of $C(P,R_P)$ with the geometric transit probability
$p_{\rm geo}$ (Equation \ref{eq:pgeo}). It is not a recovery fraction and has no denominator of its own. Section~\ref{sec:occurrence} uses its star-by-star counterpart, $p_{\rm geo}\,p_{\rm det}$ (Equation~\ref{eq:neff}).}
\end{deluxetable*}

\subsubsection{Selecting Representative Stars for Injection}
\label{sec:repstars}

Injecting into every surveyed star-sector is computationally prohibitive. \NEMESIS injects each transit inside the photometry step, into the background-subtracted aperture flux, so each injection reruns the full extraction of its star-sector from the image cutout (quality masking, aperture selection, background and centroid estimation, contamination correction, PLD, smoothing, and outlier removal) before the transit search and vetting. Following Paper~I \citep{Feliz:2021}, we instead inject into a smaller sample of stars chosen so that their photometric noise properties span those of the survey. Sensitivity at fixed $(P, R_P)$ is governed primarily by the photometric precision of the light curve rather than by which sector produced it \citep[e.g.][]{Kunimoto:2025}, so a sample stratified in noise spans the survey's range of sensitivity without reproducing its size.

Candidate representative stars are drawn from the searched star-sectors and must satisfy three simultaneous cleanliness cuts, so that an injected signal is not competing with a pre-existing one:

\begin{enumerate}
\item \emph{No threshold crossing event.} The star-sector fails the Stage~A criterion of Section~\ref{sec:transit_search}.
\item \emph{Low aperture contamination.} Flux contamination ratio
      $\mathrm{FCR} \le \repFCRmax{}$ (Section~\ref{subsubsec:SAP}).
\item \emph{No strong rotational modulation.} \leovetter{} sinusoidal
      significance $\le \repSineMax{}$, the threshold of its \texttt{sinusoidal} test.
\end{enumerate}

A star with a missing value for any of these metrics is excluded, because cleanliness cannot be confirmed in that dimension. Surviving stars are cross-matched against the TOI and CTOI catalogs and any known or candidate planet host is removed. Because the first cut uses the same rule that defines the blind catalog, no injection is placed on top of a real signal that the pipeline would have detected.

\begin{figure*}%[H]
\centering
\includegraphics[width=\textwidth,keepaspectratio]{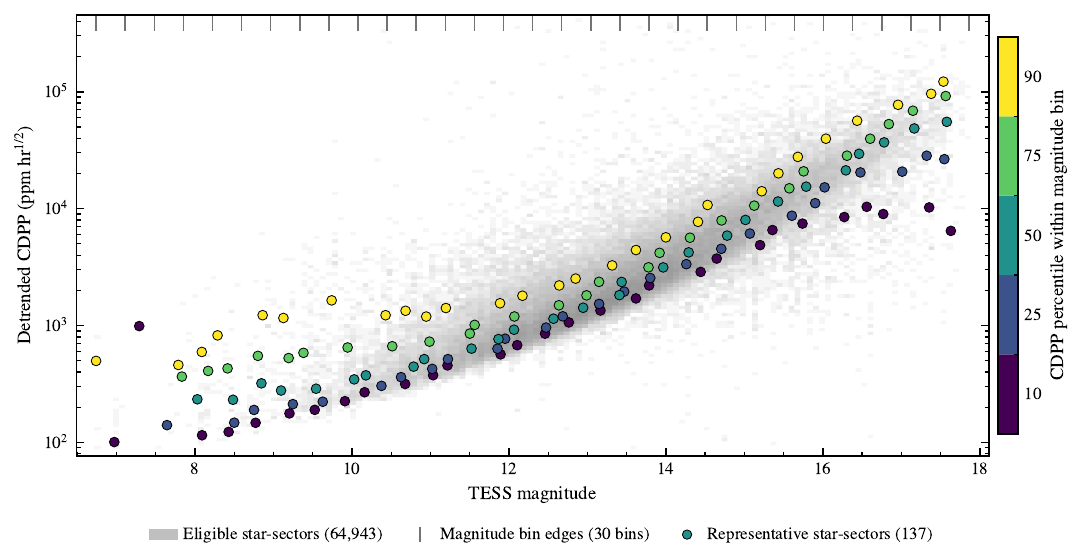}
\caption{
Selection of the \NRepStars{} representative star-sectors. Detrended CDPP against TESS magnitude for the star-sectors eligible for selection (grey; Section~\ref{sec:repstars}), with the \repNbins{} magnitude-bin edges marked along the top axis. Filled circles are the representative star-sectors, colored by the in-bin CDPP quantile that each was chosen to match.
}
\label{fig:repstars}
\end{figure*}

From the surviving pool we select deterministically rather than at random, so that the sample is reproducible across reruns. We divide the TESS magnitude range into \repNbins{} equal-width bins and, within each populated bin, select the star-sector whose detrended CDPP lies closest to each of \repNquantiles{} target quantiles of the in-bin CDPP distribution. This brackets the noise distribution at fixed brightness rather than sampling it, and guarantees coverage of the well-behaved and poorly-behaved tails at every magnitude. After de-duplication and removal of known or candidate planet hosts, the procedure yields \NRepStars{} representative star-sectors spanning $T = \NRepTmagMin{}$ to \NRepTmagMax{}. Each contributes one Primary Mission sector; sector diversity is obtained by drawing stars from many sectors rather than by injecting repeatedly into the same star. The selection is shown in Figure~\ref{fig:repstars}.

One consequence of the contamination cut must be stated plainly, because it is a mismatch between the completeness sample and the catalog that completeness is applied to. We restrict the injection sample to isolated stars because the purpose of the injections is to measure how well the pipeline detects a transit on the target star itself, separately from the added difficulty of a crowded aperture. Requiring $\mathrm{FCR} \le \repFCRmax{}$ removes \FCRcutFrac{} of searched star-sectors, whereas the blind survey applies no contamination cut. A neighbor in the aperture adds photon noise that the contamination correction of Section~\ref{subsubsec:SAP} cannot remove: the correction restores the transit depth but scales the noise by the same factor, so a transit on a contaminated host has lower signal-to-noise than the same transit on an isolated host of the same brightness. 

\subsubsection{NEMESIS Detection Efficiency as a Function of Orbital Period
and Planet Radius}
\label{sec:detection_efficiency}

\paragraph{Injection grid design }~~Synthetic transits are injected into the background-subtracted aperture flux of each representative star-sector, after the aperture is selected and before normalization and contamination correction, so that the injected signal passes through contamination correction, PLD correction, detrending, outlier rejection, the transit search, and vetting on the same footing as a real astrophysical signal. Each injected model is diluted by the flux contamination ratio of its host, the same ratio the contamination correction later removes, so the injections test every stage downstream of aperture photometry but not the aperture, the contamination estimate, or the removal of centroid-excursion cadences, which a flux-level injection cannot trigger, and the pixel time series that PLD regresses against do not contain the injected signal.

Transit models are computed with \texttt{batman} \citep{Kreidberg:2015} using the stellar radius and mass of the host and quadratic limb-darkening coefficients from \citet{Claret:2017}. We divide $P \in [\minInjP{}, \maxInjP{}]$~d into \NInjPeriodSteps{} logarithmically equal intervals and $R_P \in [\minInjRP{}, \maxInjRP{}]$~\RE{} into \NInjRadiusSteps{}, and draw one period and one radius log-uniformly within each interval with a fixed random seed, so the grid is exactly reproducible. The period range spans the regime accessible within a single $\sim$27~d sector; the radius range spans sub-Earth to Jupiter size and matches the occurrence-rate parameter space of Section~\ref{sec:occurrence} and of Paper~I. For each of the \NInjCells{} grid cells and each of the \NRepStars{} representative star-sectors, one mid-transit epoch $T_0$ is drawn uniformly from $[0, P)$ with a second fixed seed (\texttt{42}), so the grid is exactly reproducible. Eccentricity is fixed at zero. The full grid comprises $\NInjPeriodSteps{} \times \NInjRadiusSteps{} \times \NRepStars{} = \NInjTotal{}$ injections, or \NInjPerCell{} per cell.  

The injected depth is $\delta=(R_P/R_\star)^2$ before limb-darkening, and the total transit duration follows from the circular-orbit expression of \citet{Winn:2010}, with the semi-major axis ($a$) obtained from Kepler's third law given $M_\star$ and $P$:
\begin{equation}
T_{14} = \frac{P}{\pi}\,\arcsin\!\left[\frac{R_\star}{a}\,
\frac{\sqrt{\left(1 + R_P/R_\star\right)^{2} - b^{2}}}{\sin i}\right]
\label{eq:t14}
\end{equation}

All injections are placed at $b = 0$ and $i = 90\degrees$. The direction of the resulting bias is worth stating explicitly, because it is easy to state backwards. A central transit is the deepest and longest transit available at fixed $(P, R_P)$, and therefore the most detectable. Averaging over the impact parameters real planets actually have would lower the measured recovery rate in every cell. The sensitivity map reported here is consequently an upper bound on the sensitivity averaged over impact parameter, and the occurrence rates of Section~\ref{sec:occurrence}, which divide by that sensitivity, are biased low by this effect. We do not marginalize over $b$ in this work; the magnitude of the effect is set by the impact-parameter distribution of detectable transits and is discussed in Section~\ref{sec:occurrence}.

Recovery fractions are reported per cell with $68\%$ Wilson score confidence intervals \citep{Wilson:1927}, which retain valid coverage at $k = 0$ and $k = n$ where a normal interval does not. At \NInjPerCell{} injections per native cell the interval half-width is \WilsonHalf{} near $\hat{p} = 0.5$.

\paragraph{Recovery criterion }~~An injection is counted as recovered when the recovered ephemeris remains in phase with the injected one across the observing baseline. Two conditions must both hold,
\begin{equation}
\frac{|P_{\rm rec} - P_{\rm inj}|}{P_{\rm inj}} \cdot
\frac{T_{\rm span}}{T_{14}} \le f_{\rm drift},
\qquad
|\Delta\phi| \le \frac{T_{14}}{2}
\label{eq:coherence}
\end{equation}
with $f_{\rm drift} = \fdrift{}$. The first requires that the timing error accumulated over the star-sector's own baseline $T_{\rm span}$ amount to less than one transit duration, so that a transit predicted from the recovered ephemeris still lands on an injected transit at the end of the sector. The second requires that the recovered epoch coincide with an injected transit rather than with a phase offset half a period away.

We adopt this criterion in place of a $z$-test on the search's reported period uncertainty, $|P_{\rm rec} - P_{\rm inj}|/\sigma_P < N_\sigma$, which we find is not usable at single-sector baselines. A box or template search over a $\sim$27~d baseline can resolve a period only to of order $T_{14}/T_{\rm span}$, or \SearchPeriodRes{} for a typical short-period M-dwarf transit. The uncertainty BLS reports is of order \BLSPeriodErr{}, about \SigmaPRatio{} tighter than the search can actually distinguish. Dividing by an uncertainty that small converts correct recoveries into multi-sigma discrepancies: among the \NztestPCnotRec{} BLS injections classified as PC detections but rejected as unrecovered by the $z$-test, \NztestPeriodOK{} in fact carried the injected period, at a \ZtestEpochMatch{} epoch-match rate.

The two searches provide an internal check. Under the $z$-test, BLS and TLS disagreed on recoverability by \RecZtestGap{} (\RecZtestBLS{} against \RecZtestTLS{}); under Equation~\ref{eq:coherence} they agree to \RecCohGap{} (\RecCohBLS{} against \RecCohTLS{}). Two independent search algorithms converging is the behavior expected of a criterion that measures the signal rather than the error model of one search. The criterion is also insensitive to $f_{\rm drift}$: doubling it changes the verdict on \NfdriftFlipTLS{} of \NInjTotal{} TLS injections and \NfdriftFlipBLS{} of \NInjTotal{} BLS injections. TLS is the deliverable arm (Section~\ref{sec:transit_search}); BLS appears here only as this cross-check. Period aliases, integer or rational multiples of the injected period, fail Equation~\ref{eq:coherence} and are not counted as recoveries.

\paragraph{Vetting and the completeness denominator }~~Every injection is carried through the same Stage~A and Stage~B labeling as the blind survey, with one deliberate difference. In the blind arm Stage~A is a filter: rows failing $\mathrm{MES} > \MESthresh{}$ or carrying fewer than \NTransitMin{} distinct transits are removed and never appear in the catalog. In the injection arm Stage~A is applied as a flag. Injections that produced no signal, or a signal too weak to qualify as a TCE, remain in the frame and count as zeros in the map. This is what makes the denominator of Equation~\ref{eq:sensitivity} the number of injections placed rather than the number that happened to be detected, and it is the difference between measuring sensitivity and measuring the vetter.

Vetting in the injection arm applies exactly the tests applied to the blind catalog, including the \NEMESIS centroid motion gate (Section~\ref{subsect:centroid_motion}). Because the injections modify the SAP flux rather than the pixels, the centroids of injected star-sectors carry no injected signal, so the map does not include how often that gate rejects a genuine on-target transit. As in the blind survey, the centroid gate replaces the pixel-level \texttt{offset} test, and we exclude \texttt{invalid\_transits} and \texttt{bad\_fit} for the reasons given in Section~\ref{sec:leovetter}. All remaining tests run at their distributed defaults.

\paragraph{Three tiers, and how they differ from Paper~I }~~We report the same three-tier structure as Paper~I with one definition changed. Paper~I defined its middle tier by requiring the recovered planet radius to agree with the injected radius. We require instead that the injection be labeled PC. The governing principle is that a completeness map must reproduce the survey's selection function, meaning the exact set of gates that determine membership in the observed catalog. \NEMESIS~II applies three such gates to real data: the search must recover the signal, the signal must satisfy the Stage~A criteria, and Stage~B must classify it as a PC. Radius agreement is not among them, and cannot be, because no true radius exists for an observed candidate. A completeness map built on radius agreement therefore imposes a gate the survey does not apply while omitting one that it does, and the two errors do not cancel: omitting the vetting stage biases the map high, since vetting removes the large majority of TCEs, while imposing radius agreement biases it low for reasons unrelated to detection. Paper~I predates \leovetter{} and had no automated vetting stage, so radius agreement served there as a reasonable proxy for detection quality. \NEMESIS~II replaces the proxy with the outcome itself.

The radius comparison is retained as a diagnostic rather than a gate. For the \NInjRecRad{} recovered injections, the search pipeline's radius estimate, $R_{P,\mathrm{rec}} = \sqrt{\delta}\,R_\star$ with $\delta$ the TLS transit depth, differs from the injected radius by a median of $\RadRecMed$\%, and 68\% of these injections lie between $\RadRecLo$\% and $\RadRecHi$\%. Binned by injected radius between \RadRecBinRlo{} and \RadRecBinRhi{}~$R_\oplus$, the median ranges from $\RadRecBinMin$\% to $\RadRecBinMax$\%. For signals the search recovers at central transit geometry, the pipeline radii are therefore accurate to a few per cent. TLS does not fit the impact parameter, which the transit refits of Section~\ref{sec:jaxoplanet} add, together with the posterior intervals reported in Table~\ref{tab:candidates}. The refits of Section~\ref{sec:jaxoplanet} are not run on the injections, so we compare them with the search radii directly. For the 335 catalog candidates whose fits are not grazing, the refit radius exceeds the TLS-depth radius by a median of 1.4\%, with 68\% of them between $-6.1\%$ and $+12.8\%$; for the 190 grazing fits the median ratio is 1.66. Because the injections place every transit at $b = 0$, this agreement does not fix the absolute scale of the refit radii.

\begin{figure*}
\centering
\includegraphics[width=\textwidth,keepaspectratio]{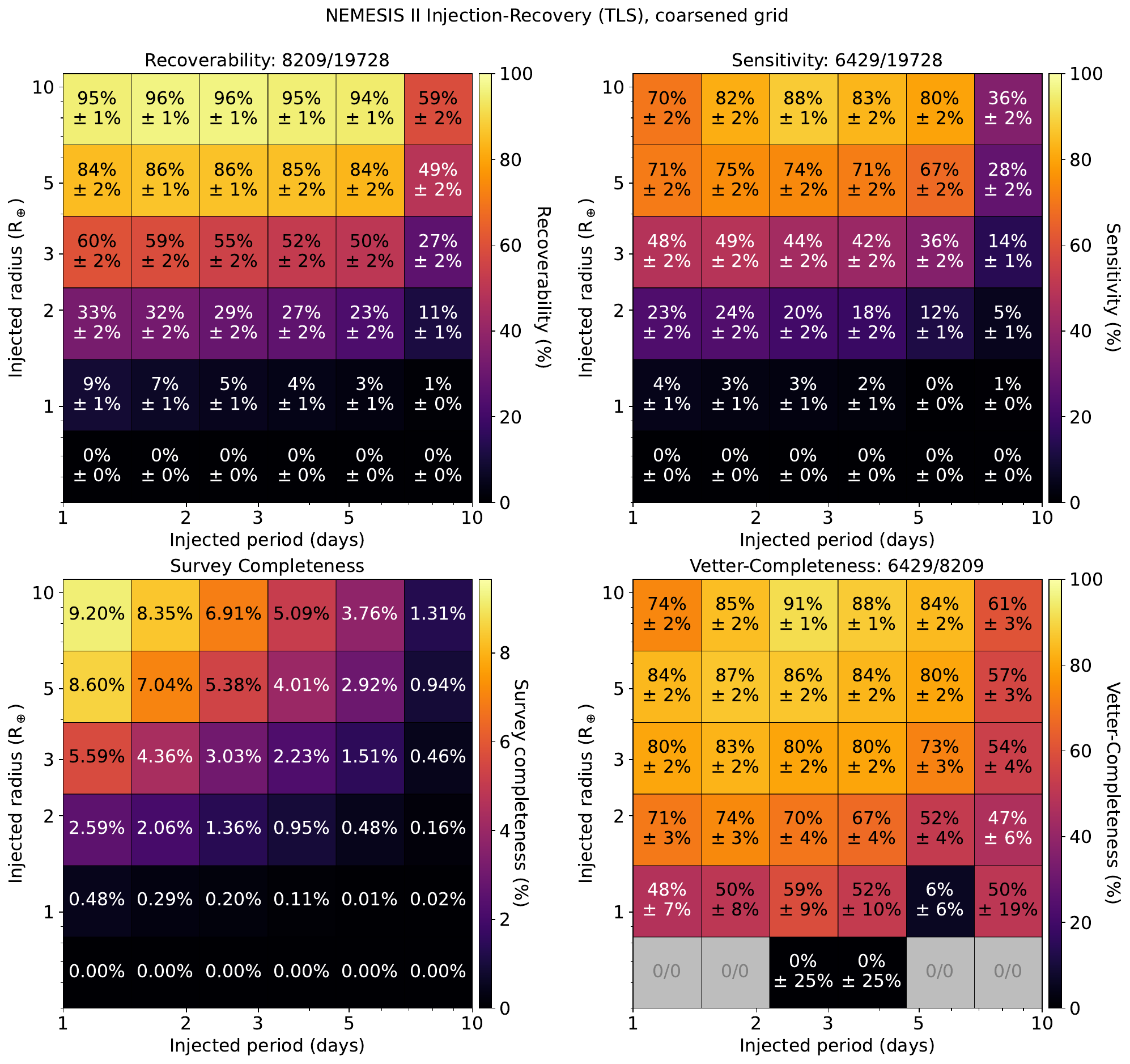}
\caption{NEMESIS~II injection-recovery maps for the TLS search, in injected orbital period and planet radius. Each displayed cell joins $2\times2$ cells of the $\NInjPeriodSteps{}\times\NInjRadiusSteps{}$ log-uniform injection grid and holds \NInjPerFigCell{} injections. Top left: Recoverability, the fraction of injections whose ephemeris satisfies Equation~\ref{eq:coherence}. Top right: Sensitivity $C(P,R_P)$, the fraction of all injections both recovered and classified as PC detections. Bottom right: Vetter-Completeness, the fraction of recovered injections classified as PC detections; cells marked 0/0 contain no recovered injections. In these three panels, each cell gives $\hat{p}=k/n$ followed by the half-width of its $68\%$ Wilson score interval ($z = 1$; \citealt{Wilson:1927}); near 0\% and 100\% the interval is asymmetric, so the printed half-width understates its extent away from the boundary. Panel titles give the totals $k/n$, and Recoverability $\times$ Vetter-Completeness $=$ Sensitivity cell by cell. Bottom left: Survey Completeness, the Sensitivity of each cell multiplied by the geometric transit probability (Equation~\ref{eq:pgeo}) averaged over the representative stars and evaluated at the cell centre. Because the transit probability is evaluated once per cell, these values depend on the cell size; the occurrence rates of Section~\ref{sec:occurrence} are computed on the native grid. Definitions and denominators are given in Table~\ref{tab:quantities}.}
\label{fig:completeness}
\end{figure*}

\paragraph{Results }~~The four maps are shown in Figure~\ref{fig:completeness}. 
Sensitivity is highest for large planets and falls toward longer periods, reflecting the small number of transits observable within a single $\sim$27~d sector and the reduced ability of the detrending stage to preserve long-duration signals. We do not quote a single aggregate recovery fraction: it depends entirely on where the injection grid places its cells and is not a property of the pipeline.

\subsubsection{Scramble Test and False Alarm Reliability}
\label{sec:scramble}

\paragraph{What the test measures, and what it cannot }~~
Scrambled light curves keep the instrumental systematics and stellar variability of the data but contain no coherent eclipse, so every detection made on them is a false alarm. A stellar signal that repeats strictly at the detected period is not a false alarm in this sense, and the scrambled data reproduce it only in part, so $R_{\rm FA}$ measures contamination by noise and systematics and may undercount false alarms of that kind. The experiment therefore measures how well the false alarm tests reject noise and cannot test the false positive tests; reliability against astrophysical false positives is addressed per candidate in Section~\ref{sec:triceratops}.

\paragraph{Construction of the scrambled sample }~~Scrambling is performed on the FFI pixel cube, before aperture photometry, by exchanging the two spacecraft orbit blocks within each sector. Orbit boundaries are taken from the TESS Data Release Notes rather than identified from gaps in the time series, because several Primary Mission sectors contain internal anomaly gaps longer than the perigee downlink gap. Because the scramble is applied to the pixel cube, each scrambled light curve requires a full photometric re-extraction rather than only a re-detrending and re-search. We therefore scramble about \ScrFrac{} of the searched star-sectors, \NScrDone{} in total, rather than the full survey. Unlike the injection arm, no cleanliness cut is applied, because the population of interest here is precisely the one that generates false alarms.

We search and label the scrambled light curves with the identical Stage~A and Stage~B rule used for the blind and injection arms, which yields \NScrTCEsRaw{} scrambled TCEs. Following \citet{Kunimoto:2025}, we then remove the \NScrCleaned{} scrambled TCEs whose host stars match known TOIs, community TOIs, or entries in the TESS Eclipsing Binary Catalog \citep{Prsa:2022}, leaving \NScrTCEs{}.

\paragraph{Effectiveness and reliability }~~Following \citet[their Eq.~15]{Kunimoto:2025}, the false alarm effectiveness $E$ is the fraction of scrambled TCEs classified as false alarms, that is, the fraction failing at least one active false alarm test,
\begin{equation}
E \approx \frac{N_{\rm FA,scrambled}}{N_{\rm scrambled}}
\label{eq:effectiveness}
\end{equation}
The numerator counts false alarm failures only. A scrambled TCE that fails only a false positive test, for example the significant secondary or odd-even test, does not contribute. The observed count $N_{\rm FA,obs}$ with which $E$ is combined below is also a false-alarm-only count, so the two come from the same partition of the catalog. The denominator is every scrambled TCE, including those that survive vetting.

Combining $E$ with the composition of the observed catalog gives the
reliability of the surviving candidate list against false alarms (Eq.~16 of \citealt{Kunimoto:2025}, following \citealt{Bryson:2020}),
\begin{equation}
R_{\rm FA} \approx 1 - \frac{N_{\rm FA,obs}}
{N_{\rm PC,obs} + N_{\rm FP,obs}}\left(\frac{1 - E}{E}\right)
\label{eq:reliability}
\end{equation}
where $N_{\rm FA,obs}$, $N_{\rm FP,obs}$, and $N_{\rm PC,obs}$ are the observed false alarm, astrophysical false positive, and planet candidate counts of Section~\ref{sec:candidates}. The denominator is the set of observed TCEs surviving the false alarm tests, the same partition used for $E$, so the two quantities are measured against consistent populations. The false-alarm-precedence convention of Section~\ref{sec:leovetter} is what makes these three counts disjoint and therefore what makes Equation~\ref{eq:reliability} well defined.

\begin{figure*}%[ht!]
\centering
\includegraphics[width=\textwidth,keepaspectratio]{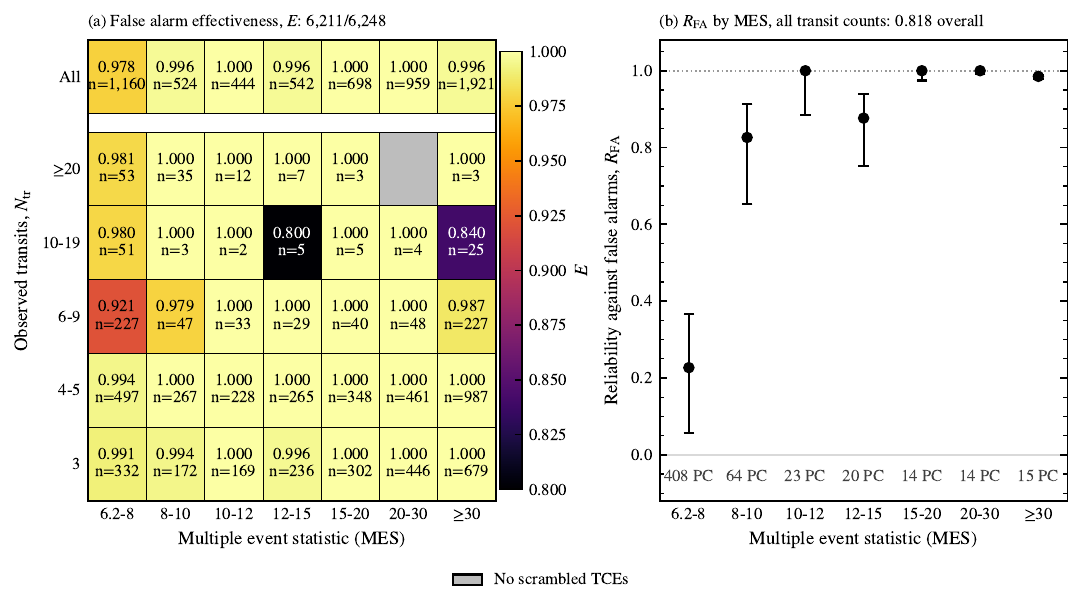}
\hfill
\caption{Left: false alarm effectiveness $E$ (Equation~\ref{eq:effectiveness}), the fraction of scrambled TCEs that the vetting rejects as false alarms, as a function of the multiple event statistic and the number of observed transits, from \NScrTCEs{} scrambled TCEs. Each cell gives $E$ and its number of scrambled TCEs; the top row pools each MES bin over all transit counts, and the grey cell contains no scrambled TCEs. Right: reliability against false alarms $R_{\rm FA}$ (Equation~\ref{eq:reliability}) in each MES bin, pooled over transit counts, with its 68\% interval and the number of PC detections in the bin. These are the reliabilities applied to each detection in Equation~\ref{eq:occrate}. The observed catalog occupies the same plane as the
left panel and is shown in Figure~\ref{fig:catalog_plane}.}
\label{fig:reliability}
\end{figure*}

We measure $E = \Effval{}$, with a $68\%$ Wilson interval of $[\Efflo{}, \Effhi{}]$, because \NScrLeak{} of the \NScrTCEs{} scrambled TCEs escape the false alarm tests, and $R_{\rm FA} = \RFAval{}$ ($68\%$ interval \RFAlo{} to \RFAhi{}). None of these lies on the host of a catalog candidate. $E$ is reported as a function of MES and number of observed transits, and $R_{\rm FA}$ as a function of MES, in Figure~\ref{fig:reliability}, alongside the distribution of the observed catalog in the same plane (Figure~\ref{fig:catalog_plane}), so that the reliability applicable to a given candidate can be read off directly rather than inferred from the aggregate. Reliability depends strongly on MES. The \NPCweakbin{} PC detections with MES below \MESbinWeakHi{} have $R_{\rm FA} \approx \RFAweakbin$, so the aggregate value overstates the reliability of a typical candidate, and for this reason Section~\ref{sec:occurrence} applies $R_{\rm FA}$ bin by bin. Summed bin by bin, the expected false alarm content of the catalog is \NexpFakePC{} of the \PCs{} detections, and all but about \NexpFakeStrong{} of these lie below MES of \MESbinWeakHi{}. The \NPCstrongbin{} detections at or above that MES therefore form a subsample in which we expect about \NexpFakeStrong{} false alarms. We do not impose this MES as a reporting floor, because a floor would also discard real planets near the threshold, which the occurrence rates of Section~\ref{sec:occurrence} recover statistically through the reliability weight of each detection. Instead, Table~\ref{tab:candidates} lists the MES of every candidate so that a reader can select the subsample. Because a candidate below MES of 8 is, on average, more likely to be a false alarm than a planet, we recommend that follow-up programs rank our candidates by MES.

\subsubsection{Verification of \leovetter{} Thresholds}
\label{sec:threshold_opt}

\citet{Kunimoto:2025} optimized the \leovetter{} pass-fail thresholds on QLP light curves, so we checked whether \NEMESIS photometry calls for different ones. A threshold cannot change whether the search found a signal, so we measure the vetting completeness $C_{\rm vet}$ only on injections that passed Stage~A and satisfy the recovery criterion of Equation~\ref{eq:coherence}, as the fraction of them classified as PC detections (Table~\ref{tab:quantities}). We require recovery because a TCE found in an injected light curve need not be the injected transit. At the published thresholds we measure $C_{\rm vet} = \Cvetval{}$ (\NInjPCrec{} of \NInjTCErec{}) and $R_{\rm FA} = \RFAval{}$, which places our completeness between the 91\% that \citet{Kunimoto:2025} report for their full sample and the \CvetQLPpm{} they report for stars observed only at 30-minute cadence. We then repeated their differential evolution search over the \NDEparams{} flux-level false alarm thresholds, minimizing their objective $L$, defined as:
\begin{equation}
    L = \sqrt{(1-C_{\rm vet})^2 + (1-R_{\rm FA})^2}
\label{eq:objective}
\end{equation}

but optimized on one random half of the scrambled sample and evaluated the result on the other half, over \NSplitRepeats{} independent splits. Of the apparent gain in $R_{\rm FA}$, \SplitShrink\% did not carry over to the held-out half, and 89\% did not when we searched only 4 thresholds. The search also favors looser vetting, because loosening moves observed false alarms into the surviving catalog and so raises $R_{\rm FA}$ in Equation~\ref{eq:reliability}. For these reasons we adopt the published thresholds without modification.

\begin{deluxetable}{lrr}
\tablecaption{NEMESIS~II disposition funnel. Each row gives the population
surviving the named stage, in unique TIC entries and in star-sectors. All
fractions quoted in the text are relative to the searched star-sector
count.\label{tab:disposition}}
\tablehead{\colhead{Stage} & \colhead{Unique TICs} & \colhead{Star-sectors}}
\startdata
\cutinhead{Selection and data availability}
Target list after selection cuts (Table~\ref{tab:selection_cuts}) & \NTargetList     & \nodata \\
Observable in Sectors 1--26 (\texttt{tess-point})         & \NObsTargets     & \NObsStarSectors \\
Light curve extracted                                     & \nodata          & \NLCss \\
Transit search completed                                  & \NSearchedTICs   & \NStarSectors \\
\leovetter{} metric set complete                          & \nodata          & \NMetricCompleteSS \\
\cutinhead{Search and vetting outcome}
Stage~A threshold crossing events                         & \TCEsTIC & \TCEs \\
\quad false alarm                                         & \FAsTIC  & \FAs \\
\quad astrophysical false positive                        & \FPsTIC  & \FPs \\
\quad PC detections 
      & \PCsTIC  & \PCs \\
\cutinhead{Candidate catalog (reconciled signals)}
Signals dropped: transit fit not converged                       & \NUnconverged  & \nodata \\
Distinct signals: catalog candidates (on \NCatalogHosts{} stars) & \PCsUnique     & \nodata \\
\quad with $R_P > \TriRadiusCeiling$~\RE{} (flagged)             & \NCatalogLarge & \nodata \\
\quad on stars with no TOI, CTOI, or transiting NASA Exoplanet Archive planet              & \NCatalogNew   & \nodata \\
\quad planet candidates (\texttt{TRICERATOPS} favored)           & \NFavored      & \nodata \\
\quad new planet candidates                                      & \NNewFavored   & \nodata \\
\quad likely planets (\texttt{TRICERATOPS})                     & \NLikelyPlanet & \nodata \\
\enddata
\tablecomments{Unique-TIC counts in the outcome rows do not sum to the TCE row, because a star can carry TCEs with different labels in different sectors. In the catalog rows, the first count gives signals rather than TICs.}
\end{deluxetable}

\section{Discussion}\label{sec:discussion}

\subsection{Planet Candidate Demographics}
\label{sec:demographics}

\begin{figure*}%[ht!]
\centering
\includegraphics[width=\textwidth,keepaspectratio]{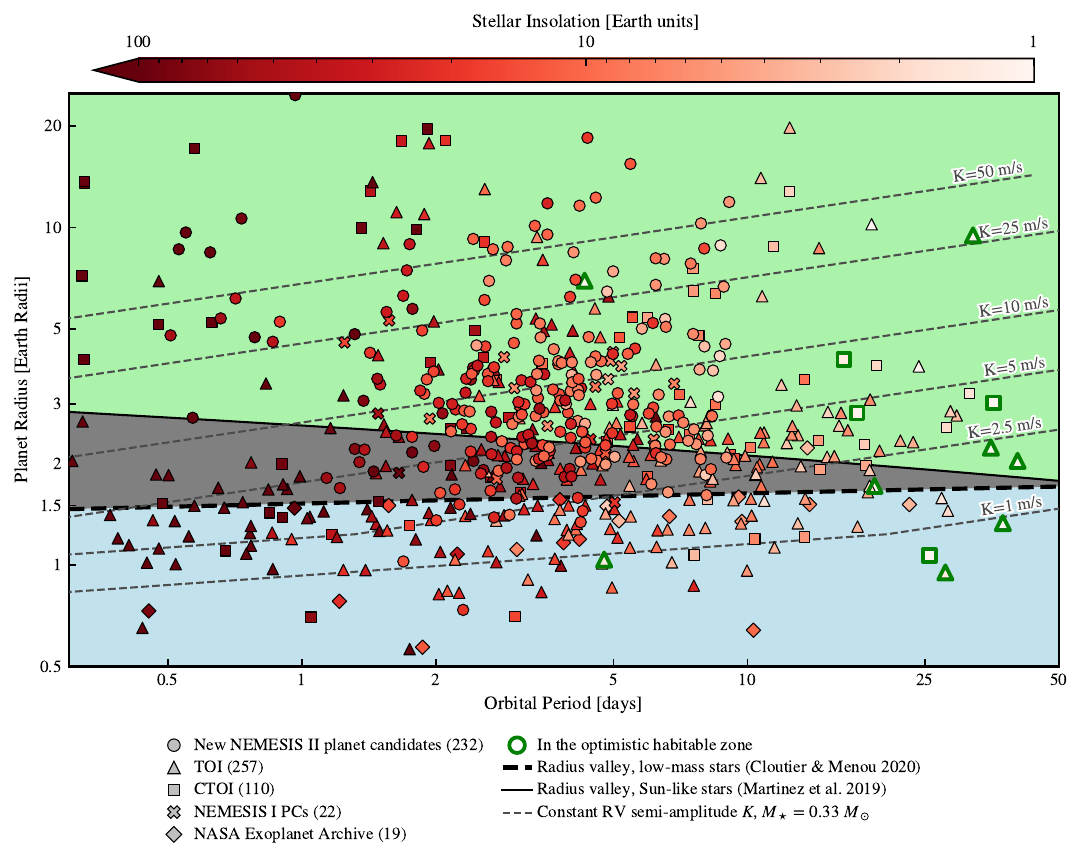}
\caption{Period-radius plane of the \NNewFavored{} new NEMESIS~II planet candidates (circles), with the TOIs with a TFOPWG disposition of PC, APC, CP, or KP (triangles), the CTOIs with a planetary or blank disposition that have not become TOIs (squares), the Paper I candidates we submitted as CTOIs, under the same rule (crosses), and the transiting NASA Exoplanet Archive planets on stars with no TOI or CTOI (diamonds), on stars in our target list in the lists of 2026 October 12 (Section~\ref{sec:comparison}). Legend counts give whole samples, including entries outside the plotted range or without a period or radius, which are not drawn; every new planet candidate is drawn. Our other planet candidates, on stars with a TOI, CTOI, or transiting NASA Exoplanet Archive entry, are not drawn; all but two of those stars appear through a drawn catalog entry, and the CTOIs of those two have no radius. Points are colored by incident stellar flux (Equation~\ref{eq:insolation}, from the sources given in Section~\ref{sec:demographics}); fluxes above 100~$S_\oplus$ share the darkest color. NEMESIS~II periods and radii are the posterior medians of Section~\ref{sec:jaxoplanet}. The dashed line is the radius valley for low-mass stars \citep{Cloutier:2020} and the solid line that for Sun-like stars \citep{Martinez:2019}; the grey band between them marks where the two estimates disagree. Objects outlined in green lie in the optimistic habitable zone of their host star, between the Recent Venus and Early Mars limits of \citet{Kopparapu:2013}: \NHZtoi{} TOIs and \NHZctoi{} CTOIs; none of our new planet candidates lies there. Grey dashed curves mark expected radial-velocity semi-amplitudes of 1, 2.5, 5, 10, 25 and 50~m~s$^{-1}$ for circular, edge-on orbits around a \KcurveMstar{}~$M_\odot$ star, the median host mass of our planet candidates, with planet masses from the mean mass-radius relation of \citet{Chen:2017}.}
\label{fig:pc_diagram}
\end{figure*}

Our catalog contains orbital periods from \minP{} to \maxP{}~days and radii from \minRP{} to \maxRP{}~\RE{}. The largest radii come from grazing fits (Section~\ref{sec:jaxoplanet}), and the largest of all belongs to a candidate on a star that already hosts a CTOI, not to a new candidate. Figure~\ref{fig:pc_diagram} places our \NNewFavored{} new planet candidates (Section~\ref{sec:triceratops}) in the period-radius plane against the radius valley, together with the TOIs, CTOIs (including our Paper I submissions), and transiting NASA Exoplanet Archive planets on stars in our target list in the catalogs of 2026 October 12 (Section~\ref{sec:comparison}), and curves of constant radial-velocity semi-amplitude that show which planets are within reach of mass measurement. We omit our own planet candidates on stars with a TOI, CTOI, or transiting NASA Exoplanet Archive planet, so that no system appears twice. We color each object by its incident stellar flux in Earth units,
\begin{equation}
\frac{S}{S_\oplus} = \left(\frac{R_\star}{R_\odot}\right)^{2}
\left(\frac{T_{\rm eff}}{5777\,\mathrm{K}}\right)^{4}
\left[\frac{1\,\mathrm{AU}}{\left(GM_\star P^{2}/4\pi^{2}\right)^{1/3}}\right]^{2}
\label{eq:insolation}
\end{equation}
following Paper~I. For \NEMESIS~II candidates we evaluate Equation~\ref{eq:insolation} with the stellar parameters adopted by our pipeline. For TOIs, CTOIs, and Archive planets, we take the flux the catalog reports; where it reports none, we evaluate Equation~\ref{eq:insolation} with the catalog's stellar parameters, deriving the mass from the surface gravity and radius where no mass is given, and where those are also missing, with the stellar parameters of our target list.

About half of our candidates are smaller than Neptune: \NSmall{} of the \NCatalog{} (\fracSmall{}) have $R_P < 4$~\RE{} in the refit radii, and \NSmallTLS{} (\fracSmallTLS{}) in the TLS radii, which do not fit the impact parameter (Section~\ref{sec:jaxoplanet}). This is not a consequence of where the survey is most sensitive. Figure~\ref{fig:completeness} shows the opposite, with Sensitivity rising toward large radii, so small candidates are about as numerous as large ones even though we are least able to find them. That pattern is expected if small planets are far more common around M dwarfs than large ones \citep{Dressing:2015}, but we do not read it as a demographic result yet, for
two reasons. First, the raw counts have not been divided by the Sensitivity, which we do in Section~\ref{sec:occurrence}. Second, false alarms concentrate at low MES, and below MES of \MESbinWeakHi{} the reliability against false alarms falls to $R_{\rm FA} \approx \RFAweakbin$ (Section~\ref{sec:scramble}). Because MES measures depth relative to noise, and our hosts span more than two orders of magnitude in noise (Figure~\ref{fig:cdpp}), low MES skews only mildly toward small radii: in TLS radii, the \NPCweakbin{} detections below MES of \MESbinWeakHi{} have a median radius of \MedRpMESlow{}~\RE, against \MedRpMEShigh{}~\RE{} for the \NPCstrongbin{} at or above it. Most expected false alarms, \ExpFAfracRpClass{}, have $R_P < 4$~\RE, but so do most detections, so false alarms are spread across the radius range rather than confined to its small end.

Two edges of the plane need care. Our search extends below one day (Section~\ref{subsec:BLS_TLS_search}), and \NUSP{} of our candidates have $P < 1$~d, but the injection grid begins at \minInjP{}~d, so detections below one day have no measured Sensitivity and do not enter the occurrence rates of Section~\ref{sec:occurrence}. At the other edge, \NCatalogLarge{} candidates have $R_P > \TriRadiusCeiling$~\RE{}. We keep them in the catalog and flag them, because their depths are also consistent with brown dwarfs and the lowest-mass stars (Section~\ref{sec:triceratops}), and Section~\ref{sec:limitations} discusses the deepest of them.

We do not attempt to measure the radius valley with this sample. At 30-minute cadence the light curve alone constrains the impact parameter poorly (Section~\ref{sec:jaxoplanet}), our radius intervals omit the uncertainty on $R_\star$, and \NexpFakePC{} of the \PCs{} detections are expected false alarms (Section~\ref{sec:scramble}). For reference only, Figure~\ref{fig:pc_diagram} shows the valley location derived for low-mass stars by \citet{Cloutier:2020}, together with the Sun-like-star relation given in its caption for comparison. Section~\ref{sec:comparison} compares the catalog with the TOI and CTOI catalogs and reports how many known planets we recover.

\subsection{Planet Occurrence Rate Estimation}
\label{sec:occurrence}

We compute the occurrence rate of short-period planets around nearby M dwarfs in period-radius bins by dividing the number of PC detections in each bin by the effective number of stars searched for them, following \citet{Dressing:2015}, and we correct the numerator for false alarms following \citet{Bryson:2020} and \citet{Kunimoto:2025}. For each bin $(P_k, R_{P,k})$ the effective number of stars surveyed is
\begin{equation}
n_{{\rm eff},k} = \sum_{i=1}^{N}
p_{\rm geo}(P_k, R_{P,k}, R_{\star,i}, M_{\star,i}) \cdot p_{{\rm det},k,i}
\label{eq:neff}
\end{equation}
where the geometric transit probability follows Paper~I,
\begin{equation}
p_{\rm geo} = \frac{R_{\star,i} + R_{P,k}}
{\left(G M_{\star,i} P_k^{2} / 4\pi^{2}\right)^{1/3}}
\label{eq:pgeo}
\end{equation}
and the sum runs over the \NMetricCompleteSS{} searched star-sectors with a complete \leovetter{} metric set (Section~\ref{sec:candidates}) rather than over unique stars, because a star observed in two sectors contributes two opportunities for detection.

The detection efficiency $p_{{\rm det},k,i}$ is the probability that a planet at the injected period and radius of cell $k$, transiting star-sector $i$, becomes a PC detection, and both factors in Equation~\ref{eq:neff} are evaluated at that period and radius. Injecting into every star-sector, as \citet{Ment:2023} do for 363 stars, is impractical for \NMetricCompleteSS{} star-sectors, so we compute $p_{{\rm det},k,i}$ star by star from the measured Sensitivity and a model,
\begin{equation}
p_{{\rm det},k,i} = \min\left(1,\ C_k\,\frac{q_{k,i}}{\langle q_k \rangle_{\rm inj}}\right)
\label{eq:pdet}
\end{equation}
where $C_k$ is the Sensitivity of cell $k$, the fraction of its \NInjPerCell{} injected planets recovered as PC detections (Section~\ref{sec:detection_efficiency}); $q_{k,i}$ is the probability that star-sector $i$ recovers a planet of cell $k$ under a detection model fitted to the injections; and $\langle q_k \rangle_{\rm inj}$ is the mean of $q_k$ over the \NRepStars{} injection hosts. The model predicts the MES (Section~\ref{sec:transit_search}) that a central transit of cell $k$ would reach in star-sector $i$,
\begin{equation}
S_{k,i} = \frac{(R_P/R_\star)^2}{\sqrt{\sigma_{w,i}^2/n_{\rm in} + \sigma_{r,i}^2/N_{\rm tr}}}\; g_T(T_{14})\, g_k(R_P/R_\star)
\label{eq:mesmodel}
\end{equation}
where $\sigma_{w,i}$ and $\sigma_{r,i}$ are the white and red noise that \leovetter{} measured for the star-sector in the blind search, $N_{\rm tr}$ is the number of transits expected to fall in data from the light curve's span and duty cycle, $n_{\rm in} = N_{\rm tr}\,T_{14}/(30~{\rm min})$ is the number of in-transit cadences, $T_{14}$ follows from the TIC radius and mass at $b = 0$, and $g_T$ and $g_k$ are empirical corrections in transit duration and radius ratio, set by the ratio of measured to predicted MES for strongly detected injections. We then take
\begin{equation}
q_{k,i} = W\,V\,A\,F(S_{k,i};\,\kappa, \theta)
\label{eq:qmodel}
\end{equation}
where $F$ is the gamma cumulative distribution \citep[as adopted for Kepler by][]{Christiansen:2015, Burke:2015} with shape $\kappa$ and scale $\theta$ and $A$ is its plateau; $W$ is the probability that at least three transits fall in data; and $V = 1 - [v_0 + (1 - v_0)\,\ell]$ is the probability of passing vetting, with a constant $v_0$ and a logistic function $\ell$ of the logarithm of the number of 30-minute exposures that a central transit spans and of $\ln(1 + R_P/R_\star)$. We fit $\kappa$, $\theta$, and $A$ by maximum likelihood on whether each injection became a PC detection, and $g_T$, $g_k$, $V$, and the coverage scaling in $W$ on subsets of the recovered injections, with each injection taking the noise, span, and duty cycle of its own host. The measured Sensitivity sets the level of $p_{\rm det}$ in each cell, and the model redistributes it: a star-sector that the model rates as easier than the average injection host receives more than $C_k$, and a harder one less. The cap at unity can only lower the level, and in a cell where $q$ vanishes for every injection host we use $p_{{\rm det},k,i} = C_k$. We use the full map rather than a single Sensitivity threshold, which would discard the gradient across the detection boundary, where most of the occurrence-rate weight sits. Stellar radii and masses, in Equation~\ref{eq:pgeo} and in the planet radii, are the TIC values throughout (Section~\ref{sec:jaxoplanet}), and we make no adjustment for an error shared by all TIC values; the radius term below includes each star's own TIC radius uncertainty, independently for each detection.

\begin{figure*}[ht!]
\centering
\includegraphics[width=\textwidth]{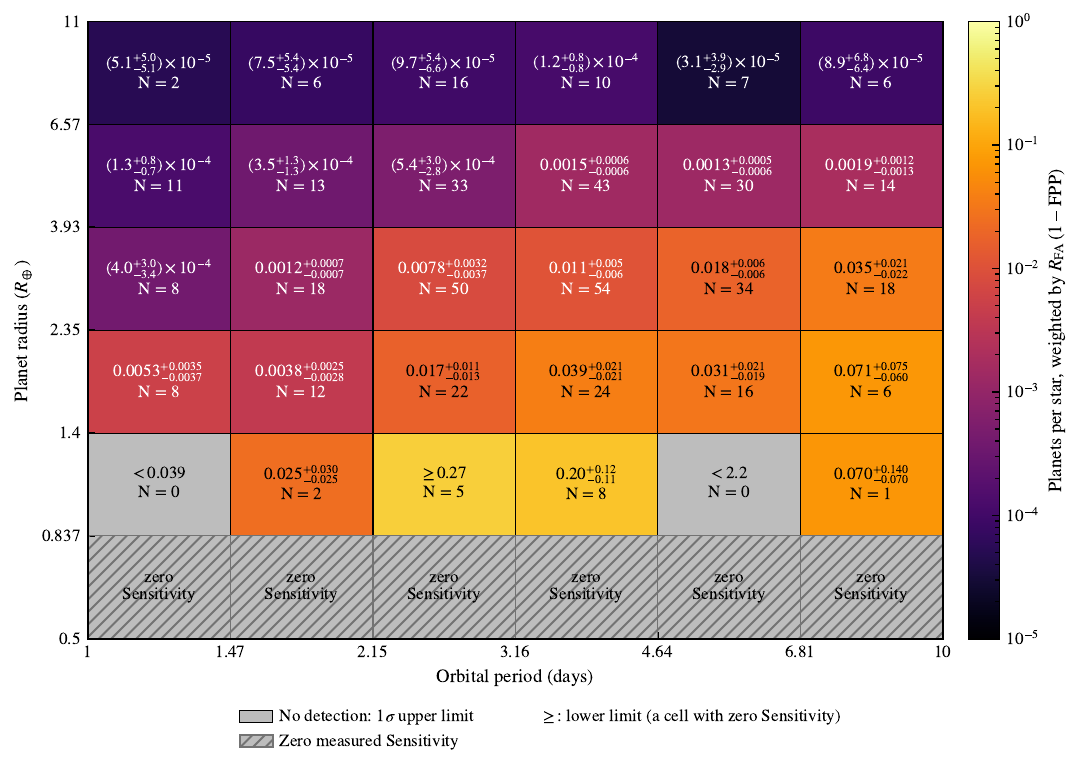}
\caption{Occurrence rates in planets per star per bin, in orbital period and planet radius, with each PC detection weighted by $R_{{\rm FA},j}\,(1 - {\rm FPP}_j)$ and each searched star-sector credited with its own detection efficiency (Section~\ref{sec:occurrence}). Each bin joins two adjacent cells of the $\NInjPeriodSteps{} \times \NInjRadiusSteps{}$ injection grid in period and two in radius, and its rate is the sum of the rates of its cells (Equation~\ref{eq:occrate}). Both axes are logarithmic, and the color scale is logarithmic from $10^{-5}$ to 1. Each bin gives its rate with the $68\%$ interval and $N$, the number of PC detections it contains. Grey bins contain no detection and give the $1\sigma$ upper limit of \citet{Gehrels:1986}, 1.841 divided by the mean $n_{{\rm eff},k}$ of their cells; the limit at $P = 4.64$ to 6.81~d and $R_P = 0.84$ to 1.40~\RE{} exceeds two planets per star and does not constrain the rate. Hatched bins, at $R_P < 0.84$~\RE{}, contain no detection, and none of their injections was recovered, so they have no rate. The value marked $\geq$ is a lower limit, because one of its cells holds two detections although none of its \NInjPerCell{} injections was recovered. Table~\ref{tab:occ_rows} gives the unweighted rate and the rate weighted by $R_{{\rm FA},j}$ alone, by radius.}
\label{fig:occurrence}
\end{figure*}

Detections outside the injected grid, with $P < \minInjP$~d or $R_P$ outside $[\minInjRP, \maxInjRP]$~\RE{}, have no measured Sensitivity, so we compute no rates there and instead report their number, \NOutGrid{} of the detections of the \PCsUnique{} catalog signals. We compute a rate in each cell of the injection grid, and the rate over any larger region is the sum of the rates of its cells. Inside the grid, cells where none of the \NInjPerCell{} injections was recovered stay in the analysis. Where such a cell holds no detection, it contributes nothing to any rate and places no constraint on it; all cells in the bins of smallest radius in Figure~\ref{fig:occurrence} are of this kind. One such cell, at $P = 2.15$ to 2.61~d and $R_P = 0.84$ to 1.08~\RE{}, holds two detections, and for it we set $C_k$ to the upper end of its $68\%$ Wilson interval, so its rate is a lower limit. In both cases the rates below are lower limits at small radii, and so are the rates integrated over the grid.

Several features of the completeness measurement and of the search propagate into the rates, and we state them because they set the direction of the result. First, all injections are placed at $b = 0$ (Section~\ref{sec:detection_efficiency}), the most favorable transit geometry, so $C_k$ is an upper bound on the Sensitivity averaged over impact parameter. Second, the injection hosts produced no TCE in the blind search, whereas \FracSurveyTCE{} of searched star-sectors did. A star-sector with a TCE carries a noise feature strong enough to cross the detection threshold, and an injected transit there would have to outrank it. Third, the search keeps only the strongest peak in each star-sector (Section~\ref{sec:reconciliation}), while each injection places a single planet on a star with no other known transiting planet. A second transiting planet in the same light curve is therefore found only in sectors where it dominates the search, and compact multi-planet systems are common around M-dwarfs \citep{Ballard:2019}. Fourth, on injection hosts held out of the fit, the detection model over-predicts recovery for the smallest and faintest hosts and where fewer than three transits are expected; for injections on the smallest quarter of hosts ($R_\star = 0.11$ to 0.21~\RS{}) whose expected MES lies where the detection curve rises from 10\% to 90\% of its plateau, it predicts 163.2 PC detections where 102 occur. All four effects overstate $p_{\rm det}$ and bias the rates low. Two further mismatches between the injection hosts and the survey enter through the model. The hosts were drawn in equal numbers from every magnitude bin, so bright stars are over-represented, and Equation~\ref{eq:pdet} corrects for this by crediting each star-sector according to the model rather than at the hosts' average. The hosts also satisfy $\mathrm{FCR} \le \repFCRmax$, whereas \FCRcutFrac{} of searched star-sectors are more crowded (Section~\ref{sec:repstars}); the model has not been tested on such hosts, and its error there is not measured. In stellar radius, TESS magnitude, and transit duration, fewer than 0.3\% of searched star-sectors lie outside the range of the injection hosts. The weights for false alarms and astrophysical false positives are expectations, and their own errors can move the rates in either direction. The unweighted rate counts every PC detection as a planet, and the rate weighted for false alarms alone counts astrophysical false positives as planets, so each is biased high in that respect by construction. The sinusoidal cut of Section~\ref{sec:repstars} removes only \FracSurveySine{} of searched star-sectors, and we neglect it. 

The occurrence rate in each bin is
\begin{equation}
f_k = \frac{1}{n_{{\rm eff},k}} \sum_{j=1}^{N_{{\rm PC},k}} w_j
\label{eq:occrate}
\end{equation}
where $N_{{\rm PC},k}$ is the number of PC detections in bin $k$ and $w_j$ is the weight of detection $j$, defined below. $N_{{\rm PC},k}$ counts star-sector detections rather than signals, to match the star-sector sum in Equation~\ref{eq:neff}, and each detection takes the TLS period adopted for its signal and the radius from its own TLS depth, because these are the quantities the injections calibrate (Section~\ref{sec:detection_efficiency}). We apply no vetting completeness correction to the numerator. Completeness and reliability enter on opposite sides of the ratio and must not both be applied on the same side: the loss of real planets, to both the search and the vetter, is already carried by $p_{\rm det}$ inside $n_{{\rm eff},k}$, while the leakage of false alarms into the candidate list is a numerator effect removed by the weights. Correcting $N_{{\rm PC},k}$ for vetting completeness in addition to dividing by $n_{{\rm eff},k}$ would correct for vetting twice.

$N_{{\rm PC},k}$ counts PC detections, not confirmed planets, and each carries two weights that answer separate questions. The false alarm reliability $R_{{\rm FA},j}$ (Equation~\ref{eq:reliability}) is the probability that detection $j$ is an astrophysical signal rather than noise; we take it from the MES bin of the detection (Figure~\ref{fig:reliability}), because the \NScrLeak{} scrambled TCEs that pass vetting are too few to resolve a further dependence on the number of transits. The quantity $1 - {\rm FPP}_j$, where ${\rm FPP}_j$ is the median \texttt{TRICERATOPS} FPP of the signal across trials (Section~\ref{sec:triceratops}), is the probability that an astrophysical signal comes from a planet transiting the target rather than from an eclipsing binary on the target or on a nearby star. Their product, as in \citet{Bryson:2020}, is the probability that the detection is both astrophysical and a planet on the target, provided the two probabilities are independent. \texttt{TRICERATOPS} assumes that every signal is astrophysical and has no noise scenario, so this need not hold. 

We therefore report three rates, each answering one question. With $w_j = 1$, the rate counts every PC detection and answers what NEMESIS~II detected, corrected for completeness. With $w_j = R_{{\rm FA},j}$, it answers how many of those signals are astrophysical rather than noise, and still counts astrophysical false positives as planets. With $w_j = R_{{\rm FA},j}\,(1 - {\rm FPP}_j)$, it answers how many are planets transiting the target. All three use the same $n_{{\rm eff},k}$, because a weight, unlike a cut, does not change the selection function, so the Sensitivity of Section~\ref{sec:detection_efficiency}, measured without \texttt{TRICERATOPS}, applies to all three. The one detection inside the grid whose signal has no converged transit fit (Section~\ref{sec:jaxoplanet}) has no FPP, and we exclude it from all three rates so that they count the same detections.

The uncertainty on each rate combines six terms, each computed alone as an upward and a downward shift; we add the terms in quadrature separately on each side (Table~\ref{tab:occ_budget}). The counting term is the uncertainty on the weighted sum, $(\sum_j w_j^2)^{1/2}$, treating detections as independent. The Sensitivity term moves each cell's $C_k$ to the ends of its $68\%$ Wilson interval, which rests on \NInjPerCell{} injections per cell (Section~\ref{sec:detection_efficiency}); the lower-limit cell, already at its upper end, contributes no Sensitivity term. These two terms are independent between cells and add in quadrature. The reliability term moves each MES bin's $R_{\rm FA}$ across its $68\%$ interval (Section~\ref{sec:scramble}). It is shared, because the same MES-bin reliabilities weight detections in every cell, so for any combination of cells we propagate it by recomputing the combined rate. The FPP term moves each detection's FPP to the 16th and 84th percentiles of its \NTriTrials{} \texttt{TRICERATOPS} trials, treating detections as independent. The detection-model term repeats the analysis on 100 bootstrap resamplings of the \NRepStars{} injection hosts, refitting the model each time and holding the measured Sensitivity fixed. The radius term redraws the planet radius of each detection 1{,}000 times from a normal distribution whose width is its TLS radius uncertainty, which combines the TIC stellar-radius uncertainty and the transit-depth uncertainty in quadrature; the redraws are independent between detections, periods are held fixed, and a redraw can move a detection between cells. A redrawn radius that falls in a cell with zero measured Sensitivity is credited with that cell's $n_{{\rm eff},k}$ at the upper end of its 68\% Sensitivity range, as for the observed lower-limit cell. For the detection-model and radius terms, the upward and downward shifts are the distances from the nominal rate to the 84th and 16th percentiles of the rate across the repetitions or redraws. Period uncertainties are negligible: the largest TLS period uncertainty of a detection in the grid is at most 4\% of the width of its cell, in period or in log period. A bin with no detection has no counting term, so for it we report the $1\sigma$ upper limit of \citet{Gehrels:1986} for zero events, 1.841 divided by the mean $n_{{\rm eff},k}$ of its cells. We report occurrence rates as planets per star per period-radius bin.

Figure~\ref{fig:occurrence} shows the rate weighted by $R_{{\rm FA},j}\,(1 - {\rm FPP}_j)$ in bins that each join two adjacent cells of the injection grid in period and two in radius, and Table~\ref{tab:occ_rows} lists all three rates by radius, summed over period. Over the whole grid, $P \in [\OccMinP{}, \OccMaxP{}]$~d and $R_P \in [\OccMinRP{}, \OccMaxRP{}]$~\RE{}, the 477 PC detections give $\IntOcc{}^{+0.42}_{-\IntOccErr{}}$ planets per star with both weights, $\OccRate{}^{+0.48}_{-0.52}$ with $R_{{\rm FA},j}$ alone, and $2.33^{+0.96}_{-0.99}$ unweighted. All three are lower limits. The row at $R_P = 0.84$ to 1.40~\RE{}, with 16 detections, supplies about 70\% of the rate with both weights (0.56 of 0.81) and is itself a lower limit; the other 461 detections, all above 1.40~\RE{}, supply the rest. Table~\ref{tab:occ_budget} breaks down the uncertainty on the three rates over the whole grid. For the rate with both weights, the counting, Sensitivity, reliability, and detection-model terms each contribute 0.1 to 0.3 planets per star, and the FPP and radius terms are smaller.

\begin{deluxetable*}{lcccc}
\tablecaption{Occurrence rates by planet radius, summed over $P = \OccMinP{}$ to \OccMaxP{}~d, in planets per star with $68\%$ ranges\label{tab:occ_rows}}
\tablehead{
\colhead{$R_P$ (\RE)} & \colhead{$N$} & \colhead{Unweighted} & \colhead{$R_{\rm FA}$} & \colhead{$R_{\rm FA}\,(1 - {\rm FPP})$}
}
\startdata
0.50 to 0.84 & 0   & \nodata & \nodata & \nodata \\
0.84 to 1.40 & 16  & $\geq 1.49^{+0.91}_{-0.94}$ & $\geq 0.65^{+0.42}_{-0.44}$ & $\geq 0.56^{+0.38}_{-0.38}$ \\
1.40 to 2.35 & 88  & $0.56^{+0.23}_{-0.16}$ & $0.19^{+0.11}_{-0.11}$ & $0.17^{+0.10}_{-0.09}$ \\
2.35 to 3.93 & 182 & $0.236^{+0.054}_{-0.051}$ & $0.084^{+0.034}_{-0.038}$ & $0.074^{+0.030}_{-0.033}$ \\
3.93 to 6.57 & 144 & $0.0353^{+0.0052}_{-0.0056}$ & $0.0129^{+0.0045}_{-0.0053}$ & $0.0057^{+0.0021}_{-0.0024}$ \\
6.57 to 11.0 & 47  & $0.0066^{+0.0013}_{-0.0015}$ & $0.0028^{+0.0010}_{-0.0010}$ & $0.00046^{+0.00015}_{-0.00017}$ \\
\hline
0.50 to 11.0 & 477 & $\geq 2.33^{+0.96}_{-0.99}$ & $\geq \OccRate{}^{+0.48}_{-0.52}$ & $\geq \IntOcc{}^{+0.42}_{-\IntOccErr{}}$ \\
\enddata
\tablecomments{$N$ is the number of PC detections. Rates marked $\geq$ are lower limits, because their row contains a cell with two detections and no recovered injection (Section~\ref{sec:occurrence}). No injection in the $R_P = 0.50$ to 0.84~\RE{} row was recovered, so it has no rate. The radius rows are those of Figure~\ref{fig:occurrence}.}
\end{deluxetable*}

\begin{deluxetable}{lccc}
\tabletypesize{\footnotesize}
\tablecaption{Uncertainty budget of the rates over the whole grid, in planets per star, as upward / downward shifts\label{tab:occ_budget}}
\tablehead{
\colhead{Term} & \colhead{Unweighted} & \colhead{$R_{\rm FA}$} & \colhead{$R_{\rm FA}\,(1 - {\rm FPP})$}
}
\startdata
Counting        & 0.55 / 0.55 & 0.26 / 0.26 & 0.24 / 0.24 \\
Sensitivity     & 0.68 / 0.27 & 0.29 / 0.11 & 0.27 / 0.10 \\
Reliability     & \nodata     & 0.24 / 0.30 & 0.18 / 0.22 \\
FPP             & \nodata     & \nodata     & 0.0003 / 0.0004 \\
Detection model & 0.32 / 0.77 & 0.12 / 0.31 & 0.11 / 0.26 \\
Planet radius   & 0.24 / 0.11 & 0.07 / 0.05 & 0.06 / 0.03 \\
\hline
Total           & 0.96 / 0.99 & 0.48 / 0.52 & 0.42 / 0.44 \\
\enddata
\tablecomments{Each term is computed alone, and the terms are added in quadrature separately on each side (Section~\ref{sec:occurrence}). The rates are 2.33 (unweighted), \OccRate{} ($R_{\rm FA}$), and \IntOcc{} (both weights), all lower limits. \nodata\ marks a term that does not apply to that weight.}
\end{deluxetable}

Where follow-up exists, it agrees with the FPP weights. Of the 27 candidates with a TFOPWG disposition of confirmed planet, known planet, or false positive (Section~\ref{sec:triceratops}), 24 are planets, against 21.8 expected from their values of $1 - {\rm FPP}$ (observed over expected 1.10, two-sided $p = 0.26$). Among the five with $R_P \geq 8$~\RE{}, 3 are planets against 1.9 expected, so the rate with both weights may be low for the largest planets, where FPP is least informative (Section~\ref{sec:triceratops}).

Computing $p_{\rm det}$ star by star matters. Applying one Sensitivity map to every star-sector, as in Paper~I, gives 0.25, 0.096, and 0.078 planets per star over the same range for the unweighted, $R_{{\rm FA},j}$, and both-weight rates, 9 to 10 times below ours. The injection hosts are uncrowded and spread evenly in brightness, so they are easier than the typical searched star-sector, and a single map credits every star-sector with their average Sensitivity. Over the cells of each radius row that hold detections, the median ratio of the single-map to the star-by-star $n_{{\rm eff},k}$ falls from 18 at $R_P = 0.84$ to 1.40~\RE{} through 7.2, 2.7, and 1.4 to 1.1 at 6.57 to 11~\RE{}. Our rates are comparable to earlier measurements for M dwarfs, although the samples and ranges differ. For $R_P = 0.84$ to 1.40~\RE{} we find at least $0.56 \pm 0.38$ planets per star with both weights at $P = 1$ to 10~d, against $0.61^{+0.24}_{-0.19}$ terrestrial planets per star larger than 0.5~\RE{} at $P = 0.4$ to 7~d around mid-to-late M dwarfs of 0.1 to 0.3~$M_\odot$ within 15~pc \citep{Ment:2023}. Summed over $R_P = 0.84$ to 3.93~\RE{}, our rate with both weights is at least $0.80^{+0.42}_{-0.44}$ planets per star, against 0.69 for $R_P = 0.5$ to 4~\RE{} and $P = 0.5$ to 10~d around the Kepler M dwarfs of \citet{Dressing:2015}, whose median stellar radius is 0.47~$R_\odot$, as quoted by \citet{Ment:2023}.

\subsection{Comparison with TOI and CTOI Catalogs}
\label{sec:comparison}

NEMESIS~II is one of several transit surveys of the TESS full-frame images,
and the value of an independent pipeline lies as much in what it recovers of
the known population as in what it adds. We therefore report the comparison in
both directions.

Of the \NCatalog{} catalog candidates, \NCatalogNew{} lie on stars with no TOI, CTOI, or transiting NASA Exoplanet Archive planet (Section~\ref{sec:candidates}). The other \NCatalogKnown{} lie on stars that host a TOI (\NmatchTOI{}), a CTOI (\NmatchCTOI{}), or both (\NmatchBoth{}); \NmatchArchive{} of these stars also host a transiting planet in the Archive, and none host a transiting NASA Exoplanet Archive planet alone. These matches use the TIC identifier alone, and they decide which candidates we call new. A shared host does not by itself mean a shared signal, so we identify a candidate with a cataloged planet only when its period and transit time also pass the test described below, as \NTriTOI{} of the \NmatchTOI{} candidates on TOI hosts do. Only these matches name a planet in the text, enter the comparison with TFOPWG dispositions in Section~\ref{sec:triceratops}, or fill the \texttt{TOI}, \texttt{CTOI}, and \texttt{P\_arch} columns of the machine-readable table (Table~\ref{tab:candidates_cols}), and their agreement in period and transit time is a practical check that our aperture photometry and detrending have not distorted these signals. The remaining candidate on a TOI host, TIC~198485881 (Sector~20, $P = 4.83$~d), matches no cataloged signal: the only TOI on its star, TOI-2257.01, has $P = 35.19$~d.

The reverse comparison is the more demanding one. We take as known planets the TOIs with a TFOPWG disposition of PC, APC, CP, or KP and the transiting NASA Exoplanet Archive planets on stars with neither a TOI nor a CTOI, with $0.5 \le P \le 10$~d, from the shortest period we search (Section~\ref{subsec:BLS_TLS_search}) to the upper edge of the injection grid (Section~\ref{sec:detection_efficiency}), on the \NSearchedTICs{} stars we searched. Of these \NknownFootprint{} planets, \NEMESIS~II recovers \NknownRecovered{} (\FracKnownRecovered{}) as PC detections under the test described below. Of the 160 we miss, 104 fall below the Stage~A floor in a single sector: for 98, the cataloged depth and duration and the host's noise in each searched sector give an expected MES at or below the floor, or fewer than \NTransitMin{} transits, and 79 of these are smaller than 2~\RE{}; the other 6 were found by TLS at the published period, or at two or three times it, with a measured MES below the floor. A further 38 were not the strongest signal in their star-sector light curves, 11 of them because another known planet on the same star was stronger, and 18 were rejected by a named vetting test. The recovery fraction therefore mixes the sensitivity of a single-sector search with our choice to keep one peak per star-sector. The sensitivity itself is measured by our injections (Figure~\ref{fig:completeness}): for the 166 planets inside the grid it predicts between 33 recoveries, if each planet is credited with one searched sector, and 55, if every searched sector counts as an independent chance; we find 27, a shortfall in the direction Section~\ref{sec:occurrence} anticipates, since every injection is a single central transit on an isolated host. The test below isolates the vetting stage.

We also asked a narrower question that isolates the vetting stage. When \NEMESIS~II detects a known planet as a TCE, does our catalog rule keep it as often as it keeps an injected planet of the same signal strength? To answer this, we matched every TCE on a star hosting a TOI or CTOI in the lists of 2026 October 12 against that star's published ephemerides. We required the periods to agree within $\tau_P$ at an integer ratio of at most three, and the epochs to agree within the largest of the shorter transit duration, three times the propagated ephemeris uncertainty, and 1.5~h (three 30-minute cadences), so that the test is modeled on the one we use for cross-sector reconciliation (Section~\ref{sec:reconciliation}). TLS reports no uncertainty on the transit epoch, and a single sector holds few transits, so the 1.5~h floor stands in for our own timing error. Of the \KPnHostTCE{} TCEs on such stars, \KPnMatch{} recover a known signal. For the comparison below we kept the \KPnCmp{} that match at the published period and carry a confirmed, known, or candidate planet disposition, and within those the \KPnCP{} confirmed and known planets. We take each disposition from the TFOPWG column of ExoFOP-TESS (retrieved 2026 October 12), or from the TOI catalog itself where ExoFOP gives none.

A direct comparison of pass rates against $C_\mathrm{vet}$ would mislead, because our injected population is dominated by strong signals. Of the recovered injections, \KPinjHighFrac{}\% lie above $\mathrm{MES}=\KPhighMES{}$, where \KPinjHighC{} of them survive vetting, while just above the Stage~A floor only \KPinjLowC{} do. Known planets detected in a single 30-minute sector sit at lower MES on average. We therefore gave each matched known planet the injection pass rate of its MES bin and summed these probabilities to obtain the expected number of planet candidates, taking the variance as the sum of $C_i(1-C_i)$. Of the \KPnCmp{} known planets and candidates, \KPnCmpPC{} are labeled PC against \KPnCmpExp{}~$\pm$~\KPnCmpSD{} expected ($z=\KPcmpZ$). Of the \KPnCP{} confirmed and known planets, \KPnCPPC{} pass against \KPnCPExp{}~$\pm$~\KPnCPSD{} expected ($z=\KPcpZ$). Known planets and candidates therefore survive our vetting at \KPratio{}~$\pm$~\KPratioErr{} of the rate our injections predict, and we find no evidence that the injection-derived completeness overstates the vetting efficiency for real planets.

Among the confirmed and known planets that fail vetting, L~98-59~c in Sector 2 and the brown dwarf LP~261-75~C fail only the \texttt{secondary} test, the first through the multi-planet effect and the second through a single outlying cadence (Section~\ref{sec:limitations}). This comparison tests the vetting stage alone. Known planets that never produced a TCE are not counted here, and those search-stage losses enter through $C(P,R_P)$. 

\subsection{Limitations}
\label{sec:limitations}

% Several features of the survey design bound what the results above can support, and we state them here rather than distributing them through the methods sections.
We collect here the limitations of the survey design, with pointers to the sections that measure them where a measurement exists.

\emph{Single-sector baseline.} \NEMESIS~II searches each star-sector independently. The longest reliably recoverable period is therefore set by the requirement of at least \NTransitMin{} distinct transits within a $\sim$27~d window, which restricts the survey to $P \lesssim \maxInjP{}$~d. Planets at longer periods are not merely under-recovered but largely inaccessible, and the occurrence rates of Section~\ref{sec:occurrence} say nothing outside the injected grid.

\emph{Multi-planet systems.} \leovetter{} vets each TCE independently and was characterized on searches that return one TCE per star \citep{Kunimoto:2025}, and our injection arm likewise places a single planet in the light curve of a star with no known planets. Neither one measures how often a genuine planet fails vetting because a second transiting planet shares its light curve. The failure we found is specific. When a companion orbits close to a low-order period ratio with the detected planet, its transits land at nearly the same phase of the detected planet's fold, where one or two of them in a single sector can mimic a secondary eclipse. TOI-270 shows this clearly \citep{Gunther:2019}. TOI-270~d orbits at close to twice the period of TOI-270~c, so in Sectors 3 and 5 two consecutive transits of d fell within about one transit duration of each other in phase, shortly after transits of c, and the unmodified \texttt{secondary} test rejects c in both. In Sector 4 the middle transit of d fell in the data gap, the two that remained landed too far apart in phase to stack, and c passed. Because the stacked event lies far from phase 0.5, the phase condition of Section~\ref{sec:leovetter} restores the Sector 5 detection, while the Sector 3 detection also fails \texttt{chases} and remains a false alarm. TOI-270~c therefore enters our catalog through Sectors 4 and 5. We found one other known planet affected, L~98-59~c in Sector 2 \citep{Kostov:2019a}, which also lies close to a 2:1 ratio with its outer neighbor. There a single transit of L~98-59~d falls 8 minutes from the event, which sits closer to phase 0.5 ($e_{\min}=\LNNemin{}$), inside the range an eclipsing binary can reach, so that detection is still rejected, but L~98-59~c passes vetting in \KPlninetyeightSectors{} other sectors. The effect costs us individual detections, not planets. Because a companion that is itself undetectable in one sector can cause the rejection, as TOI-270~d does with only two transits per sector, masking detected planets and searching again would not recover
these cases. The search compounds this, because it keeps only the strongest peak in each star-sector, so a second planet is recovered only in sectors where it dominates; Paper~I missed TOI-270~b for this reason \citep{Feliz:2021}.

\emph{Stray events near phase 0.5.} The phase condition cannot separate a stray event from a binary's secondary eclipse when the event happens to fall near phase 0.5. We lose the transiting brown dwarf LP~261-75~C \citep{Irwin:2018} this way: a single outlying cadence fails \texttt{secondary}, and at $e_{\min}=\LPemin{}$ it lies inside the range an eclipsing binary can reach. A test that asks whether the secondary event recurs at every epoch, as an eclipsing binary's must, would recover this case and the multi-planet case above, and we leave it to future work (Section~\ref{sec:future}). 

\emph{Candidates too deep to be planets.} Vetting asks whether a signal is consistent with a transit on the target, not whether the transiting body is a planet, nor whether the eclipse occurs on the target at all. TIC~159344139 passes every active test in Sectors 14 and 26, the former through the phase condition of Section~\ref{sec:leovetter}, with a deblended depth of \TICdeepDepth~ppt, and its fitted $R_P/R_\star$ reaches the upper limit of the prior (Section~\ref{sec:jaxoplanet}). The target supplies only \TICdeepShare{} of the flux in its aperture (Section~\ref{subsubsec:SAP}), so in the aperture as recorded the dip is \TICdeepDip{}. On the target it would require an eclipse of about half of the star's light or more, whereas \texttt{TRICERATOPS} attributes it to a nearby source (median NFPP $= \TICdeepNFPP$), most probably an eclipsing binary at twice our period on the neighbor TIC~159345581. TIC~251076512 ($R_P \approx \TICbigRp$~\RE) is false positive favored (FPP $= \TICbigFPP$), most probably an eclipsing binary at twice the period on the target itself; \texttt{TRICERATOPS} has no brown-dwarf scenario, so above about \TriRadiusCeiling{}~\RE{} it cannot separate a transiting brown dwarf from a stellar companion. We keep both in the catalog, where their labels and the large-radius flag of Section~\ref{sec:triceratops} identify them.

\emph{Nearby sources.} At the TESS pixel scale, the open question for most candidates is which star hosts the signal, not whether the signal is planetary.   \NLikelyNFPall{} of the \NTriEvaluated{} evaluated candidates are likely nearby false positives and a further \NTriNearbyOpen{} cannot exclude a resolved neighbor, together \NTriNearbyFrac{} of the sample (Section~\ref{sec:triceratops}). Seeing-limited photometry that resolves the neighbors, and high-resolution imaging for companions the TIC does not resolve, are therefore the most informative follow-up for this catalog.

\emph{Eclipsing binaries that pass vetting.} \NFPfavored{} candidates that pass every active vetting test are nonetheless false positive favored by \texttt{TRICERATOPS}. 

\emph{Cadence.} At 30-minute cadence the exposure time is a substantial
fraction of a short-period M-dwarf transit duration. Ingress and egress are
poorly resolved, which limits how well the transit shape constrains the impact
parameter and the scaled semi-major axis, and which is the reason the
transit-model fits of Section~\ref{sec:jaxoplanet} integrate over the exposure
window rather than evaluating at cadence midpoints.

\emph{Detrending and transit depth.} The PLD systematics correction and the
subsequent smoothing are tuned to preserve short-duration signals, but any
detrending applied without knowledge of the transit ephemeris will absorb some
transit depth. Section~\ref{sec:detection_efficiency} measures this effect with injections that pass through the same detrending: the radius the search recovers differs from the injected radius by a median of $\RadRecMed$\%, and injected planets smaller than about \RadRecSmallRmax{}~\RE{} are recovered \RadRecSmallMin\% to \RadRecSmallMax\% small.

\emph{Aperture selection in crowded fields.} \NEMESIS constructs a per-target aperture from the FFI cutout. For isolated stars this performs well, but in crowded fields, and particularly for faint targets, the selected aperture can center on a brighter neighbor. Paper~I identified $T > \PaperICrowdMag$ as the regime where this becomes common. 

\emph{No difference imaging.} \NEMESIS does not produce difference images, so we screen off-target eclipses with the centroid motion test of Section~\ref{subsect:centroid_motion}, which locates the source of a transit less precisely than difference imaging does. Neither control arm measures what this screen misses: the injections place every transit on the target, and the scrambled data contain no eclipses.

\emph{The scramble test bounds one failure mode, not both.} As set out in
Section~\ref{sec:scramble}, scrambled light curves contain no eclipses, so
$R_{\rm FA}$ constrains contamination by noise and systematics and says
nothing about contamination by eclipsing binaries. The latter is addressed
per candidate by \texttt{TRICERATOPS} and not at the population level. The
catalog therefore carries a quantified reliability against false alarms and an
unquantified population-level reliability against astrophysical false
positives.

\emph{Impact parameter.} All injections are placed at $b = 0$, so $C(P, R_P)$ is an upper bound and the occurrence rates are biased low (Sections~\ref{sec:detection_efficiency} and \ref{sec:occurrence}).

\emph{Reliability near the detection floor.} Most of our candidates lie just
above the Stage~A floor, where reliability against false alarms is lowest:
\NPCweakbin{} of the \PCs{} detections have MES below \MESbinWeakHi{}, where
$R_{\rm FA} \approx \RFAweakbin$ (Section~\ref{sec:scramble}). The occurrence
rates correct for this statistically, through the reliability weight of each
detection in Equation~\ref{eq:occrate}, but an individual candidate in that
range is, on average, more likely to be a false alarm than a planet. We
therefore recommend that follow-up programs rank our candidates by MES.

\emph{Stellar parameters.} We adopt the TIC stellar radii, masses, and effective temperatures as fixed values (Section~\ref{sec:jaxoplanet}), so the intervals on $R_P$ in Table~\ref{tab:candidates} omit the stellar uncertainties, and any error in a host's TIC radius carries directly into the radius of its candidate.

\emph{Completeness sample composition.} The representative sample is drawn from \NRepSectors{} sectors, gives each magnitude bin the same number of stars, and satisfies three cleanliness cuts that the survey does not apply. The contamination cut and the equal weighting by magnitude both create selection-function mismatches, discussed in Sections~\ref{sec:repstars} and \ref{sec:occurrence}.

\subsection{Prospects for Follow-up Characterization}
\label{sec:followup}

\begin{figure*}
\centering
\includegraphics[width=\textwidth,keepaspectratio]{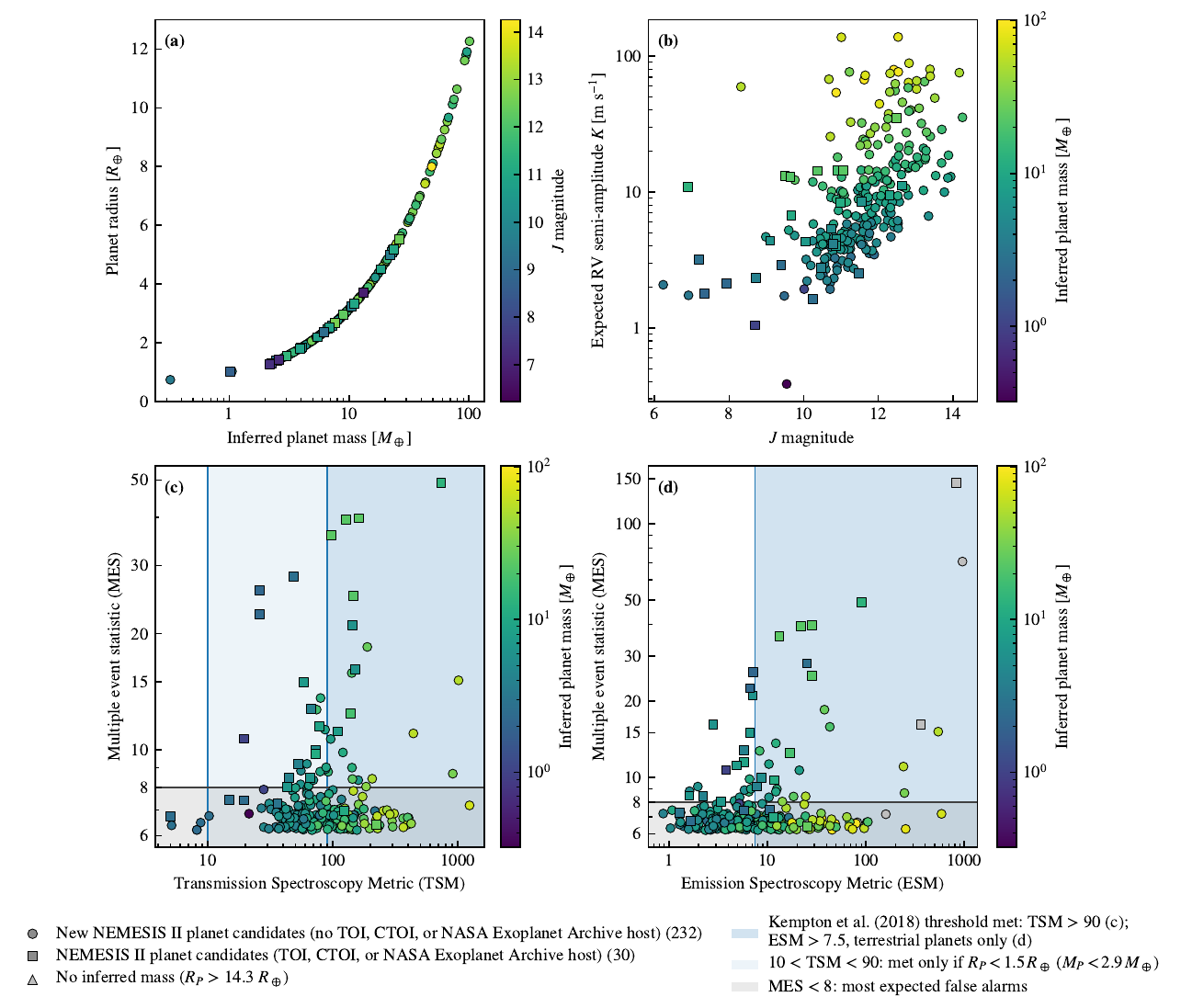}
\caption{Follow-up metrics for the \NFavored{} planet candidates. Circles mark the \NNewFavored{} new planet candidates and squares the others, whose hosts have a TOI, CTOI, or transiting NASA Exoplanet Archive planet. Grey markers are the \NFollowNoMass{} candidates larger than \CKRmax{}~\RE, for which the mean mass-radius relation of \citet{Chen:2017} gives no mass. (a) Planet radius against inferred mass, colored by $J$ magnitude; the points follow that relation by construction. (b) Expected radial-velocity semi-amplitude $K$ against $J$, for a circular, edge-on orbit and neglecting the planet's mass, colored by inferred mass. (c) MES against the transmission spectroscopy metric (TSM) and (d) MES against the emission spectroscopy metric (ESM) of \citet{Kempton:2018}, colored by inferred mass; the TSM is not defined at or above \TSMRmax{}~\RE. In (c), the darker region marks TSM $> \TSMs$, the threshold \citet{Kempton:2018} recommend from \RpTerr{} to \TSMRmax{}~\RE, and the lighter region marks $\TSMsTerr <$ TSM $< \TSMs$, which meets their threshold only below \RpTerr{}~\RE, an inferred mass below \MpTerr{}~$M_\oplus$. In (d), the shaded region marks ESM $> \ESMs$, a threshold \citet{Kempton:2018} set for terrestrial planets alone. In (c) and (d), the grey band marks MES below \MESbinWeakHi{}, where most expected false alarms lie (Section~\ref{sec:scramble}). Grazing fits inflate $R_P$, and with it the TSM, because their radii are set largely by the prior (Section~\ref{sec:jaxoplanet}).}
\label{fig:followup}
\end{figure*}

In Paper~I, we assessed which of our candidates were best suited to mass measurement and atmospheric characterization, and we repeat that assessment here for our \NFavored{} planet candidates. Because none of them has a measured mass, we infer each mass from the planet radius with the mean mass-radius relation of \citet{Chen:2017}, the relation behind the constant-$K$ curves of Figure~\ref{fig:pc_diagram}, here applied with each host's TIC mass. The relation cannot be inverted above \CKRmax{}~\RE, so \NFollowNoMass{} candidates receive no mass. From the inferred mass we compute the expected radial-velocity semi-amplitude $K$ for a circular, edge-on orbit, neglecting the planet's mass against the star's (Paper~I, Equation~11). We then evaluate the transmission and emission spectroscopy metrics (TSM and ESM) of \citet[their Equations 1 and 4]{Kempton:2018}, with the TIC $J$ and $K_S$ magnitudes and an equilibrium temperature for zero albedo and full heat redistribution. Figure~\ref{fig:followup} shows the results against MES, the statistic by which we recommend ranking candidates (Section~\ref{sec:scramble}). Because these values depend on the assumed mass relation, we leave them out of Table~\ref{tab:candidates} and will release them for every catalog candidate with the High Level Science Product (Section~\ref{sec:Vetting_reports}).

We focus on the \NNewFavoredHighMES{} new planet candidates with MES of at least \MESbinWeakHi{}, which Section~\ref{sec:triceratops} identifies as the strongest targets for follow-up and several of which appear in Figure~\ref{fig:gallery}. \PCsToFollowUp{} of them exceed the TSM thresholds that \citet{Kempton:2018} recommend for high-quality targets, TSM $> \TSMsTerr$ below \RpTerr{}~\RE{} and TSM $> \TSMs$ from \RpTerr{} to \TSMRmax{}~\RE, or \NFollowTSMclean{} if we set aside grazing fits. \NFollowTSMsmall{} of these \NFollowTSMclean{} have radii from \FollowRpLo{} to \FollowRpHi{}~\RE. The other \NFollowTSMlarge{} lie above \TriRadiusCeiling{}~\RE{} at periods below 1~d, where FPP is less informative (Section~\ref{sec:triceratops}), and a few RV spectra would distinguish a planet from a stellar companion. None of the \NNewFavoredHighMES{} is smaller than \FollowRpMinHigh{}~\RE, so our strongest new candidates include no terrestrial targets, and the ESM threshold, which \citet{Kempton:2018} set for terrestrial planets alone, applies to none of them. Terrestrial-size candidates that pass the TSM threshold do exist in our catalog, but the \NFollowTerrLowMES{} on new hosts all have MES below \MESbinWeakHi{}, where $R_{\rm FA} \approx \RFAweakbin$ (Section~\ref{sec:scramble}).

For mass measurement, we find that the \NFollowKmass{} of these \NNewFavoredHighMES{} candidates with an inferred mass have $K$ of at least \FollowKmin{}~m~s$^{-1}$, with a median of \FollowKmed{}~m~s$^{-1}$. Their hosts are faint, however, with a median $J$ of \FollowJmed{}, which favors spectrographs that work in the red optical or near-infrared.

We attach three cautions to these numbers. First, grazing fits inflate $R_P$, and with it the TSM, because their radii are set largely by the prior (Section~\ref{sec:jaxoplanet}). Most of the largest TSM values in Figure~\ref{fig:followup} belong to grazing fits, most of those below MES of \MESbinWeakHi{}, and \citet{Kempton:2018} list near-grazing transits among the reasons to deprioritize a target. Second, a mean mass-radius relation ignores the scatter of real planets about it, so a candidate near a threshold may land on either side of it once its mass is measured. Third, \NFollowNbr{} of the \NNewFavoredHighMES{} are inconclusive with a nearby star not ruled out (Table~\ref{tab:triceratops}). For those, the first step is seeing-limited photometry of their neighbors (Section~\ref{sec:limitations}), and RV and atmospheric observations should wait until that step confirms which star hosts the signal.

\subsection{Multi-Sector and Extended Mission Prospects}
\label{sec:future}

NEMESIS~II is deliberately restricted to single-sector searches of the Primary
Mission. That choice bounds the accessible period range
(Section~\ref{sec:limitations}) but keeps the selection function simple enough
to characterize with the two control analyses of
Section~\ref{sec:injection_analysis}, which is the requirement an occurrence
rate imposes.

Three extensions follow naturally. First, the \NEMESIS pipeline already supports multi-sector light curve construction, and stitching consecutive Primary Mission sectors extends the recoverable period range beyond \maxInjP{}~d. This requires re-deriving $C(P, R_P)$ on multi-sector baselines rather than reusing the single-sector map, because both the number of observed transits and the systematics structure change; the survey and its completeness must be extended together. Second, Extended Mission FFIs are observed at finer cadence than the 30-minute Primary Mission products, which directly relieves the ingress and egress smearing described in Section~\ref{sec:jaxoplanet}. Vetting improves as well: \citet{Kunimoto:2025} find that \leovetter{} completeness rises from \CvetQLPpm{} for stars observed only at 30-minute cadence to \CvetQLPemOne{} for stars observed only at 10-minute cadence. Third, the candidates reported here become priors for a targeted re-search: Extended Mission coverage of a Primary Mission candidate provides an independent ephemeris check that single-sector data cannot supply.

The vetting chain can also improve. A test that asks whether a secondary event recurs at every epoch would recover the planets that \texttt{secondary} still rejects (Section~\ref{sec:limitations}). A MES-dependent form of \texttt{bad\_fit} could restore the test where it does separate noise from transits: above MES of \BfHighMES{} it fails \BfHighLeakCaught{} of the \BfHighLeakN{} scrambled TCEs that pass every false-alarm test, at a cost of \BfHighInjPct{} of the injected planets labeled PC in that range (Section~\ref{sec:leovetter}). Finally, because we measure $R_{\rm FA}$ as a function of MES (Section~\ref{sec:scramble}), the MES at which reliability crosses a chosen threshold is a measured quantity, so a future search can set its Stage~A floor on evidence rather than convention.

Each of these is deferred to a companion paper. The completeness and
reliability measurements reported here provide the single-sector baseline
against which those extensions will be measured.

\section{Conclusion}
\label{sec:conclusion}

We have presented \NEMESIS~II, a blind single-sector transit survey of nearby
M dwarfs in the TESS Primary Mission full-frame images. The survey extends
Paper~I \citep{Feliz:2021} from Sectors 1--5 to Sectors 1--26 and replaces its
manual vetting with a fully automated disposition chain, so that the
selection function is reproducible in the control analyses that the
occurrence rates depend on. Our main results are as follows.

\begin{enumerate}
%1
\item We extracted light curves and completed transit searches for \NSearchedTICs{} unique M dwarfs across \NStarSectors{} star-sectors, covering \CoverageFrac{} of the observable star-sector list (Section~\ref{sec:results}).
%2
\item The search produced \TCEs{} threshold crossing events. Automated disposition classifies \PCs{} as PC detections, \FPs{} as astrophysical false positives, and \FAs{} as false alarms (Section~\ref{sec:candidates}). After cross-sector reconciliation these PC detections correspond to \PCsUnique{} distinct signals with converged transit fits, which form our candidate catalog and span $P \in [\minP, \maxP]$~d.
%3
\item \NCatalogNew{} of the catalog candidates lie on stars with no TOI, CTOI, or transiting NASA Exoplanet Archive planet (Section~\ref{sec:candidates}). \texttt{TRICERATOPS} favors a planetary origin for \NFavored{} of the \NTriEvaluated{} candidates, which we report as planet candidates (\NNewFavored{} of them new), and also excludes an eclipse on a resolved neighbor for \NLikelyPlanet{} of these (Section~\ref{sec:triceratops}). At 30-minute cadence we claim none of them as a statistically validated planet. 
% \NoteToSelf{Add the CTOI sentence once the submission rule is decided.}
%4
\item Every vetting label in this catalog is produced by a single rule applied identically to the blind survey and to both control analyses (Section~\ref{sec:injection_analysis}). No candidate was promoted, demoted, or removed by visual inspection.
%5
\item From \NInjTotal{} flux-level transit injections into \NRepStars{} representative star-sectors, we measured the pipeline's sensitivity $C(P, R_P)$ on a $\NInjPeriodSteps{} \times \NInjRadiusSteps{}$ grid (Section~\ref{sec:detection_efficiency}). 
%6 
\item From \NScrTCEs{} threshold crossing events recovered from orbit-scrambled pixel data, we measured a false alarm effectiveness $E = \Effval{}$ and a reliability against false alarms $R_{\rm FA} = \RFAval{}$ (Section~\ref{sec:scramble}), which falls to $R_{\rm FA} \approx \RFAweakbin$ below MES of \MESbinWeakHi{}, so that about \NexpFakePC{} of the \PCs{} detections are expected to be false alarms. 

\item Combining the PC detections with a detection probability for each searched star-sector and the weights for false alarms and astrophysical false positives gives an integrated occurrence rate of $\IntOcc{} \pm \IntOccErr{}$ planets per star over $P \in [\OccMinP{}, \OccMaxP{}]$~d and $R_P \in [\OccMinRP{}, \OccMaxRP{}]$~\RE{} (Section~\ref{sec:occurrence}). It is a lower limit at small radii, and our assumption of central transits biases it low further; the other effects of Section~\ref{sec:occurrence} can move it in either direction.

\item Of the \NNewFavoredHighMES{} new planet candidates with MES of at least \MESbinWeakHi{}, \PCsToFollowUp{} exceed the transmission spectroscopy thresholds of \citet{Kempton:2018}, none is smaller than \FollowRpMinHigh{}~\RE, and \NFollowNbr{} first need seeing-limited photometry of their neighbors (Section~\ref{sec:followup}).

%8
%POST-PUBLICATION VERSION (AJ)
% \item All \NEMESIS light curves, best-fit tables, vetting reports, and injection-recovery products are available at MAST as a High Level Science Product via \dataset[10.17909/zqyx-zg73]{\doi{10.17909/zqyx-zg73}}\footnote{\url{https://archive.stsci.edu/hlsp/nemesis}}, and the pipeline is publicly available at \url{https://github.com/daxfeliz/NEMESIS}.
%
%PRE-PUBLCIATION VERSION (arxiv)
\item All \NEMESIS light curves, the TLS vetting report and best-fit results of every searched star-sector, the threshold crossing event and candidate catalogs, the transit-fit posterior samples, TRICERATOPS results, and follow-up metrics (Section~\ref{sec:followup}) of every candidate, and the injection-recovery and orbit-scrambling results will be released as a MAST High Level Science Product upon publication, and the pipeline is publicly available at \url{https://github.com/daxfeliz/NEMESIS}.

\end{enumerate}

\begin{acknowledgments}

We thank Drs.~David Hogg and Megan Bedell for meaningful discussions throughout this project, Dr.~Luke Bouma for useful insights about the TESS noise model, and the members of the Astronomical Data Group at the Flatiron Institute Center for Computational Astrophysics for many enriching discussions and feedback.
D.L.F. acknowledges funding support from NASA under award No.~80NSSC21K0636.~D.L.F. was also supported by the Flatiron Research Fellowship at the Flatiron Institute, a division of the Simons Foundation. The computations reported in this paper were (in part) performed using resources
made available by the Flatiron Institute. M.A. acknowledges support from the AMNH/CUNY Astrophysics REU Program (NSF award EAR-2050797). M.M.G. acknowledges support from the Barnard College Summer Research Institute (SRI). 
% TODO after Oct 19: student and co-author funding statements from the co-author replies
The software presented in this work is open source under the MIT License and is available at \\ \NEMESISGitHub, together with installation instructions and example notebooks hosted at \NEMESISDocs.~We acknowledge the use of public TESS data from pipelines at the TESS Science Office and at the TESS Science Processing Operations Center.

% \acknowledgments
\facilities{TESS}

\software{
Astroquery \citep{Astroquery:2019},
Astropy \citep{Astropy:2018},
batman \citep{Kreidberg:2015},
Beautiful Soup \citep{beautifulsoup4},
fbpca \citep{fbpca},
jaxoplanet \citep{jaxoplanet} and its dependencies
   \citep{exoplanet:kipping13, exoplanet:luger18, exoplanet:agol19},
\leovetter{} \citep{Kunimoto:2025},
Lightkurve \citep{LightKurve:2018},
Matplotlib \citep{Matplotlib},
Numpy \citep{Numpy},
numpyro \citep{numpyro:2019},
Pandas \citep{Pandas},
photutils \citep{Bradley:2024},
Python \citep{Python},
Rasterio \citep{Rasterio},
Scipy \citep{Scipy},
Shapely \citep{Shapely},
TESS-POINT \citep{Burke:2020},
TESSCut \citep{TESScut:2019},
TRICERATOPS \citep{Giacalone:2021},
Transit Least Squares \citep{Hippke:2019},
W\={o}tan \citep{Wotan}
}

% \NoteToSelf{DOUBLE CHECK THESE STATEMENTS FOR TESS, MAST, STScI, ExoFOP, DSS, AMNH/CCA, ADS}
This paper includes data collected with the TESS mission, obtained from the MAST data archive at the Space Telescope Science Institute (STScI). Funding for the TESS mission is provided by the NASA Explorer Program. STScI is operated by the Association of Universities for Research in Astronomy, Inc., under NASA contract NAS 5–26555. This research has made use of the Exoplanet Follow-up Observation Program (ExoFOP; DOI: 10.26134/ExoFOP5) website, which is operated by the California Institute of Technology, under contract with NASA under the Exoplanet Exploration Program. This work has made use of data from the European Space Agency (ESA) mission \emph{Gaia} (\url{https://www.cosmos.esa.int/gaia}), processed by the \emph{Gaia} Data Processing and Analysis Consortium (DPAC, \url{https://www.cosmos.esa.int/web/gaia/dpac/consortium}). Funding for the DPAC has been provided by national institutions, in particular the institutions participating in the \emph{Gaia} Multilateral Agreement. This publication makes use of data products from the Two Micron All Sky Survey, which is a joint project of the University of Massachusetts and the Infrared Processing and Analysis Center/California Institute of Technology, funded by the National Aeronautics and Space Administration and the National Science Foundation. This research has made use of the NASA Exoplanet Archive, which is operated by the California Institute of Technology, under contract with the National Aeronautics and Space Administration under the Exoplanet Exploration Program. The Digitized Sky Surveys were produced at the Space Telescope Science Institute under U.S. Government grant NAG W-2166. The images of these surveys are based on photographic data obtained using the Oschin Schmidt Telescope on Palomar Mountain and the UK Schmidt Telescope. The plates were processed into the present compressed digital form with the permission of these institutions. Portions of this work were conducted using data housed at the computational facilities of the American Museum of Natural History, with additional computations performed on resources supported by the Scientific Computing Core at the Flatiron Institute, a division of the Simons Foundation. This research has also made use of the ADS bibliographic service. 
\end{acknowledgments}

% Appendix (after acknowledgements but before references)
\appendix\label{sec:appendix}

\section{Example Vetting Report}\label{apndx:vetting_report}

\begin{figure*}[p]
\centering
\includegraphics[width=\textwidth]{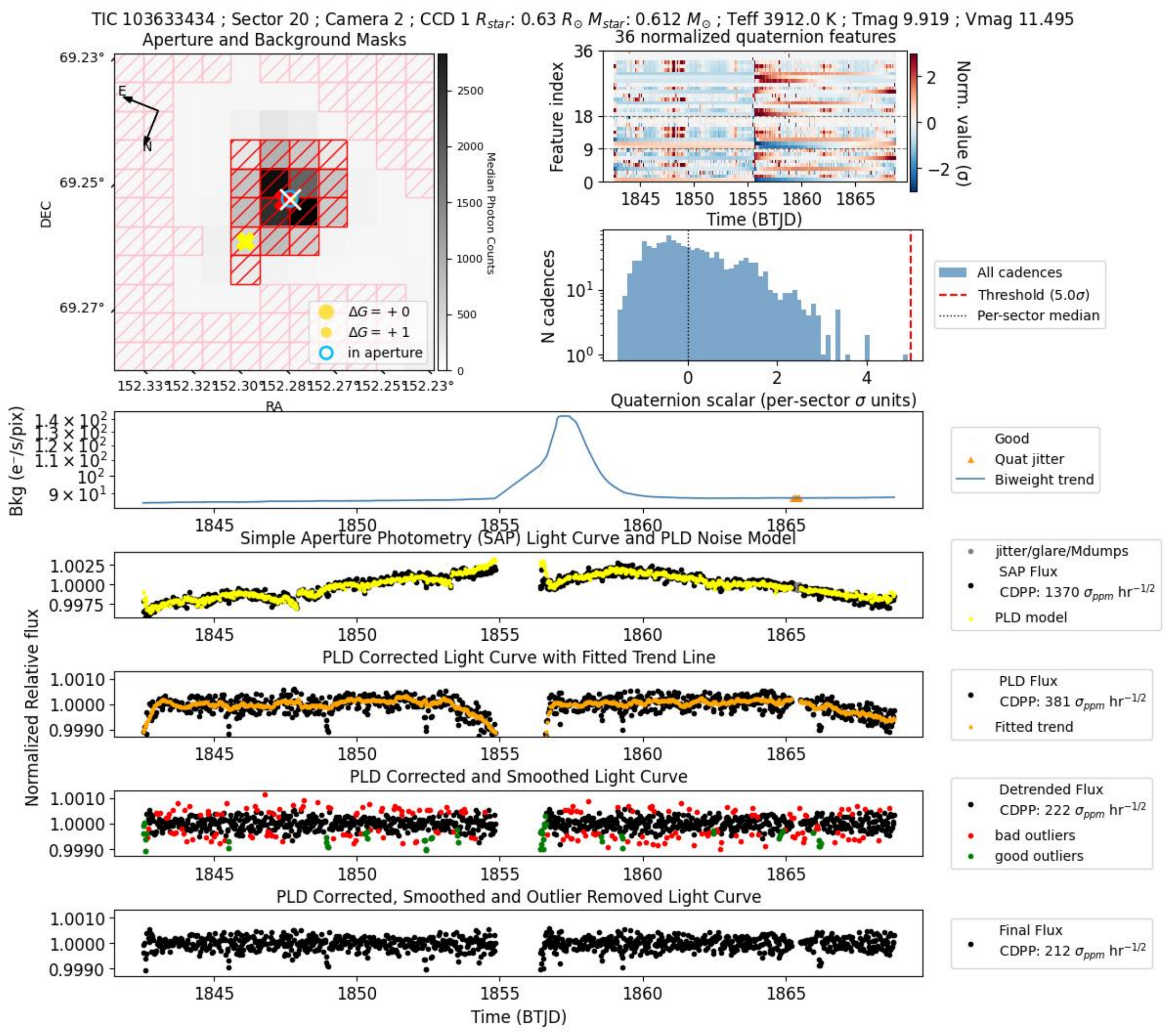}
\caption{Page~2 of the vetting report of Figure~\ref{fig:vetting_report_example}, which follows the light curve through the processing of Section~\ref{subsec:tess}. Top left: the aperture and background masks, as on page~1. Top right: the normalized quaternion features stored for display (Section~\ref{subsubsec:masks}) against time, and the distribution of the per-cadence jitter in units of its per-sector scatter, with the per-sector median and the masking threshold marked. Below, from top to bottom: the background flux with its biweight trend and the cadences flagged for centroid excursion, anomalous background, and quaternion jitter (Section~\ref{subsubsec:SAP}); the SAP light curve with the PLD model, with grey points (legend ``jitter/glare/Mdumps'') marking cadences that \NEMESIS flagged for spacecraft pointing jitter or a background excursion, which are excluded from normalization, deblending, and the transit search; cadences removed by the TESS QUALITY bitmask, including momentum dumps, are dropped before photometry and do not appear (Section~\ref{subsubsec:SAP}); the PLD-corrected light curve with the smoothing trend; the smoothed light curve, with removed outliers in red and runs of consecutive low cadences, kept as possible transits, in green (Section~\ref{subsubsec:outlier_removal}); and the light curve passed to the transit search. Each light-curve panel gives its CDPP.}
\label{fig:vetting_report_p2}
\end{figure*}

\begin{figure*}[p]
\centering
\includegraphics[width=\textwidth]{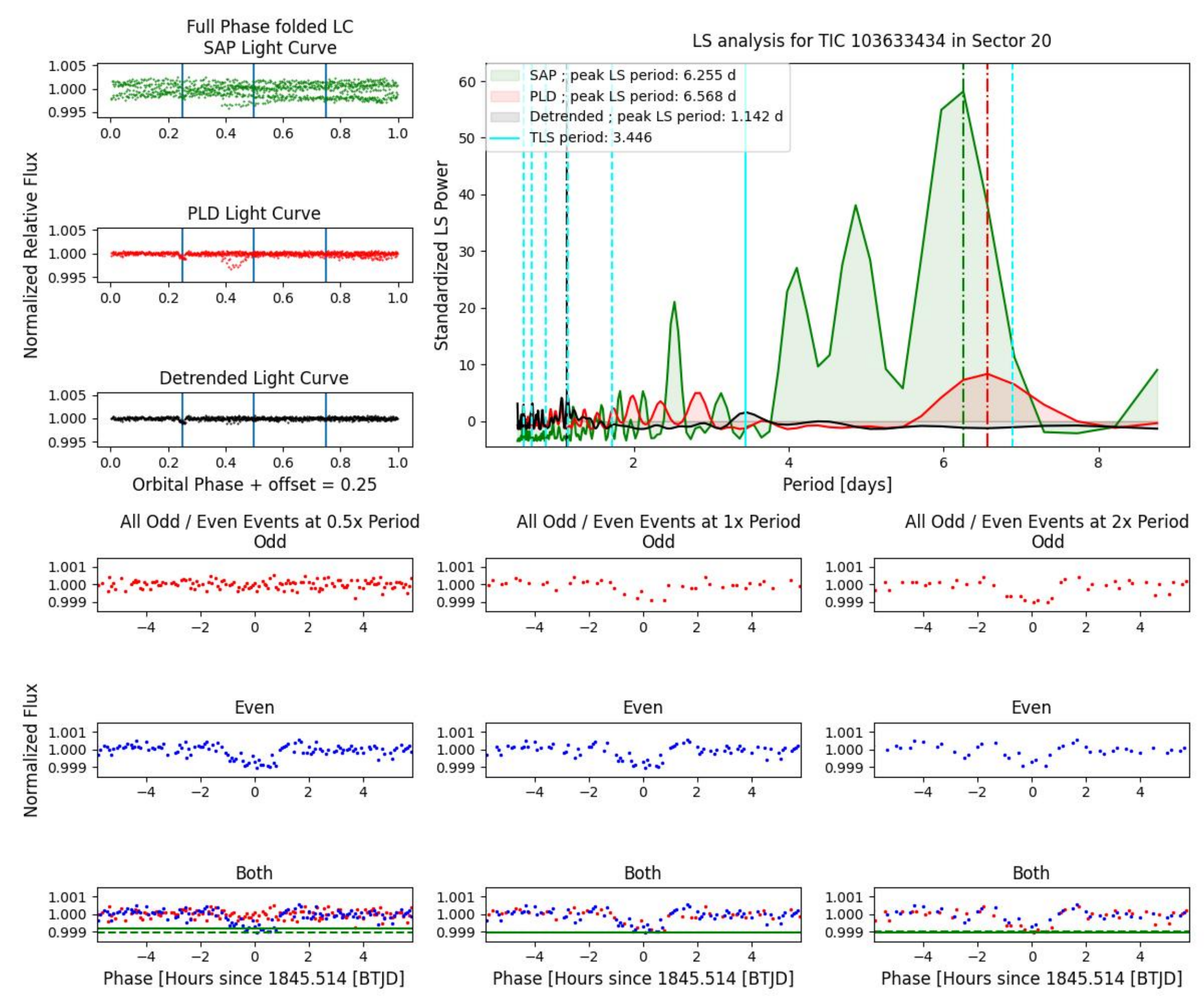}
\caption{Page~3 of the vetting report of Figure~\ref{fig:vetting_report_example}. Top left: the SAP, PLD-corrected, and detrended light curves folded over the full orbit at the TLS period. Top right: Lomb-Scargle periodograms of the same three light curves (Section~\ref{subsect:LS}), with the peak period of each in the legend and the TLS period and its harmonics marked. Bottom: odd (red) and even (blue) transits, and both together, folded at half, one, and two times the TLS period. At half the period, a transit at the TLS period appears in only one of the two sets, as it does here.}
\label{fig:vetting_report_p3}
\end{figure*}

\begin{figure*}[p]
\centering
\includegraphics[width=\textwidth]{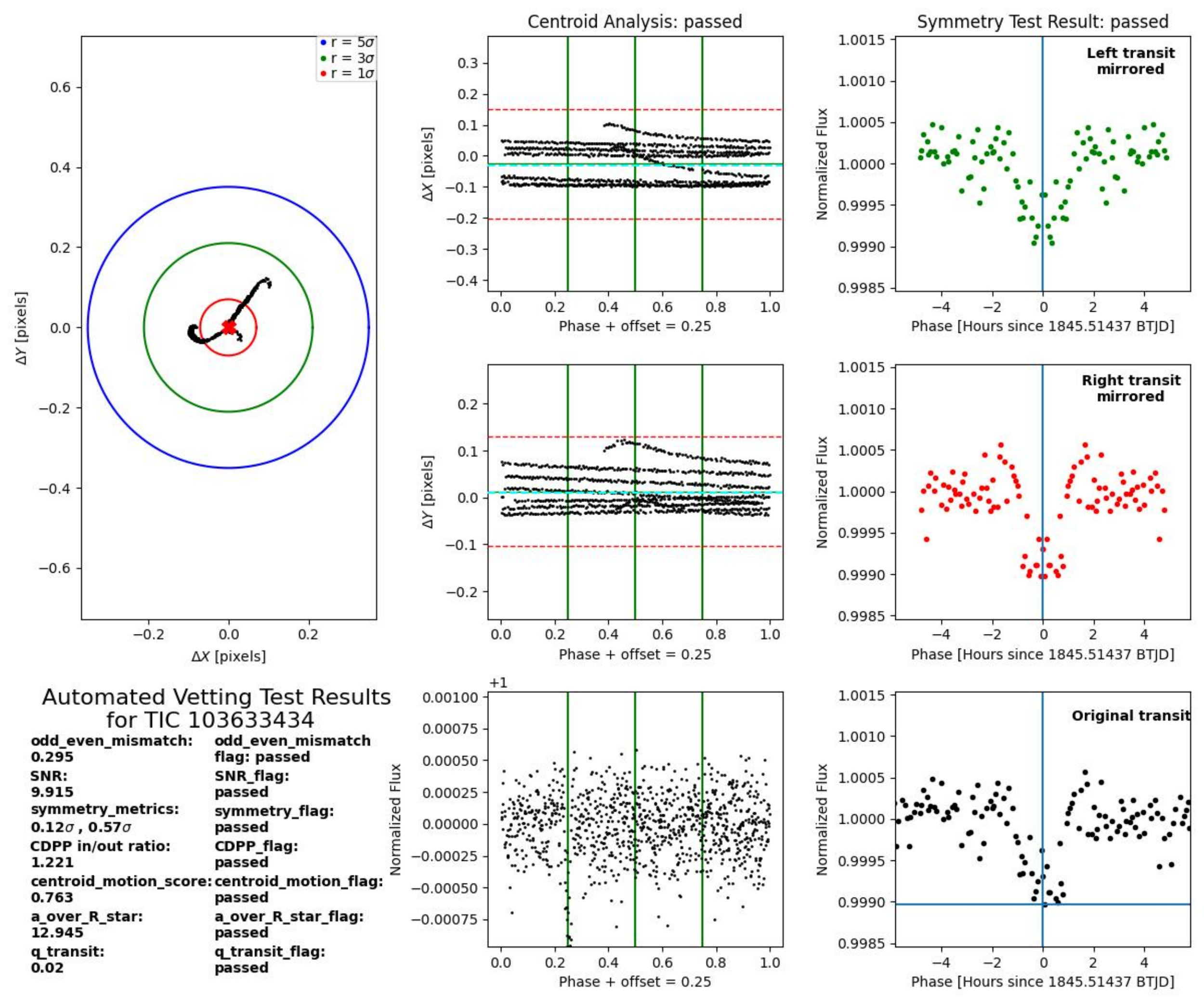}
\caption{Page~4 of the vetting report of Figure~\ref{fig:vetting_report_example}. Top left: column against row centroid offsets, with circles at 1, 3, and 5 times the out-of-transit scatter for visual reference (Section~\ref{subsect:centroid_motion}). Center: the column and row offsets folded at the TLS period, titled with the verdict of the centroid motion test, and below them the detrended light curve folded over the full orbit. Right: the transit symmetry test (Section~\ref{sec:symmetry}), showing the left and right halves of the folded transit, each mirrored about mid-transit, and the original. Bottom left: the \NEMESIS diagnostics for this star-sector and their verdicts, including the odd-even mismatch, the TLS SNR, the two symmetry significances, the comparison of in-transit and out-of-transit CDPP (Section~\ref{sec:intransit_noise}), the centroid motion score, the scaled semi-major axis $a/R_\star$, and the transit duty cycle $q$. Of these, only the centroid motion verdict enters the disposition rule (Section~\ref{sec:internal_diagnostics}).}
\label{fig:vetting_report_p4}
\end{figure*}

\begin{figure*}[p]
\centering
\includegraphics[width=\textwidth]{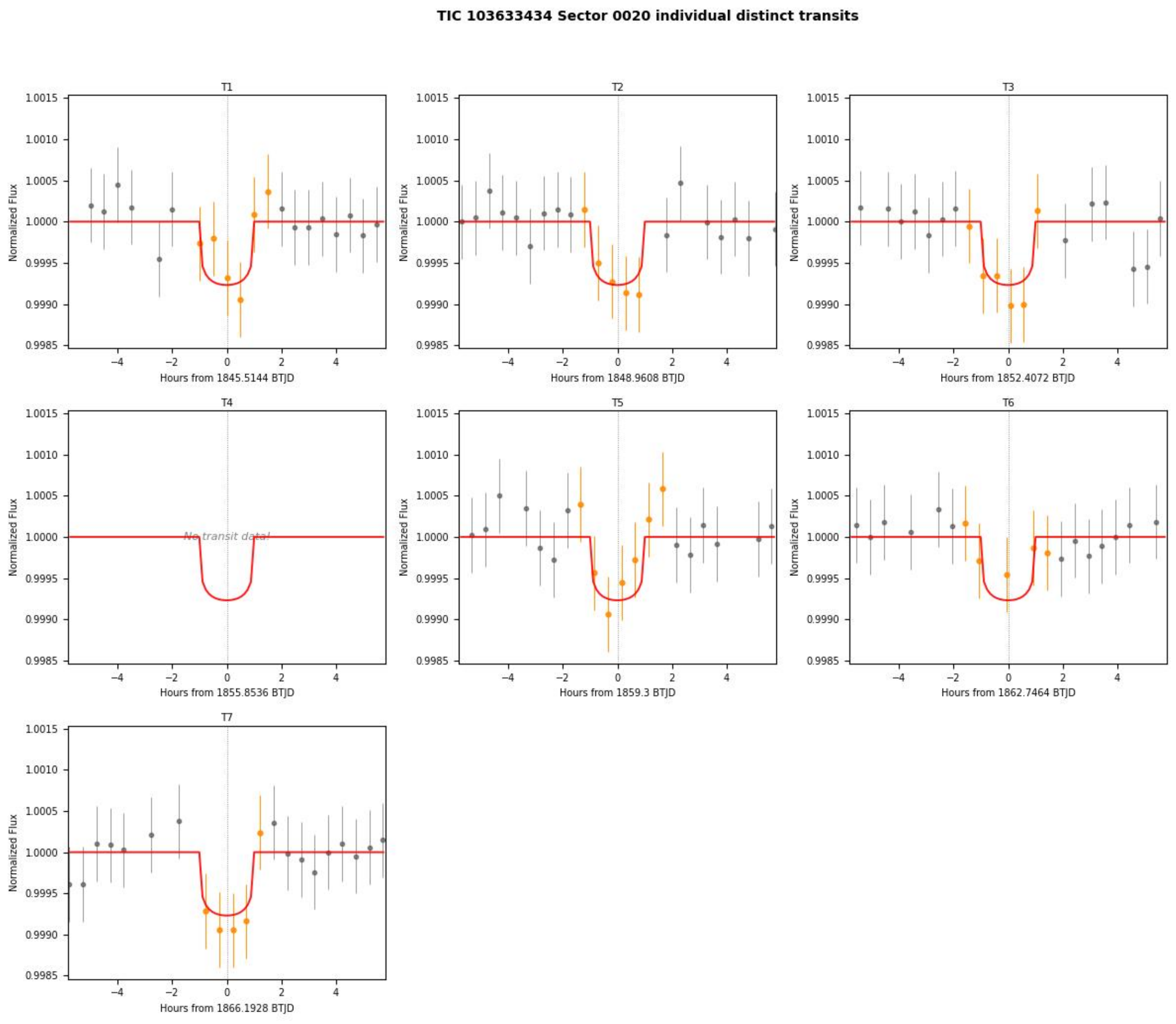}
\caption{Page~5 of the vetting report of Figure~\ref{fig:vetting_report_example}: individual transits, with cadences within one transit duration of each predicted mid-transit time in orange, the other cadences in grey, and the TLS model in red. The page shows every predicted epoch when the sector holds nine or fewer, counting epochs that fall in data gaps. With more, it shows the three with the highest, the three nearest the median, and the three with the lowest single-transit SNR, on blue, grey, and red backgrounds; this SNR is the median flux of the grey cadences minus that of the orange ones, divided by the scatter of the grey ones. Here the sector holds seven epochs, six with data, and the page shows all seven; the fourth falls in the mid-sector data gap and is marked as having no transit data.}
\label{fig:vetting_report_p5}
\end{figure*}

\begin{figure*}[p]
\centering
\includegraphics[width=\textwidth]{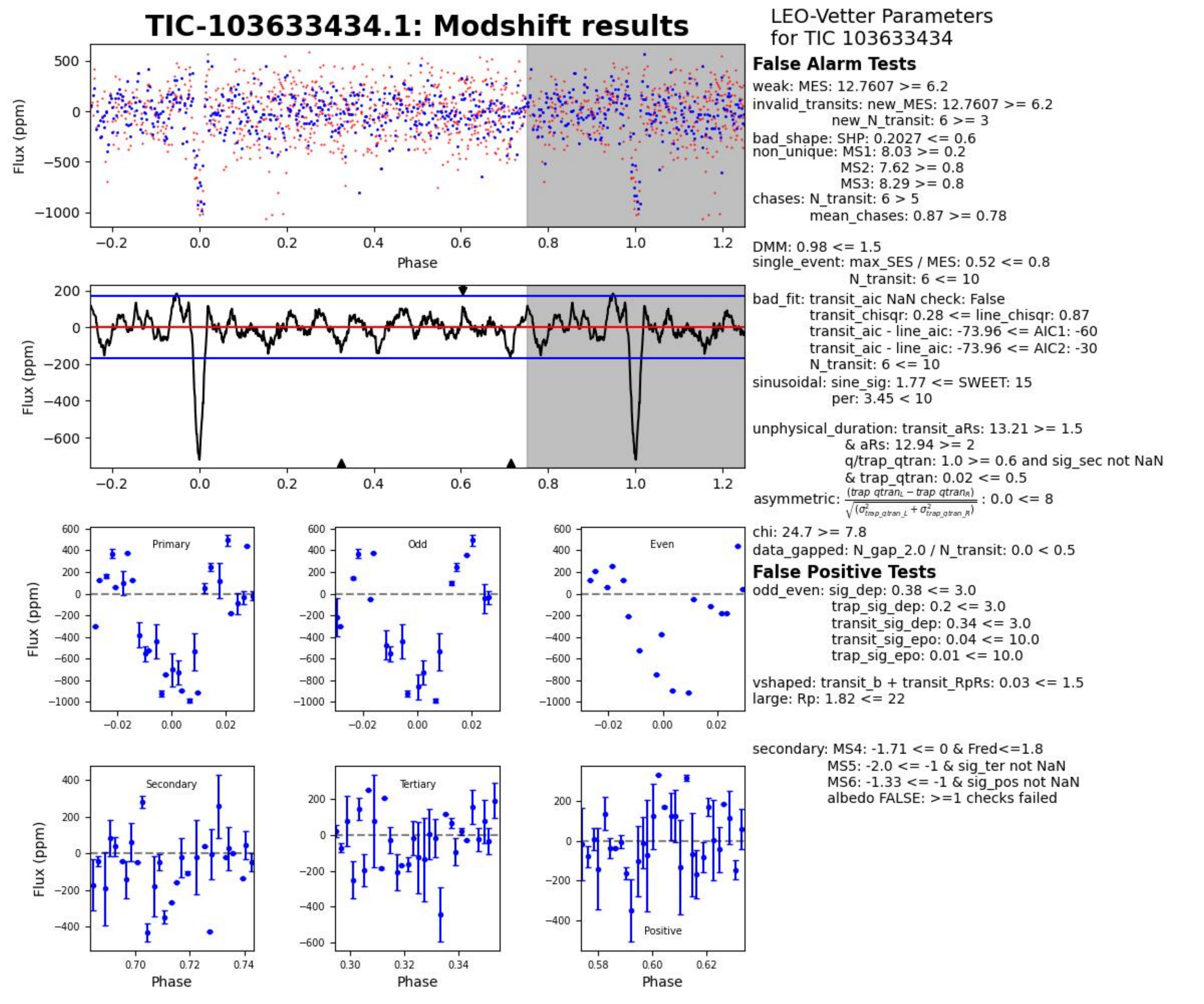}
\caption{Page~6 of the vetting report of Figure~\ref{fig:vetting_report_example}, the \leovetter{} model-shift analysis \citep{Kunimoto:2025}. Top: the light curve folded at the TLS period, as individual cadences (red) and binned values (blue). Second row: the folded light curve convolved with the transit model, with the primary event and the strongest secondary, tertiary, and positive events marked. Bottom: close views of the primary transit, the odd and even transits, and the secondary, tertiary, and positive events. Right: every \leovetter{} false alarm and false positive statistic for this star-sector with its threshold, including the two tests we do not apply, \texttt{invalid\_transits} and \texttt{bad\_fit}. The \texttt{secondary} criteria shown are the distributed ones; the phase condition of Section~\ref{sec:leovetter} is applied afterward, in the disposition rule. The albedo line reports \leovetter{}'s planetary-reflection exemption to the secondary-eclipse test, which is consulted only when a significant secondary eclipse is detected; none is detected here.}
\label{fig:vetting_report_p6}
\end{figure*}

\begin{figure*}[p]
\centering
\includegraphics[width=\textwidth]{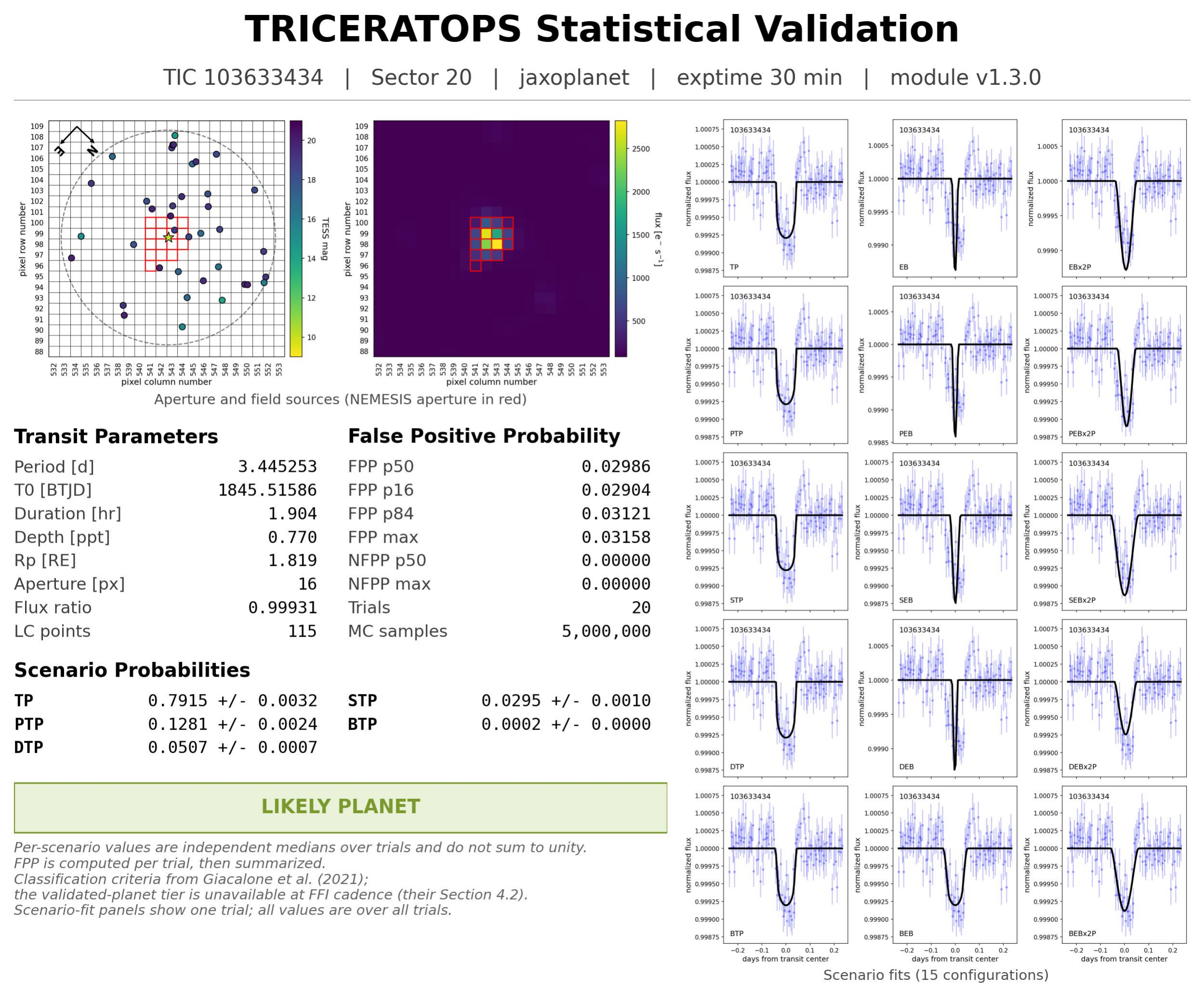}
\caption{Page~7 of the vetting report of Figure~\ref{fig:vetting_report_example}, the \texttt{TRICERATOPS} analysis of Section~\ref{sec:triceratops}, which we append only to the reports of catalog candidates. Top: the TESS magnitudes of the field sources that \texttt{TRICERATOPS} considers, and the pixel image with the \NEMESIS aperture in red; the transit parameters we supply, the ephemeris and duration from the fit of Section~\ref{sec:jaxoplanet} and the TLS depth; the FPP (median, 16th and 84th percentiles, and maximum) and NFPP across the \NTriTrials{} trials, each with \TriMCsamples{} samples; the probability of each scenario; and the resulting label; the subtitle gives the version of our \texttt{TRICERATOPS} wrapper. Below: the fit of each of the 15 scenarios in one trial. For TOI-1235.01, FPP $= 0.030$ and NFPP $= 0$, and nearly all of the false positive probability comes from the scenario of a planet transiting an unresolved bound companion rather than the target, so the candidate is labeled a likely planet; at 30-minute cadence we assign no validated tier \citep[][Section~4.2]{Giacalone:2021}.}
\label{fig:vetting_report_p7}
\end{figure*}

Figures~\ref{fig:vetting_report_p2} to \ref{fig:vetting_report_p7} show pages 2 to 7 of the TLS vetting report whose first page is Figure~\ref{fig:vetting_report_example}, for TOI-1235.01 (TIC~103633434) in Sector~20; Section~\ref{sec:Vetting_reports} describes the report. The verdicts printed on pages 4 and 6 are the diagnostics of Sections~\ref{sec:internal_diagnostics} and \ref{sec:leovetter} for this star-sector: every \NEMESIS diagnostic on page~4 passes, and every \leovetter{} statistic on page~6 meets its pass condition. Of the \NEMESIS diagnostics on page~4, only the centroid motion test enters the disposition rule.

\section{LEO-Vetter Threshold Verification}\label{apndx:thresholds}

\begin{figure}%[ht]
\centering
\includegraphics[width=\columnwidth,keepaspectratio]{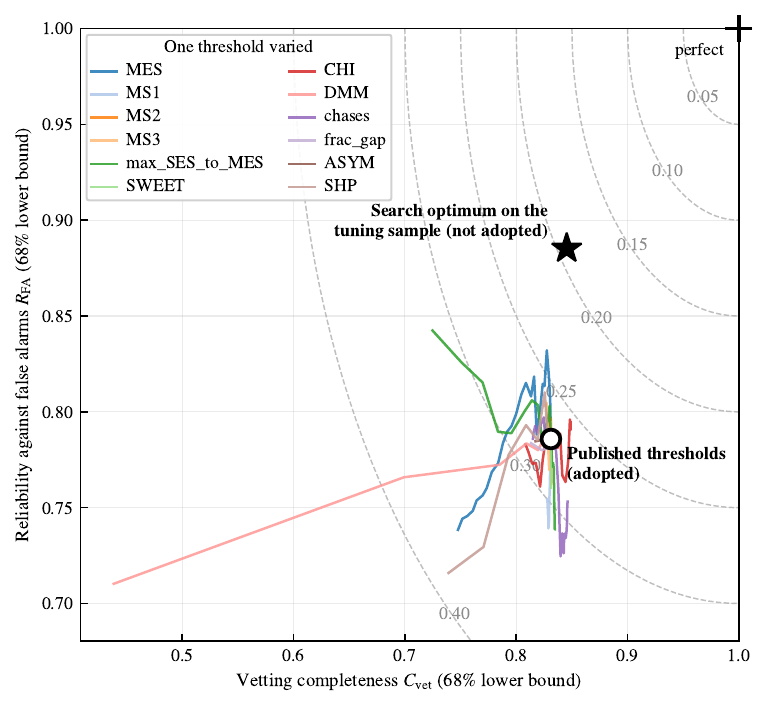}
\caption{Vetting completeness $C_{\rm vet}$ against reliability against false alarms $R_{\rm FA}$ (Equation~\ref{eq:reliability}) for the \leovetter{} flux-level thresholds, both at the lower ends of their 68\% intervals so that a threshold set cannot score well through counting noise alone. Perfect vetting lies at the upper right, and grey arcs mark constant values of the objective $L$ (Equation~\ref{eq:objective}), the distance from that corner. The open circle marks the published thresholds of \citet{Kunimoto:2025}, which we adopt. Each colored track moves one threshold across its search range with the others held at their published values. The star marks the optimum of a differential evolution search over all \NDEparams{} thresholds, scored on the same scrambled sample it was tuned on; on held-out halves of that sample \SplitShrink\% of its gain in $R_{\rm FA}$ disappears, so we do not adopt it (Section~\ref{sec:threshold_opt}).}
\label{fig:ce_plane}
\end{figure}

\begin{figure*}[ht]
\centering
\includegraphics[width=\textwidth]{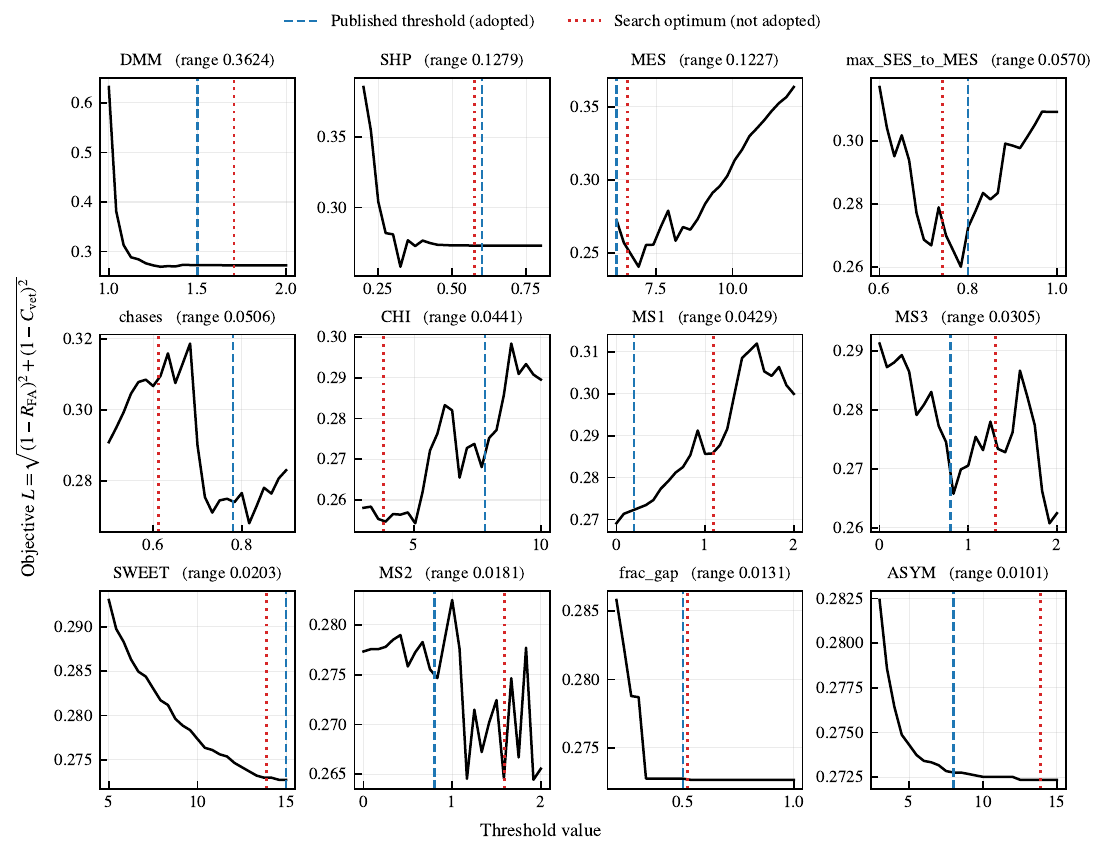}
\caption{The objective $L$ (Equation~\ref{eq:objective}) as each \leovetter{} threshold is moved across its search range with all others held at their published values, the one-at-a-time sweeps of Figure~\ref{fig:ce_plane}. Panels are ordered by the range of $L$ they span. Dashed blue lines mark the published thresholds, which we adopt, and dotted red lines the search optimum of Figure~\ref{fig:ce_plane}, which we do not. Flat curves mark thresholds our data do not constrain; the jagged structure in the others arises from individual scrambled false alarms entering or leaving the surviving set.}
\label{fig:threshold_scan}
\end{figure*}

Figures~\ref{fig:ce_plane} and \ref{fig:threshold_scan} show the threshold search of Section~\ref{sec:threshold_opt} in detail. Figure~\ref{fig:ce_plane} places each set of thresholds in the plane of vetting completeness $C_{\rm vet}$ and reliability against false alarms $R_{\rm FA}$, both at their 68\% lower bounds, where the objective $L$ of Equation~\ref{eq:objective} is the distance from perfect vetting. Moving one threshold at a time away from the published values traces the colored
tracks, and the star marks the optimum of the full differential evolution
search. Figure~\ref{fig:threshold_scan} shows the same sweeps as $L$ against
each threshold's value. Thresholds with flat curves are not constrained by our
data, and the jagged structure in the others arises from individual scrambled
false alarms entering or leaving the surviving set, of which there are
\NScrLeak{} in total. The optimum's apparent gain is measured on the same
scrambled sample it was tuned on; on held-out halves of that sample
\SplitShrink\% of the gain disappears (Section~\ref{sec:threshold_opt}),
which is why we retain the published thresholds.

 % Flush all remaining appendix figures before references
\clearpage

\bibliographystyle{aasjournalv7}
\bibliography{main}

%% This command is needed to show the entire author+affiliation list when
%% the collaboration and author truncation commands are used.  It has to
%% go at the end of the manuscript.
%\allauthors

%% Include this line if you are using the \added, \replaced, \deleted
%% commands to see a summary list of all changes at the end of the article.
%\listofchanges

\end{document}